\documentclass[
  aps,prl,reprint,superscriptaddress,
  longbibliography,nofootinbib,floatfix
]{revtex4-2}

\newcommand{\papertitle}{Gaussian Is All You Need: R\'enyi Optimizers and Optical Channel Simulation}

\usepackage[T1]{fontenc}
\usepackage{amsmath,amssymb,amsthm}
\usepackage{txfonts}
\usepackage{mathrsfs}
\usepackage{bm}
\usepackage{graphicx}
\usepackage[expansion=false]{microtype}
\usepackage[colorlinks=true,allcolors=blue]{hyperref}
\usepackage{orcidlink}
\hypersetup{
 pdftitle=\papertitle,
 pdfauthor={Yuan Gao and Gerardo Adesso}
}

\newcommand{\Tr}{\operatorname{Tr}}
\newcommand{\id}{\mathbb I}
\newcommand{\dd}{\mathrm d}
\newcommand{\Sep}{\mathrm{Sep}}
\newcommand{\SN}{\mathrm{SN}}
\newcommand{\Fid}{\mathsf F}
\newcommand{\ket}[1]{|#1\rangle}

\newcommand{\norm}[1]{\left\|#1\right\|}
\newcommand{\cI}{\mathcal I}
\newcommand{\ensnorm}{\Upsilon}
\DeclareMathOperator{\atanh}{atanh}

\newtheorem{theorem}{Theorem}
\newtheorem{lemma}[theorem]{Lemma}
\newtheorem{corollary}[theorem]{Corollary}

\begin{document}

\title{\papertitle}

\author{Yuan Gao}
\email{willgao@alumni.harvard.edu}
\affiliation{First Principle Science, San Francisco, California, USA}

\author{Gerardo Adesso\orcidlink{0000-0001-7136-3755}}
\email{gerardo.adesso@nottingham.ac.uk}
\affiliation{Algorithmiq S.r.L., Via della Chiusa 15, 20123 Milano, Italy}
\affiliation{
School of Mathematical Sciences and Centre for Mathematics and
Theoretical Physics of Quantum Non-Equilibrium Systems,
University of Nottingham, University Park,
Nottingham NG7 2RD, United Kingdom
}

\date{November 19, 2026}

\begin{abstract}
Preparing an entangled state faithfully requires an amount of
resource given by its entanglement cost. With less entanglement,
the best achievable fidelity is governed by the R\'enyi entanglement
of formation of orders above one. We prove that, for any finite-energy
multimode bosonic Gaussian state, all these quantities are additive and attained by a
Gaussian pure-state decomposition. It follows that the closest separable state
in fidelity to any Gaussian state is Gaussian, and that the
strong converse exponent of entanglement dilution, which
sets how fast the optimal fidelity decays below the cost, is exactly
computable. For channels, we prove that pure
Gaussian inputs minimize the output R\'enyi entropies  of every multimode Gaussian channel at all orders
$\alpha\ge1$, and we obtain the optimal fidelity
of simulating pure-loss and quantum-limited amplifier channels with
finite entanglement, for any number of uses.
\end{abstract}

\maketitle

Entanglement is the prime resource that quantum technologies consume,
and like any resource it has a price. The \emph{entanglement cost} of a
bipartite state counts the maximally entangled qubit pairs, or ebits,
per copy that two distant laboratories must spend to prepare it by
local operations and classical communication (LOCC), with vanishing
error over many copies, a task known as entanglement dilution
\cite{BennettConcentration,BennettFormation}. In general, the cost
equals the regularized entanglement of formation
\cite{HaydenHorodeckiTerhal,Yamasaki}: one must optimize over every
decomposition of the state into pure states, and over many copies at
once. For Gaussian states, which describe squeezed light, lossy fibers
and thermal oscillators in continuous-variable optics
\cite{Weedbrook,SerafiniBook,ARLReview},
two questions stood in the way: whether non-Gaussian decompositions
could ever beat Gaussian ones, and whether the entanglement of
formation is additive \cite{Giedke,Wolf}. Both have recently been
settled. Gaussian decompositions are optimal, as proven first for symmetric
\cite{Giedke} and general \cite{AdessoEoF} two-mode states and then
for arbitrary multimode states \cite{GA}; 
the entanglement of formation is moreover strongly superadditive and
therefore equals the entanglement cost \cite{AdessoEoF,GA}. The price of Gaussian
entanglement is \emph{de facto} now known.

However, the entanglement cost answers an idealized question: it
prescribes the entanglement needed for asymptotically perfect
preparation, but says nothing about what can be achieved with less.
That is the situation of any real device, from a teleportation link
with finite squeezing \cite{LiuzzoScorpoTeleportation} to a source
whose entanglement must be certified against the best separable
impostor \cite{GuhneWitness,AolitaWitness}. When dilution
runs at a rate below the cost, or in the one-shot regime
\cite{BuscemiDatta}, the figure of merit is the achievable \emph{fidelity},
which is governed by the R\'enyi
entropies of order $\alpha>1$ \cite{TheurerFangGour,HayashiExponents}. Each order defines its own
variant of the entanglement of formation, with its own optimization problem. In the limit $\alpha\to\infty$ it
yields the largest fidelity with a separable state, which determines the
geometric entanglement \cite{Shimony,WeiGoldbart,Streltsov}. Nothing guarantees that the decomposition optimal for the cost ($\alpha \rightarrow 1$)
remains optimal at other orders. 

Gaussian optimality for R\'enyi-$\alpha$ entanglement of Gaussian states has been conjectured \cite{ARLReview,LamiThesis}, yet even for the simplest case of two-mode symmetric Gaussian states \cite{MarianBures} this problem has remained open. 
The companion problem for
channels, whether Gaussian inputs minimize the output R\'enyi entropies
of an arbitrary Gaussian channel, is a case of Holevo's \emph{Gaussian
optimizers} conjecture \cite{GiovannettiConjecture,HolevoReview}. It has been established for
gauge-covariant and contravariant channels \cite{GHGP,MGH,GHM}, and
hence for single-mode channels reducible to them by Gaussian unitaries
\cite{HolevoOneMode}, but it has remained open for general multimode
channels \cite{HolevoNorm}. For Gaussian
systems, the whole regime below the entanglement cost, which is where
experiments actually operate, has thus remained uncharted.

In this Letter we conquer such territory conclusively. We prove that for every finite-energy
Gaussian state of an arbitrary finite number of modes on either side,
and at every
R\'enyi order $\alpha>1$ including $\alpha=\infty$, the optimal
decomposition is Gaussian. It consists of random local displacements of
a pure Gaussian state, which depends on the order $\alpha$, and the resulting R\'enyi entanglement of
formation measures are additive on tensor products
(Theorem~\ref{thm:formation}). The proof rests on a new
Gaussian optimality theorem for the Schatten norms of the Schmidt
coefficients of bipartite vectors. The same result shows that pure Gaussian
inputs are optimal, additively, for the minimum output R\'enyi entropies of every
finite-mode Gaussian channel at all orders $\alpha\ge1$, including multimode channels with
correlated phase-sensitive noise (Corollary~\ref{cor:channels}).

The consequences reach well beyond the optimization itself. First, the
closest separable state in fidelity to any Gaussian state can always be chosen Gaussian, obtained by
removing the squeezing from the optimal pure state. This proves the
conjecture in \cite{MarianBures}, extends it to any multimode bipartition, and
yields an efficiently computable entanglement-certification threshold, robust to non-Gaussian imperfections.
Second, we obtain the exact strong converse exponent of entanglement
dilution, the rate at which the optimal fidelity decays when the
entanglement supplied per copy is below the cost. 
Third, we find the exact optimal fidelity of
simulating $n$ uses of pure-loss and
quantum-limited amplifier channels with an entangled resource of any Schmidt rank, against
adaptive tests: it equals the total probability of the $M$ most likely photon-number
strings of the teleportation resource~\cite{BraunsteinKimble,PLOB}, refining the asymptotic simulation cost
of Ref.~\cite{WildeCost}. 

\emph{How to prepare a Gaussian state.}---A Gaussian state is
specified by first and second moments of its canonical position and
momentum operators \cite{Weedbrook,SerafiniBook}. Let $d$ be the total number of modes
shared by $A$ and $B$. In units $\hbar=1$ their quadratures
$\boldsymbol R=(\hat q_1,\ldots,\hat q_d,\hat p_1,\ldots,\hat p_d)^{\top}$
obey $[R_j,R_k]=i\Omega_{jk}$, where
$\Omega=\left(\begin{smallmatrix}0&\id_d\\-\id_d&0\end{smallmatrix}\right)$
and $\id_d$ is the $d$-dimensional identity. First moments can be
set to zero by local displacements, which do not affect entanglement or entropy, so we generally take all
states centered throughout. The covariance matrix $V$ of a centered Gaussian state
$\rho_V$ has entries $V_{jk}=\Tr[\rho_V(R_jR_k+R_kR_j)]$ and has to obey the uncertainty relation
 $V+i\Omega\ge0$ \cite{SimonMukundaDutta,SerafiniBook}. By Williamson's
theorem, a symplectic transformation brings $V$ to a diagonal form
whose entries, the symplectic eigenvalues $\nu_j$, obey $\nu_j \geq 1$  for $j=1,\ldots,d$
\cite{Williamson,SerafiniBook,ARLReview}. We call a covariance matrix $\Gamma$ pure when $\Gamma\Omega \Gamma=\Omega$, meaning that the corresponding Gaussian state $\psi_\Gamma$ is a pure state, with all $\nu_j=1$.

Choose any pure $\Gamma\le V$, where the inequality means that
$V-\Gamma$ is positive semidefinite. Alice and Bob prepare the
corresponding pure seed state $\psi_\Gamma$, sample a joint Gaussian vector of quadrature shifts with
covariance $(V-\Gamma)/2$, and displace their own modes by their
respective components. Averaging over the displacements gives exactly a preparation of
$\rho_V$  \cite{Wolf}. Since entanglement is invariant under local displacements, an entire ensemble has the entanglement of the single
pure state $\psi_\Gamma$, which we call its seed.  Minimizing the marginal R\'enyi entropy of order $\alpha$ over these Gaussian
preparations defines the Gaussian R\'enyi-$\alpha$ entanglement of formation
\cite{AGS,Wolf},
\begin{equation}
 G_\alpha(V)=\min_{\substack{\Gamma\le V\\ \Gamma\ {\rm pure}}}
            S_\alpha(\psi_{\Gamma,A}),
 \label{eq:Gaussianroof}
\end{equation}
where $\psi$ denotes a pure-state projector, with
$\psi_A=\Tr_B\psi$ its reduced state on $A$. Here the R\'enyi entropy
\cite{Renyi} for $1<\alpha<\infty$ is $S_\alpha(\omega)=(1-\alpha)^{-1}\log\Tr\omega^\alpha$, while
$S_\infty(\omega)=-\log\|\omega\|_\infty$ amounts to the min-entropy; the same minimization of Eq.~\eqref{eq:Gaussianroof} with the von Neumann entropy $S_1(\omega)=-\Tr(\omega\log
\omega)$ defines $G_1(V)$ as in Refs.~\cite{Wolf,AdessoEoF,GA}. We use natural logarithms, so entropies and rates are in nats; one ebit is $\log2$ nats.

\emph{Gaussian optimality.}---Every Gaussian preparation in
Eq.~\eqref{eq:Gaussianroof} is one way of writing $\rho_V$ as a
mixture of pure states. In principle, a mixture that includes
non-Gaussian components could prepare the same state more economically. To define the unrestricted R\'enyi entanglement of formation, we must therefore optimize over all pure-state ensembles $\mu$ with average
$\rho=\int\psi\,\dd\mu$. This can be done in two ways. The conventional convex roof minimizes the
mean R\'enyi entropy of the components
\cite{BennettFormation,VidalMonotones,KimSanders}. The conditional definition
instead uses the R\'enyi entropy of $A$ conditioned on the classical
label of the component, which amounts to averaging the norms of the
marginals before taking the logarithm \cite{SBW}: 
\begin{align}
 &E^{\rm av}_{F,\alpha}(\rho)=\inf_\mu\int S_\alpha(\psi_A)\,\dd\mu ,
 \label{eq:arithroof}\\
 &E^{\rm c}_{F,\alpha}(\rho)=-t_\alpha^{-1}\log\ensnorm_\alpha(\rho),\quad
   \ensnorm_\alpha(\rho)=\sup_\mu\int\|\psi_A\|_\alpha\,\dd\mu .
 \label{eq:conditionalroof}
\end{align}
Here $\|\omega\|_\alpha=(\Tr\omega^\alpha)^{1/\alpha}$ at finite
$\alpha$ and $t_\alpha=1-1/\alpha$, with  $t_\infty=1$, so that $S_\alpha(\omega)=-t_\alpha^{-1}\log\|\omega\|_\alpha$
for all $\alpha\in\cI \equiv (1,\infty]$. By concavity of the logarithm,
$E^{\rm c}_{F,\alpha}\le E^{\rm av}_{F,\alpha}$ for every state, and the Gaussian preparations give
$E^{\rm av}_{F,\alpha}(\rho_V)\le G_\alpha(V)$.
We now show that both inequalities are saturated for Gaussian states.

\begin{theorem}[Gaussian optimality of R\'enyi entanglement]\label{thm:formation}
For every finite-energy bipartite Gaussian state $\rho_V$ and all
$\alpha\in\cI$,
\begin{equation}
 E^{\rm av}_{F,\alpha}(\rho_V)=E^{\rm c}_{F,\alpha}(\rho_V)=G_\alpha(V).
 \label{eq:mainformation}
\end{equation}
Furthermore, the quantity in Eq.~\eqref{eq:mainformation} is additive on product states
and strongly superadditive across groups of modes.
\end{theorem}

The proof is  given in the Supplemental Material \cite{SM}.%
\nocite{Lieb,Bargmann,HaydenWinter,FrankLieb,DPT,Banchi,StreltsovAlgorithm,HouQi,Sion,Nielsen,PeschelEisler,CampoParentaniEntropy08,SteaneIshizaka2026,WolfExtremality,ChenWitness,KuboAndo,AndoHiai,HolsteinPrimakoff,Kato,Fekete,BerryEsseen,Esseen}
Additivity holds even against ensembles that correlate the copies,
and strong superadditivity means that a correlated target on
$A_1A_2{:}B_1B_2$ is worth at least the sum of its $A_1{:}B_1$ and
$A_2{:}B_2$ marginals. At $\alpha=2$, properties established for the Gaussian
minimum $G_2$, namely monogamy, equality with the Gaussian R\'enyi-2 squashed
entanglement \cite{AGS,Lami16,Lami17}, and, for normal states including all
two-mode states, equality with the Gaussian intrinsic entanglement
\cite{LamiMistaAdesso}, now hold, on Gaussian states, for the unrestricted R\'enyi-$2$ entanglement of formation.

The mechanism behind Theorem~\ref{thm:formation} is a norm
inequality valid for every pure state (Gaussian or not),
Eq.~\eqref{eq:weightedbound} in the End Matter. The covariance
minimization yields a positive Gaussian operator $K$ with
$\Tr(\rho_VK)=1$, whose expectation in any pure state bounds the norm
of its reduced state. Every ensemble of the target obeys the same
bound on average, and the Gaussian preparation saturates it.

The same norm inequality settles the minimum output R\'enyi entropy of
Gaussian channels. Every finite-mode Gaussian channel $\mathcal N$ has a
Gaussian Stinespring isometry with a pure Gaussian environment
\cite{GaussianDilation}, so the output spectrum of a pure input is the
Schmidt spectrum of a bipartite output--environment vector, to which
the inequality applies \cite{SM}.

\begin{corollary}[Gaussian optimizers]\label{cor:channels}
For every Gaussian channel $\mathcal N$ with finitely many modes and all
$\alpha\in[1,\infty]$,
\begin{equation}
 \inf_{\rho}S_\alpha[\mathcal N(\rho)]
 =\inf_{\psi\ \mathrm{pure\ Gaussian}}S_\alpha[\mathcal N(\psi)],
 \label{eq:minoutput}
\end{equation}
and this minimum output R\'enyi entropy $S^{\min}_\alpha(\mathcal N)$ is
additive, $S^{\min}_\alpha(\mathcal N_1\otimes\mathcal N_2)=S^{\min}_\alpha(\mathcal N_1)+S^{\min}_\alpha(\mathcal N_2)$.
\end{corollary}

For $\alpha>1$, equivalently, the $1\to\alpha$ norms of all Gaussian channels are
attained or approached by Gaussian inputs and are multiplicative. This was previously known for gauge-covariant and contravariant
channels \cite{GHGP,MGH,GHM}, for specific noise channels at integer
orders \cite{GL04}, and within Gaussian inputs \cite{SEW}; a related
one-shot single-mode communication task was treated in
Ref.~\cite{JiOneShot}. Here it holds for every multimode Gaussian channel, including those
with correlated phase-sensitive noise that cannot be reduced by Gaussian unitaries to a
phase-insensitive form \cite{CarusoMultimode}, and with no constraint
on the input energy. 

The von Neumann case $\alpha=1$ follows by a limit. An $m$-mode
Gaussian output is unitarily equivalent to a product of thermal states,
which gives $0\le S_1-S_\alpha\le m[1-\log\alpha/(\alpha-1)]
\xrightarrow[\alpha\to1]{}0$. Since $S_1\ge S_\alpha$ for every output,
Eq.~\eqref{eq:minoutput} and additivity extend to $\alpha=1$
\cite{SM}, generalizing the finite-energy bound from \cite{GA} to unrestricted inputs. This settles the
minimum-output-entropy case of the Gaussian optimizers conjecture
\cite{GiovannettiConjecture,HolevoReview} at all orders $\alpha\ge1$.

\emph{The closest separable state.}---How well can a Gaussian state
$\rho_V$ be imitated without entanglement? We measure closeness by the
squared Uhlmann fidelity $\Fid(\rho,\sigma)=\|\sqrt\rho\sqrt\sigma\|_1^2$,
where $\|\cdot\|_1$ is the trace norm \cite{Uhlmann,Jozsa}, and define
the separable fidelity $F_{\Sep}(\rho)=\sup_{\sigma\in\Sep}\Fid(\rho,\sigma)$,
where $\Sep$ is the set of separable states. 
The $\alpha=\infty$ case of
Theorem~\ref{thm:formation} determines $F_{\Sep}(\rho_V)$ and the closest separable state for every Gaussian  $\rho_V$.

Consider first a pure state with Schmidt decomposition
$|\psi\rangle=\sum_k\sqrt{p_k}\,|a_k\rangle|b_k\rangle$, ordered so
that $p_0\ge p_1\ge\dots$. Its closest product state is the leading
term $|a_0\rangle|b_0\rangle$, with fidelity
$p_0=\|\psi_A\|_\infty$ \cite{Shimony,WeiGoldbart}. Local Gaussian
unitaries bring every pure Gaussian state to a product of two-mode
squeezed vacuum states, one for each entangled pair of modes, and local vacua
\cite{BoteroReznik}. A two-mode squeezed vacuum has Schmidt
probabilities $(1-\theta)\theta^k$, $0\le\theta<1$, in the photon-number basis
$|k\rangle|k\rangle$, so its leading term is the two-mode vacuum. Let $\Gamma_\infty$ attain the minimum in Eq.~\eqref{eq:Gaussianroof} at $\alpha=\infty$. The closest product state to the optimal seed $\psi_{\Gamma_\infty}$ is
thus the local vacuum $\phi$ in these coordinates; let
$\Gamma_\phi$ be its covariance in the original coordinates. Applying to $\phi$
the same random displacements that prepare $\rho_V$ from the seed $\psi_{\Gamma_\infty}$ gives
a separable Gaussian state $\sigma_*$ with covariance
\begin{equation}
 V_{\sigma_*}=V-\Gamma_\infty+\Gamma_\phi .
 \label{eq:separablecovariance}
\end{equation}
Each displaced pair of seed and product state has fidelity
$\|\psi_{\Gamma_\infty,A}\|_\infty=e^{-G_\infty(V)}$, and the root
fidelity is jointly concave, so $\Fid(\rho_V,\sigma_*)\ge e^{-G_\infty(V)}$.
Conversely, the separable fidelity of any state is the largest
ensemble average of the leading Schmidt probability
\cite{Streltsov,WeiGoldbart,SM}, which is $\ensnorm_\infty$ in
Eq.~\eqref{eq:conditionalroof}. By Theorem~\ref{thm:formation} it
equals $e^{-G_\infty(V)}$ for $\rho_V$, so the two bounds meet:

\begin{corollary}[Closest separable state]\label{cor:closest}
For every finite-energy Gaussian state $\rho_V$, across any
bipartition,
\begin{equation}
 F_{\Sep}(\rho_V)=\Fid(\rho_V,\sigma_*)=e^{-G_\infty(V)} ,
 \label{eq:separablefidelity}
\end{equation}
so the Gaussian state $\sigma_*$ of
Eq.~\eqref{eq:separablecovariance} is a closest separable state.
\end{corollary}

This determines every entanglement
measure built on the separable fidelity, including the geometric
entanglement $1-F_{\Sep}$
\cite{Shimony,BarnumLinden,WeiGoldbart,Streltsov} and the Bures
distance of entanglement \cite{VedralPlenio}, and proves, for every
Gaussian state and multimode cut, a conjecture first raised
for symmetric two-mode states \cite{MarianBures}. This instance of Gaussian optimality is far from obvious. For the relative entropy of entanglement,
the closest separable state to a two-mode squeezed vacuum is
not Gaussian \cite{VedralPlenio}, and Gaussian-restricted values
differ from unrestricted ones \cite{MishraWilde}. Likewise, Gaussian
measurements are in general not optimal for distinguishing Gaussian
states \cite{OhFidelityMeasurement,TurnerDiscrimination}.

The separable fidelity also sets an experimental threshold: any source
whose output has fidelity above $F_{\Sep}(\rho_V)$ with the target is
entangled. The threshold is computable from $V$ alone. The largest
eigenvalue of a Gaussian state is its joint vacuum probability in
Williamson coordinates, $\prod_j2/(1+\nu_j)$. Relaxing the purity of
$\Gamma$ to the uncertainty relation does not change the minimum
\cite{SM}, so with $\mathcal C(V)$ the set of real symmetric $X$
satisfying $X+i\Omega\ge0$ and $X\le V$, Eq.~\eqref{eq:Gaussianroof}
at $\alpha=\infty$ becomes
\begin{equation}
 -\log F_{\Sep}(\rho_V)
    =\min_{X\in\mathcal C(V)}\sum_j\log\frac{1+\nu_j(X_A)}2 ,
 \label{eq:fidelityprogram}
\end{equation}
where $\nu_j(X_A)$ are the symplectic eigenvalues of the reduced block of $X$
on $A$.
Both constraints are linear matrix inequalities; a sequence of linear
semidefinite programs converges to the global minimum and bounds the
remaining error \cite{SM}.

The same threshold certifies an imperfect source. If a source emits a
state $\omega$ with fidelity $f>c=F_{\Sep}(\rho_V)$ to the target, the
triangle inequality for the Bures angle gives
\begin{equation}
 1-F_{\Sep}(\omega)\ \ge\
 \bigl(\sqrt{f(1-c)}-\sqrt{c(1-f)}\bigr)^2 ,
 \label{eq:imperfectfidelity}
\end{equation}
tight given $f$ and $c$ alone \cite{Kondra,SM}. Corollary~\ref{cor:closest}
makes this certificate exact for every Gaussian target, whatever
non-Gaussian noise the source adds.

Since $G_\infty$ is additive (Theorem~\ref{thm:formation}), it follows that the
separable fidelity of Gaussian states is multiplicative,
\begin{equation}
 F_{\Sep}(\rho_V\otimes\rho_{V'})=F_{\Sep}(\rho_V)F_{\Sep}(\rho_{V'}),
 \label{eq:fidproduct}
\end{equation}
so correlated separable competitors offer no advantage, unlike for
general mixed states \cite{ZhuAdditivity}. Whenever local transformations factor a target into
independent pairs of modes, its separable fidelity is thus computable
pair by pair, as the End Matter illustrates for a thermal ring of
oscillators.

\emph{Entanglement dilution below the cost.}---Suppose Alice and Bob
now share a maximally entangled state $\Phi_M$ of Schmidt rank $M$ and
can use local operations and classical communication (LOCC). Write
$\|\omega\|_{(M)}$ for the sum of the $M$ largest eigenvalues of $\omega$,
its Ky Fan $M$-norm \cite{BhatiaMatrix}, and $F_V^*(n,M)$ for the optimal dilution fidelity, the best fidelity
with which they can prepare $n$ copies of the target $\rho_V$. For a pure state the best fidelity is
$\|\psi_A\|_{(M)}$, attained by normalizing its first $M$ Schmidt terms
\cite{VidalJonathanNielsen}.
For a mixed target the corresponding optimum is the supremum of that
quantity averaged over its pure-state ensembles
\cite{BowenDatta,BuscemiDatta,TheurerFangGour,SM}. By H\"older's
inequality, the sum of the $M$ largest Schmidt probabilities obeys
$\|\psi_A\|_{(M)}\le M^{t_\alpha}\|\psi_A\|_\alpha$ for every $\alpha\in\cI$.
Averaging over ensembles therefore bounds the fidelity by
$M^{t_\alpha}\ensnorm_\alpha$, the averaged norm of
Eq.~\eqref{eq:conditionalroof}, which Theorem~\ref{thm:formation}
evaluates as $e^{-t_\alpha G_\alpha(V)}$. For $n$ copies additivity
gives
\begin{equation}
 F_V^*(n,M)\le \min\!\left\{1,\inf_{\alpha\in\cI}
               M^{t_\alpha}e^{-nt_\alpha G_\alpha(V)}\right\}.
 \label{eq:dilution}
\end{equation}
For $M_n=\lfloor e^{n\mathcal R}\rfloor$ at rate $\mathcal R\ge0$
nats per copy, the bound above is asymptotically tight \cite{SM},
which gives the exact
single-letter \emph{strong converse exponent} of entanglement dilution for a Gaussian state $\rho_V$,
\begin{equation}
 \mathscr E_V(\mathcal R):=\lim_{n\to\infty}-\tfrac1n\log F_V^*(n,M_n)
       =\sup_{\alpha\in\cI}t_\alpha[G_\alpha(V)-\mathcal R].
 \label{eq:generalexponent}
\end{equation}
It is positive at every rate below the entanglement cost $G_1(V)$ and zero at or above it, so every rate below the cost incurs
an exponential strong converse. At zero rate the exponent is $\mathscr E_V(0)=G_\infty(V)=-\log
F_{\Sep}(\rho_V)$, matching the closest separable state.
Conversely, the exponent curve determines every  $G_\alpha(V)$, through
$t_\alpha G_\alpha(V)=\inf_{\mathcal R\ge0}[\mathscr E_V(\mathcal R)+t_\alpha\mathcal R]$
\cite{SM}.

Equation~\eqref{eq:generalexponent} also specifies an optimal
protocol. At a given rate the parties choose a single pure Gaussian
seed for that rate \cite{SM}, prepare from $\Phi_{M_n}$ the normalized
truncation of its $n$th tensor power to the $M_n$ most likely Schmidt
strings, generally a non-Gaussian state, and apply the original
random displacements. If the
seed has Schmidt probabilities $p_i$, the retained strings are typical
of the distribution tilted to order $\alpha$, proportional to $p_i^\alpha$
\cite{HayashiExponents}, whose entropy equals the rate at an
interior maximizing order. Rates thus select their own order and, in general, their own seed: the cheapest high-fidelity preparation and
the closest separable approximation of a state can be
physically different (see the End Matter).

\emph{Simulating optical links with limited resources.}---Replacing a physical channel by
shared entanglement and classical communication underlies
teleportation-based networks and the converse bounds on what any
channel can transmit
\cite{BennettReverseShannon,BertaChannelCost,WildeCost,PLOB}. To
simulate a channel $\mathcal N$, Alice receives an unknown input and
Bob must return the corresponding output, using LOCC and a shared
maximally entangled state $\Phi_M$. Unlike teleportation benchmarks,
which average the fidelity over a fixed set of inputs
\cite{HammererBenchmark,NamikiBenchmark,LiuzzoScorpoTeleportation},
we let a tester choose inputs of any energy, keep a reference entangled
with them, and adapt each input to earlier outputs
\cite{ChiribellaCombs}. Let $F^{\rm sim}_{n,M}(\mathcal N)$ be the
largest squared fidelity, guaranteed for every such test of $n$ uses,
between the tester's final states with the simulator and with
$\mathcal N$.

An optical fiber coupled to vacuum implements the pure-loss channel
$\mathcal L_\eta$ with transmissivity $0<\eta<1$. Its counterpart in
amplification is the quantum-limited amplifier $\mathcal A_\kappa$ of
intensity gain $\kappa>1$ \cite{HolevoOneMode,Weedbrook}. Both can be
realized by continuous-variable teleportation with the appropriate gain
and a two-mode squeezed resource
\cite{BraunsteinKimble,LiuzzoScorpoTeleportation,LaurenzaFinite,WildeCost}.
The number of photons in either arm of that resource has the
thermal distribution
\begin{equation}
 \tau_\theta=(1-\theta)\sum_{k=0}^\infty\theta^k|k\rangle\langle k|,
 \qquad \theta=\eta\ \text{or}\ 1/\kappa .
 \label{eq:thermaltheta}
\end{equation}
These probabilities, which govern the preparation of the resource,
also fix the best possible \emph{simulation} of the channel.

\begin{theorem}[Optical channel simulation]\label{thm:simulation}
For all positive integers $n$ and $M$, the pure-loss channel
$\mathcal L_\eta$ and the quantum-limited amplifier $\mathcal A_\kappa$
have optimal simulation fidelities
\begin{align}
 F^{\rm sim}_{n,M}(\mathcal L_\eta)=\|\tau_\eta^{\otimes n}\|_{(M)},\qquad
 F^{\rm sim}_{n,M}(\mathcal A_\kappa)=\|\tau_{1/\kappa}^{\otimes n}\|_{(M)}.
 \label{eq:exactsimulation}
\end{align}
\end{theorem}

We outline the proof here and give it
in full in the Supplemental Material \cite{SM}. To attain the optimum,
list the photon-number strings $(k_1,\ldots,k_n)$ of the $n$ ideal
resources in order of decreasing probability, keep the first $M$ terms
of their superposition, normalize, and use this state in the
teleportation circuit. Its squared overlap with the ideal
resource is $\|\tau_\theta^{\otimes n}\|_{(M)}$, and since the same circuit and test act on either
resource, the output fidelity is at least this large. For the
converse, feed half of a squeezed pair into each channel use. As the
probes brighten, amplifier output majorization bounds the fidelity of
every simulator supplied with $\Phi_M$ by the same sum
\cite{MGH,GHGP,GHM}. The optimal fidelity is thus a finite sum over
total photon numbers, although the rank-$M$ state that attains it is
generally non-Gaussian and keeps coherence among the retained strings.

\begin{figure}[tb]
 \centering
 \includegraphics[width=\columnwidth]{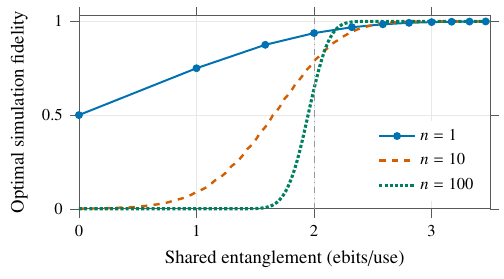}
 \caption{\label{fig:simulation}
 Simulating a $50\%$ loss channel with a given amount of shared
 entanglement per use. The one-use fidelity (blue, solid) increases
 with Schmidt rank. Its symbols mark integer ranks, with a line to guide
 the eye. Pooling the resource over ten uses
 (orange, dashed) or one hundred (green, dotted) sharpens the
 transition near two ebits per use. The horizontal coordinate is $(\log_2 M)/n$.}
\end{figure}

For one use the geometric series gives
\begin{equation}
 F^{\rm sim}_{1,M}(\mathcal L_\eta)=1-\eta^M,\qquad
 F^{\rm sim}_{1,M}(\mathcal A_\kappa)=1-\kappa^{-M}.
 \label{eq:oneuse}
\end{equation}
For two uses even a small resource is better spent jointly. Take a
$50\%$ loss channel and two maximally entangled qutrit pairs. Separate
simulations retain photon numbers $0,1,2$ in each use, with fidelity
$7/8$ each and $49/64$ jointly. Used together, the same resource can
keep any nine strings: replacing $(2,2)$ by $(3,0)$, which has one
fewer photon and is twice as probable, reaches the optimum $50/64$.
Figure~\ref{fig:simulation} shows how this advantage grows with the
number of uses.

Two familiar rates appear at the ends of this tradeoff. With no shared
entanglement only the all-vacuum string is retained, and the simulator
is an entanglement-breaking (measure-and-prepare) channel
\cite{HorodeckiShorRuskai}: Alice heterodynes each input and communicates her outcome $\beta$ to Bob, who
prepares a coherent state of amplitude $\sqrt\eta\,\beta$ for loss, or
$\sqrt\kappa\,\beta$ for amplification. Its
fidelity for either channel
\begin{equation}
 F^{\rm sim}_{n,1}=(1-\theta)^n=2^{-nK_{\leftrightarrow}}
 \label{eq:classicalchannel}
\end{equation}
decays at the rate $-\log_2(1-\theta)$, the min-entropy of either arm of
the resource, which coincides with the channel's two-way secret-key
capacity $K_{\leftrightarrow}$ in bits per use \cite{PLOB}. For pure
loss tested with a squeezed probe of finite brightness, the best
measure-and-prepare simulator does better, as the End Matter shows.
Pirandola recently used teleportation simulation restricted to
photon-number sectors to obtain the energy-constrained two-way
capacities of these channels  \cite{PirandolaEnergy} (also independently reported for pure-loss in~\cite{FalcoEnergy}). At the other end, Wilde showed that faithful
simulation needs the entropy of $\tau_\theta$ in shared entanglement
per use \cite{WildeCost}. In between, $F^{\rm sim}_{n,M}$ is the
dilution fidelity of the pure resource, so
Eq.~\eqref{eq:generalexponent} with $G_\alpha$ replaced by
$S_\alpha(\tau_\theta)$ gives the exact exponent
\begin{equation}
 \mathscr E_\theta(\mathcal R)=\sup_{\alpha\in\cI}
             t_\alpha[S_\alpha(\tau_\theta)-\mathcal R]
 \label{eq:exactexponent}
\end{equation}
at $M_n=\lfloor e^{n\mathcal R}\rfloor$. For a half-transmitting fiber
the resource has photon statistics $p_k=2^{-(k+1)}$, one photon per arm
on average. Each use halves the best classical fidelity ($p_0=1/2$),
and the entropy $\sum_kp_k(k+1)\log2=2\log2$ nats (two ebits) sets the
asymptotic cost at two ebits per use. Figure~\ref{fig:simulation} shows this threshold emerging
from finite blocks: at exactly two ebits per use ($M=4^n$) the fidelity
tends to $1/2$, above it to $1$, and the transition narrows as
$1/\sqrt n$, set by the fluctuations of the total photon number
\cite{HayashiSecondOrder,SM}.

\emph{Discussion.}---We have shown that, below the entanglement cost,
Gaussian states have Gaussian optimal decompositions at every R\'enyi order,
with additive optimal values, and that Gaussian inputs minimize the output
R\'enyi entropies of every Gaussian channel for $\alpha\ge1$, settling the
minimum-output-entropy case of a longstanding conjecture
\cite{GiovannettiConjecture,HolevoReview}. 
These results determine the closest separable state to every Gaussian state
and the exact decay rate of the
optimal dilution fidelity, both computable
from the covariance matrix alone. We have also established the optimal simulation of the two most prominent optical communication channels, namely pure-loss and quantum-limited amplifiers, with
finite entanglement.

The separable fidelity is a certification threshold robust to
non-Gaussian imperfections of the source [Eq.~\eqref{eq:imperfectfidelity}].
Inferring the entanglement of an imperfect source from its covariance
alone requires more care. At order $\alpha=1$, the Gaussian value lower-bounds
the entanglement of formation of every state with the same covariance
\cite{AdessoEoF,GA}, but at each order $\alpha>1$ there are non-Gaussian
states with the same covariance on opposite sides of the Gaussian R\'enyi-$\alpha$ value
\cite{SM}. One can nevertheless provide quantitative witnesses that bound the separable
fidelity and all R\'enyi entanglements of formation of a general state
from second moments alone \cite{WernerWolfBound,HyllusEisert,SM}.

Teleportation simulation underlies the ultimate limits of quantum communication  \cite{PLOB,Azuma}. 
Pure-loss and quantum-limited amplifier channels can be simulated
exactly with finite squeezing, but only with a resource of unbounded
Schmidt rank
\cite{LiuzzoScorpoTeleportation,KaurWilde,LaurenzaFinite,LaurenzaTight}.
Our results reveal how close we can get with finitely many ebits:
the optimal resource is squeezed light truncated in photon number,
which may be prepared by quantum scissors  \cite{PeggScissors,WinnelScissors},
paving the way to practical finite-entanglement experimental simulations of optical links. 
Determining the corresponding optimal fidelity for thermal-loss and general multimode
Gaussian channels remains open.

More broadly, Gaussian states are where quantum optics, many-body
physics, gravity and cosmology meet \cite{Weedbrook,ARLReview,SerafiniBook,AgulloHawking22,MartinVenninFields21}, and our
results equip these areas with an exact and computable account of how much
entanglement their systems hold and how closely an entanglement-free world can
imitate them. In cosmology, this measures how faithfully a separable state could reproduce the primordial fluctuations, which are born as
squeezed pairs
\cite{MartinVenninDiscord,BelfiglioCosmology25,SM}. In holography, the
entanglement of purification, a proposed dual of the entanglement
wedge cross section \cite{TakayanagiUmemotoPurification,NguyenPurification}, is computed
for free fields under the assumption that Gaussian purifications
suffice \cite{BhattacharyyaTakayanagiUmemoto,CamargoPurification,WindtGaussian}. We have proved the analogous statement for the entanglement of
formation, which lower bounds the entanglement of purification
\cite{TerhalPurification}. Whether our methods can now settle the
purification conjecture is an exciting open question.

\begin{acknowledgments}
\section*{ACKNOWLEDGMENTS}
This work was supported by the Engineering and Physical
Sciences Research Council (EPSRC) Grant No.~UKRI6536. AI tools based on OpenAI and Anthropic
models assisted with mathematical development, literature
review, and manuscript
preparation, under the authors' creative direction.
\end{acknowledgments}

\setcounter{equation}{0}
\renewcommand{\theequation}{E\arabic{equation}}
\renewcommand{\theHequation}{E\arabic{equation}}
\section*{End Matter}

\emph{Geometry of Gaussian optimality.}---An ordinary mixture of
two pure states generally loses purity. Feasible pure covariances
$X,Y\le V$ can instead be joined through pure covariances by their
matrix geometric mean,
\begin{equation}
 X\#_uY=X^{1/2}(X^{-1/2}YX^{-1/2})^uX^{1/2},\qquad 0\le u\le1 .
 \label{eq:geometricmean}
\end{equation}
Every point on this path still lies below $V$, and the R\'enyi-$\alpha$
entropy of the seed's marginal on $A$ is convex along it \cite{SM}, extending the
$\alpha=1$ geodesic convexity of Ref.~\cite{GA}; related
covariance-geometric methods appear in Refs.~\cite{Lami17,MishraWilde}.
As a result the optimized $G_\alpha$ is convex in $V$. At an interior
covariance $V+i\Omega>0$, a supporting plane determines a positive
Gaussian operator $K$ normalized by $\Tr(\rho_VK)=1$. For a pure-state
projector $\psi=|\psi\rangle\langle\psi|$ one has
\begin{equation}
 \|\psi_A\|_\alpha\le e^{-t_\alpha G_\alpha(V)}
                         \|K^{1/2}|\psi\rangle\|^2 .
 \label{eq:weightedbound}
\end{equation}
A Gaussian optimality theorem for Schatten norms
(Theorem~\ref{thm:schatten} \cite{SM}), proved with tools from Refs.~\cite{Heinavaara,BeigiMehrabi},
extends this inequality from Gaussian vectors to every vector in the
form domain of $K$; outside it the right-hand side is infinite. Averaging over an arbitrary
ensemble gives $\ensnorm_\alpha(\rho_V)\le e^{-t_\alpha G_\alpha(V)}$.
Jensen's inequality for the negative logarithm gives the same lower
bound $G_\alpha(V)$ on the average entropy. The displaced seed attains
both bounds. Continuity extends the result to covariances on the boundary of the
uncertainty relation, including those of pure Gaussian states.

\begin{figure}[tb]
 \centering
 \includegraphics[width=\columnwidth]{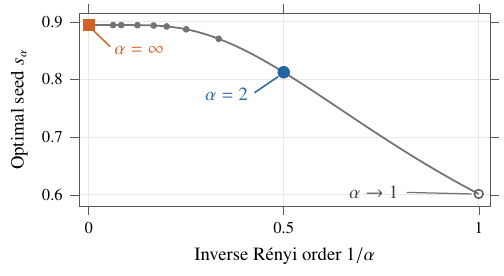}
 \caption{\label{fig:optimizer}
 Fraction $s_\alpha$ of the added position covariance carried by the
 optimal Gaussian seed of a noisy four-oscillator target, as a function
 of the R\'enyi order. Open circle: von Neumann limit; blue circle:
 $\alpha=2$; orange square: $\alpha=\infty$; dots: other integer
 orders.}
\end{figure}

\emph{An order-dependent optimal seed.}---Across a $1{:}m$ cut, where every
pure seed has a single entangled pair of modes, every order favors
the least feasible squeezing, so all orders share one optimal seed
\cite{SM}. With two pairs, reducing the squeezing of one can increase
that of the other, and the orders can disagree. In
Fig.~\ref{fig:optimizer}, two oscillators on each side, in the ground
state of a coupled quadratic Hamiltonian, receive one common random
position kick along a fixed direction. The optimal seeds can then be
taken on a one-parameter path, labeled by the fraction $s_\alpha$ of
the added position covariance they carry \cite{SM}. The von Neumann
cost selects $s_1\simeq0.60$ and the closest separable state
$s_\infty\simeq0.89$, while the order $\alpha=2$, selected at about
$0.36$ nats ($0.53$ ebits) per copy, gives exactly $s_2=13/16$.

\emph{A thermal ring of oscillators.}---Consider $2L$ unit-mass
oscillators on a ring with periodic boundary conditions and Hamiltonian
$H=\frac12\sum_i[p_i^2+\omega_0^2q_i^2+\gamma(q_{i+1}-q_i)^2]$, where
$\omega_0,\gamma>0$. The ring is in a Gibbs state at temperature $T$, and
let $A$ hold the even sites and $B$ the odd ones
\cite{AudenaertHarmonic,AndersWinter}. Separate
normal-mode transformations of the two sublattices are local, and they
split the state into $L$ independent pairs whose two normal modes have squared frequencies
$\omega_{j,\pm}^2=\omega_0^2+2\gamma\pm2\gamma|\cos(\pi j/L)|$. Each pair is a
symmetric two-mode Gaussian state whose optimal seed has squeezing
$r_j=\max\{0,-\frac12\log z_j\}$, where
\begin{equation}
 z_j^2=\frac{\omega_{j,-}}{\omega_{j,+}}
       \coth\frac{\omega_{j,-}}{2T}\coth\frac{\omega_{j,+}}{2T}
 \label{em:ringz}
\end{equation}
in units $\hbar=k_B=1$ \cite{SM}. Additivity and multiplicativity then
give, against all decompositions and all separable competitors,
\begin{equation}
 G_\alpha(V_T)=\sum_jf_\alpha(r_j),\qquad
 F_{\Sep}(\rho_{V_T})=\prod_j\operatorname{sech}^2r_j ,
 \label{em:ringfamily}
\end{equation}
where $f_\alpha(r)$ is the order-$\alpha$ entanglement of a two-mode
squeezed vacuum with squeezing $r$, for instance
$f_\infty(r)=2\log\cosh r$. All orders vanish together at the
temperature $T_*$ where $z_0=1$, the even--odd separability threshold
\cite{CavalcantiThermal}. In the thermodynamic limit $L\to\infty$, taken before
$T\uparrow T_*$, and with $\epsilon=1-T/T_*$, only a fraction of order $\sqrt\epsilon$ of the
pairs remains entangled, each with $r_j=O(\epsilon)$. Hence
$-\log F_{\Sep}$ per pair of sites vanishes as $\epsilon^{5/2}$, every finite
order $\alpha>1$ gives the same power with a coefficient
$\alpha/(\alpha-1)$ times larger, and the entanglement cost acquires
an extra factor that grows as $2|\log\epsilon|$ \cite{SM}. At every $T<T_*$ the best
separable approximation to the whole ring therefore has fidelity
exponentially small in its length.

\emph{A probe of finite brightness.}---Take as probe a two-mode squeezed
pair with mean photon number $N$ per arm. Its photon-number
distribution is geometric, $(1-q)q^k$ with $q=N/(N+1)$. One arm is sent
through the link and the other, $R$, is kept as a reference. Coherent-state
benchmarks for lossy links instead average the fidelity over a Gaussian
ensemble of inputs \cite{HammererBenchmark,NamikiKoashiImoto}. Among
measure-and-prepare (entanglement-breaking, EB) channels acting on the transmitted arm, the largest
fidelity with the output of the pure-loss channel of transmissivity
$0\le\eta\le1$ is
\begin{equation}
 F^{\rm EB}_\eta(N)=
 \left[\frac{\sqrt{1-\eta q^2}+q\sqrt{1-\eta}}{1+q}\right]^2 .
 \label{eq:finiteenergyEB}
\end{equation}

Alice heterodynes the input and sends the outcome $\beta$ to Bob, who
prepares the coherent state $|\sqrt\eta\,q\beta\rangle$. The
Supplemental Material \cite{SM} proves optimality even for collective
measure-and-prepare strategies on independent probes, with optimum
$F^{\rm EB}_\eta(N)^n$. For an ideal link ($\eta=1$), Eq.~\eqref{eq:finiteenergyEB}
reduces to $F^{\rm EB}_1(N)=(1-q)/(1+q)=1/(2N+1)$, which has a short direct proof.
Every separable output $\sigma_{RB}$ that preserves the reference
marginal $\rho_R=\tau_q$ of Eq.~\eqref{eq:thermaltheta} obeys $\sigma_{RB}\le\rho_R\otimes\id_B$
\cite{ReductionCriterion}. Its overlap with the initial pair is
therefore at most $\Tr\rho_R^2=1/(2N+1)$, a value also attained when
Alice counts the photons in the transmitted arm and Bob prepares the
corresponding Fock state. For $N=1$ this gives $1/3$, whereas the best
separable approximation to the same pair, whose reference marginal is
not fixed, reaches $1/2$. The fixed reference marginal accounts for the
difference: a simulator of the link has access only to the transmitted
arm. At fixed $\eta<1$, increasing the probe brightness brings the
benchmark down to $1-\eta$, the single-use value of
Eq.~\eqref{eq:classicalchannel}.

\makeatletter
\immediate\write\@auxout{\string\citation{PRLbibControl}}
\makeatother
\bibliographystyle{apsrev4-2}
\bibliography{references}

\clearpage
\onecolumngrid
\clearpage
\setcounter{page}{1}
\renewcommand{\thepage}{S\arabic{page}}
\setcounter{equation}{0}
\renewcommand{\theequation}{S\arabic{equation}}
\renewcommand{\theHequation}{S\arabic{equation}}
\setcounter{theorem}{0}
\renewcommand{\thetheorem}{S\arabic{theorem}}
\providecommand{\theHtheorem}{}
\renewcommand{\theHtheorem}{S\arabic{theorem}}
\setcounter{section}{0}
\setcounter{secnumdepth}{2}
\renewcommand{\thesection}{S\Roman{section}}

\begin{center}
{\large\bfseries Supplemental Material}\\[4pt]
{\large \papertitle}\\[4pt]
Yuan Gao and Gerardo Adesso
\end{center}

This Supplemental Material contains complete proofs of the results
stated in the Letter, together with the calculations behind its
examples. Section~\ref{sm:norm} proves a Gaussian optimality theorem
for Schatten norms of Gaussian maps, together with its
multiplicativity, and Sec.~\ref{sm:channels} deduces from it
Corollary~\ref{cor:channels}, including the von Neumann limit.
Section~\ref{sm:geometry} develops the covariance geometry of pure
seeds, the order-dependent seed of Fig.~\ref{fig:optimizer}, and the
algorithm for Eq.~\eqref{eq:fidelityprogram}.
The positive Gaussian operator $K$ of the End Matter is constructed in
Secs.~\ref{sm:operator} and~\ref{sm:weighted}, which complete the proof of
Theorem~\ref{thm:formation}. In Sec.~\ref{sm:dilution} we prove
Corollary~\ref{cor:closest}, the dilution bound~\eqref{eq:dilution},
and the strong converse exponent~\eqref{eq:generalexponent};
Theorem~\ref{thm:simulation} and the finite-brightness benchmark of
the End Matter follow in Sec.~\ref{sm:simulation}.
Section~\ref{sm:physics} derives the thermal-ring results of the End
Matter and treats field modes. Finally, non-Gaussian states with given
second moments, including the certification bound for an imperfect
source, are discussed in Sec.~\ref{sm:nongaussian}.
Equation numbers without the prefix S refer to the Letter, and those
with prefix E to its End Matter. Logarithms are natural, and
entropies and rates are in nats.
Throughout, $\SN_M$ denotes the trace-norm closed convex hull of
projectors onto pure states of Schmidt rank at most $M$, so that
$\SN_1=\Sep$. Norms without a subscript are Hilbert norms on vectors
and operator norms on matrices or operators.

\section{Gaussian stationarity implies global optimality}
\label{sm:norm}

We show that Gaussian inputs maximize the Schatten norms of the
coefficient operators of the output vectors of a Gaussian map, and
that these maxima are multiplicative. Lemma~\ref{lem:coherenttransfer}
bounds an explicit Gaussian coherent map, and Lemma~\ref{lem:stationary}
uses this bound to show that every stationary Gaussian input is a
global maximizer. Heat regularization supplies stationary Gaussian
inputs and is removed in the limit, while the Euler
equation~\eqref{eq:euler} carries stationarity over to tensor
products. Theorem~\ref{thm:schatten} collects the results.

For an operator $T$, $\|T\|_p=(\Tr|T|^p)^{1/p}$ denotes the Schatten $p$
norm, with the operator norm at $p=\infty$.
For a vector $|v\rangle=\sum_{ab}v_{ab}|a\rangle_A|b\rangle_B$,
let $T_v=(v_{ab})$ be its coefficient operator in fixed local
orthonormal bases. Thus $T_vT_v^\dagger=\Tr_B|v\rangle\langle v|$
and $\|T_v\|_2=\|v\|$. The split is the physical $A:B$ bipartition,
so the singular values of $T_v$ are the Schmidt coefficients of $|v\rangle$.

In the Bargmann representation \cite{Bargmann}, a vector of $d$ modes
is an entire function $f$ on $\mathbb C^{d}$, with inner product
$\langle f,g\rangle=\pi^{-d}\int_{\mathbb C^{d}}\overline{f(z)}\,g(z)\,
e^{-|z|^2}\,\dd^{2d}z$. The creation operator $a_j^\dagger$ acts as
multiplication by $z_j$, and the annihilation operator
$a_j=(\hat q_j+i\hat p_j)/\sqrt2$ as $\partial/\partial z_j$. The vacuum
is the constant function $1$, and the unnormalized coherent vector
$e(v)=\exp(v\cdot a^\dagger)|0\rangle$, $v\in\mathbb C^d$, is the
function $z\mapsto e^{v\cdot z}$. Thus
$\langle e(v'),e(v)\rangle=e^{v'^\dagger v}$ and
$\|e(v)\|=e^{|v|^2/2}$. Finite linear combinations of coherent vectors
are dense, so a bounded operator is determined by its action on them.
A Gaussian map $\mathcal G$ is a Hilbert-space operator with a Gaussian
Bargmann kernel: on coherent vectors it acts as
$\mathcal G e(v)=c_0\,e^{b^{\top}v+\frac12v^{\top}\Sigma v}
e^{\frac12 a^{\dagger\top}Za^\dagger+(Lv+w)\cdot a^\dagger}|0\rangle$
for a constant $c_0$, complex vectors $b,w$, a complex matrix $L$, and
complex symmetric $\Sigma,Z$ with $\norm Z<1$; here $a^\dagger$ denotes
the vector of output creation operators.
Gaussian isometries are included, and so is the bounded inverse
square root $K_C^{-1/2}$ of Lemma~\ref{lem:positiveinverse} (constructed in
Sec.~\ref{sm:operator}): the vector
$K_C^{-1/2}|0\rangle$ is a nonzero Gaussian vector, and
$e^{\frac12 a^{\dagger\top}Za^\dagger+w\cdot a^\dagger}|0\rangle$ is
normalizable only if $\norm Z<1$.
For a bounded Gaussian map $\mathcal G$ with $d_{\rm in}$ input modes,
put
\[
 a_{\alpha}(\mathcal G)=\sup_{\|f\|=1}\|T_{\mathcal Gf}\|_{2\alpha}.
\]
Lieb's Gaussian-kernel theorem~\cite{Lieb} answers the analogous
question for norms on function spaces.
Until Eq.~\eqref{eq:uniformnorm} we fix a finite $\alpha>1$, set
$p=2\alpha>2$, and let $\mathcal S_p$ be the class of operators with finite
Schatten $p$-norm.

For $k$ pairs of output modes $(A_j,B_j)$ set $r_j=\atanh \lambda_j$,
$0<\lambda_j<1$, and
$S_\lambda =\prod_j\exp[r_j(a_{A_j}^\dagger a_{B_j}^\dagger
                         -a_{A_j}a_{B_j})]$, a product of two-mode
squeezers. Applied to the vacuum, it gives the product of two-mode
squeezed vacua
\[
 |\Psi_\lambda \rangle=S_\lambda|0\rangle=\bigotimes_{j=1}^k
       \left(\sqrt{1-\lambda_j^2}\sum_{n\ge0}\lambda_j^n|n,n\rangle\right).
\]
Write the coherent label of these $2k$ modes as $z=(z_+,z_-)$, with
$z_+$ for Bob's modes $B_j$ and $z_-$ for Alice's modes $A_j$, and define
\begin{equation}
 \varkappa_j=\frac{\lambda_j-\lambda_j^{p-1}}{1-\lambda_j^p},\qquad
 Q(z)=\sum_j \varkappa_jz_{+,j}z_{-,j}=\tfrac12z^{\top}\Xi z,\qquad
 g_p=\|T_{\Psi_\lambda}\|_p
     =\prod_j\frac{\sqrt{1-\lambda_j^2}}{(1-\lambda_j^p)^{1/p}}.
 \label{eq:realQ}
\end{equation}
Here $\Xi$ is the symmetric $2k\times2k$ matrix whose only nonzero entries are
$\Xi_{+j,-j}=\Xi_{-j,+j}=\varkappa_j$.

The first step is a bound on the following coherent map, to which every
stationary Gaussian input will be reduced.

\begin{lemma}[Gaussian coherent-map bound]
\label{lem:coherenttransfer}
Let $L:\mathbb C^{d_{\rm in}}\to\mathbb C^{2k}$ be any complex matrix with
\begin{equation}
 \|Lv\|^2+\Re[(Lv)^{\top}\Xi(Lv)]\le\|v\|^2
 \qquad(v\in\mathbb C^{d_{\rm in}}).
 \label{eq:coherentcontraction}
\end{equation}
The linear map defined on coherent vectors by
\begin{equation}
 \mathcal V e(v)=e^{Q(Lv)}T_{S_\lambda  e(Lv)}
 \label{eq:realcoherentmap}
\end{equation}
extends to a bounded operator from the input Hilbert space to $\mathcal S_p$ of operator norm $g_p$.
Zero Schmidt ratios $\lambda_j=0$ and unpaired local output modes in
the vacuum are included by continuity.
\end{lemma}

\begin{proof}
The proof is a Lindeberg-type site replacement along the lines of
Ref.~\cite{BeigiMehrabi}. We realize the input modes as a limit of
$N$ sites, each a copy of $\mathbb C^{1+d_{\rm in}}$, and process every
site either by a random phase, which yields a scalar and preserves the
squared norm on average, or by a qubit map. The two agree to first
order. Provided the second-order terms compare in the right direction
(the local comparison) and the higher orders stay small (the uniform
remainder), replacing phases by qubits one site at a time does not
increase the averaged squared Schatten $p$-norm, up to errors that
vanish as $N\to\infty$. In the same limit the qubits belonging to each
pair form a collective spin that becomes an oscillator, and the
weighted $p$-norm of the all-qubit output tends to $\|\mathcal Vf\|_p/g_p$. Relative to
Ref.~\cite{BeigiMehrabi}, the new ingredients are the comparison
inequality~\eqref{eq:realdivideddifference} for Schatten norms of
coefficient operators, the local comparison built on it, and the
uniform remainder~\eqref{eq:realtaylor}.
Until the last two paragraphs, $\mathcal V$ acts on Fock polynomials,
finite combinations of number states, through the Taylor coefficients
of Eq.~\eqref{eq:realcoherentmap}.

\emph{The local comparison.}
For $x,y\ge0$, define the divided differences
\[
 h_p(x,y)=\frac{x^p-y^p}{x^2-y^2},\qquad
 d_p(x,y)=xy\frac{x^{p-2}-y^{p-2}}{x^2-y^2},
\]
extended continuously to $x=y$, where they equal $px^{p-2}/2$ and
$(p-2)x^{p-2}/2$. Both are nonnegative, homogeneous of degree $p-2$,
and zero at the origin.
For $D=\operatorname{diag}(x_i)\ge0$ with $\|D\|_p=1$ and a matrix $Z$,
both written in singular bases of $D$, the coefficient of $t^2$ in $\|D+tZ\|_p^2$ is
\begin{equation}
 \sum_{i,l}\left[h_p(x_i,x_l)|Z_{il}|^2
             +d_p(x_i,x_l)\Re(Z_{il}Z_{li})\right]
 -(p-2)\left(\Re\sum_i x_i^{p-1}Z_{ii}\right)^2.
 \label{eq:realhessian}
\end{equation}
This follows by differentiating $\Tr|D+tZ|^p$ through the
Hermitian dilation $\bigl(\begin{smallmatrix}0&D+tZ\\(D+tZ)^\dagger&0\end{smallmatrix}\bigr)$
and then taking the outer power $2/p$;
continuity covers zero singular values, and padding with zeros covers
rectangular matrices.

Each qubit site carries $k$ qubits, one per pair, with the
product qubit weight state
\begin{equation}
 \omega=\bigotimes_j\frac{\operatorname{diag}(1,\lambda_j^p)}{1+\lambda_j^p},
 \qquad c_j=\frac{1-\lambda_j^p}{1+\lambda_j^p},
 \qquad y_j=\sqrt{\frac{\lambda_j}{1-\lambda_j^2}}.
 \label{eq:qubitreference}
\end{equation}
An input site is $\mathbb C^{1+d_{\rm in}}$ with orthonormal letters:
a vacuum letter $\varepsilon_0$ and tangent letters $\varepsilon_a$,
$a=1,\ldots,d_{\rm in}$. The one-site qubit map $\mathcal E$ sends
$\varepsilon_0$ to $\omega^{1/p}$ and a tangent
$v=\sum_av_a\varepsilon_a$ to
\[
 \omega^{1/(2p)}
 \left[\sum_j\frac{(Lv)_{+,j}\sigma_j+(Lv)_{-,j}\sigma_j^\dagger}
                         {y_j\sqrt{c_j}}\right]
 \omega^{1/(2p)},\qquad \sigma=|0\rangle\langle1|,
\]
with $\sigma_j$ acting on qubit $j$. In the replacement, the processed site
is tensored with an auxiliary matrix space carrying the rest of the
system, so it holds $D\otimes\varepsilon_0+\sum_aY_a\otimes\varepsilon_a$
with a vacuum coefficient matrix $D$ and tangent coefficient matrices
$Y_a$. Write $U_{\pm,j,il}$ for the entries of
$U_{\pm,j}=\sum_aL_{(\pm,j),a}Y_a$. Applying $\mathcal E$ gives
$D\otimes\omega^{1/p}$, whose $p$-norm is $\|D\|_p=1$ because $\Tr\omega=1$,
plus $Z=\sum_aY_a\otimes\mathcal E(\varepsilon_a)$. Every term of $Z$
flips one qubit, so $Z$ has zero diagonal in the product singular
basis and the linear coefficient vanishes.
Substituting into Eq.~\eqref{eq:realhessian}, we find
the quadratic coefficient
\begin{equation}
 \sum_{i,l,j}\frac{1-\lambda_j^2}{1-\lambda_j^p}
 \left\{h_p(x_i,\lambda_jx_l)(|U_{+,j,il}|^2+|U_{-,j,li}|^2)
       +2d_p(x_i,\lambda_jx_l)\Re(U_{+,j,il}U_{-,j,li})\right\}.
 \label{eq:spinquadratic}
\end{equation}
The prefactor follows from Eq.~\eqref{eq:qubitreference}:
$\lambda_j/[(1+\lambda_j^p)y_j^2c_j]=(1-\lambda_j^2)/(1-\lambda_j^p)$.

The classical alternative contracts the site with a random linear
form $\zeta$, a ``random row,'' with $\zeta(\varepsilon_0)=1$ and
$\zeta(\varepsilon_a)=\zeta_a$ i.i.d.\ uniform on $\{\pm1,\pm i\}$. It turns the site into
$D+\sum_a\zeta_aY_a$ and preserves squared norms on average, as
$\mathbb E\zeta_a=\mathbb E\zeta_a\zeta_b=0$ and
$\mathbb E\bar\zeta_a\zeta_b=\delta_{ab}$. With $Z=\sum_a\zeta_aY_a$
the $d_p$ term of Eq.~\eqref{eq:realhessian} averages to zero, and
the averaged coefficient of $t^2$ is
\[
 \sum_{i,l,a}h_p(x_i,x_l)|Y_{a,il}|^2
 -\frac{p-2}{2}\sum_a\left|\sum_i x_i^{p-1}Y_{a,ii}\right|^2.
\]
Cauchy--Schwarz and $\sum_ix_i^p=1$ bound the subtracted term by
$(p-2)\sum_{i,a}x_i^{p-2}|Y_{a,ii}|^2/2$.
Since $h_p(x,x)=px^{p-2}/2$, the averaged classical coefficient is
at least $\sum_{i\ne l}h_p(x_i,x_l)\|Y_{il}\|^2+\sum_ix_i^{p-2}\|Y_{ii}\|^2$,
with $Y_{il}=(Y_{a,il})_a\in\mathbb C^{d_{\rm in}}$.

To bound Eq.~\eqref{eq:spinquadratic} by this, we use, for $x,y\ge0$
and $0\le\lambda<1$,
\begin{equation}
 \frac{1-\lambda^2}{1-\lambda^p}[h_p(x,\lambda y)+d_p(x,\lambda y)]\le h_p(x,y).
 \label{eq:realdivideddifference}
\end{equation}
To prove it, use homogeneity to set $x=1$, and put
$F(t)=(1-t^{p-1})/(1-t)$, $H(t)=(1-t^p)/(1-t^2)$.
Since $h_p(1,t)+d_p(1,t)=F(t)$, $h_p(1,t)=H(t)$, and
$(1-\lambda^2)/(1-\lambda^p)=1/H(\lambda)$, the desired
inequality is $F(\lambda y)\le H(\lambda)H(y)$.
Now $F(t)=(p-1)\int_0^1(1-r+rt)^{p-2}\,\dd r$ and
$(1-r+r\lambda y)^2\le(1-r+r\lambda^2)(1-r+ry^2)$, so the integral
Cauchy--Schwarz inequality gives
$F(\lambda y)^2\le F(\lambda^2)F(y^2)$, and it remains to note that
$H(t)^2-F(t^2)=t^2(1-t^{p-2})^2/(1-t^2)^2\ge0$.
Continuity covers vanishing denominators and $x=0$.

For $i\ne l$ and fixed $j$, the terms of Eq.~\eqref{eq:spinquadratic}
in $U_{+,j,il}$ and $U_{-,j,li}$ form a two-by-two quadratic form. As
$d_p\ge0$ and $2\Re(U_+U_-)\le|U_+|^2+|U_-|^2$, this form is at most
$|U_{+,j,il}|^2+|U_{-,j,li}|^2$ times its largest eigenvalue
$(1-\lambda_j^2)(h_p+d_p)(x_i,\lambda_jx_l)/(1-\lambda_j^p)$, which
Eq.~\eqref{eq:realdivideddifference} bounds by $h_p(x_i,x_l)$.
Summing over $j$ and exchanging $i$ and $l$ in the minus terms bounds
the off-diagonal part by $\sum_{i\ne l}h_p(x_i,x_l)\|LY_{il}\|^2$,
and $\|L\|\le1$, which follows by averaging
Eq.~\eqref{eq:coherentcontraction} over $v\to e^{i\phi}v$, gives the
classical off-diagonal bound.
For $i=l$, since $(1-\lambda_j^2)h_p(1,\lambda_j)=1-\lambda_j^p$ and
$(1-\lambda_j^2)d_p(1,\lambda_j)=(1-\lambda_j^p)\varkappa_j$, the
exact coefficient is
\[
 x_i^{p-2}\left[\|LY_{ii}\|^2+
                  \Re((LY_{ii})^{\top}\Xi(LY_{ii}))\right]
 \le x_i^{p-2}\|Y_{ii}\|^2,
\]
by Eq.~\eqref{eq:coherentcontraction} with $v=Y_{ii}$. Hence the qubit
quadratic coefficient never exceeds the averaged classical one, for
every nonzero $D$ by homogeneity.

\emph{A uniform remainder.}
Let $\varsigma=\min\{3,p\}$. For $D\ne0$, let $R(D,Z)$ be
$\|D+Z\|_p^2$ minus the second-order Taylor polynomial of
$t\mapsto\|D+tZ\|_p^2$ at $t=0$, evaluated at $t=1$. Then
\begin{equation}
 |R(D,Z)|\le C_p\|D\|_p^{\,2-\varsigma}\|Z\|_p^\varsigma ,
 \label{eq:realtaylor}
\end{equation}
with $C_p$ depending only on $p$.
For $p>0$, the real span of any two operators in $\mathcal S_p$ is
isometric to a subspace of an $L_p$ space \cite{Heinavaara}; the
reduction from complex matrices to Hermitian ones is part of that
result. Hence there are real functions $\phi_1,\phi_2$ in an $L_p$
space such that $\|sD+tZ\|_p=\|s\phi_1+t\phi_2\|_{L_p}$ for all
real $s,t$. Normalize $\|\phi_1\|_{L_p}=1$.
Pointwise, the second-order Taylor remainder of
$t\mapsto|\phi_1+t\phi_2|^p$ at $t=0$, evaluated at $t=1$, is bounded
by $C_p|\phi_2|^p$ for $2<p\le3$ and by
$C_p(|\phi_1|^{p-3}|\phi_2|^3+|\phi_2|^p)$ for $p>3$. Integrating and
applying H\"older's inequality bounds the corresponding remainder of
$\|\phi_1+t\phi_2\|_{L_p}^p$ by $C_p\|\phi_2\|_{L_p}^\varsigma$
when $\|\phi_2\|_{L_p}\le1/2$. This norm power then stays in an
interval bounded away from zero, where the outer power $2/p$ has
bounded derivatives.
For $\|\phi_2\|_{L_p}>1/2$, the triangle inequality and the bounds
on the linear and quadratic coefficients give the same estimate,
and homogeneity restores the factor $\|D\|_p^{2-\varsigma}$.

\emph{Replacing sites.}
Let $f=\sum_\gamma f_\gamma|\gamma\rangle$ be a Fock polynomial of
degree $d_0$. Since $e(v)=\sum_\gamma v^\gamma|\gamma\rangle/\sqrt{\gamma!}$,
applying $\sum_\gamma f_\gamma\partial_v^\gamma/\sqrt{\gamma!}$ at
$v=0$ to $e(v)$ gives $f$; applied instead to the $N$-site analog
\begin{equation}
 \left(\varepsilon_0+N^{-1/2}\sum_{a=1}^{d_{\rm in}}
                       v_a\varepsilon_a\right)^{\otimes N}
 \label{eq:fluctuationinput}
\end{equation}
it defines $F_N\in(\mathbb C^{1+d_{\rm in}})^{\otimes N}$.
Two such vectors with labels $h$ and $v$ have inner product
$(1+h^\dagger v/N)^N\to e^{h^\dagger v}=\langle e(h),e(v)\rangle$
coefficientwise, so $\|F_N\|\to\|f\|$.
Let $\zeta_s$, $s=1,\ldots,N$, be independent copies of the random
row, with $\zeta_s(\varepsilon_0)=1$ and
$\zeta_s(\varepsilon_a)=\zeta_{s,a}$, and write
$\omega^{[m]}=\omega^{\otimes m}$, with $\omega^{[0]}=1$.
With the first $m$ sites processed by $\mathcal E$ and the others by
rows, the random output $O_m$ and the compared quantity $W_m$ are
\[
 O_m=(\mathcal E^{\otimes m}\otimes \zeta_{m+1}\otimes\cdots
               \otimes \zeta_N)F_N,\qquad
 W_m=\mathbb E\big\|O_m\oplus\delta(\omega^{[m]})^{1/p}\big\|_p^2 ,
 \quad \delta>0 .
\]
The auxiliary block, of $p$-norm $\delta$, keeps the base norm in
Eq.~\eqref{eq:realtaylor} away from zero. To compare $W_m$ with
$W_{m-1}$, condition on $\zeta_{m+1},\ldots,\zeta_N$, process sites
$1,\ldots,m-1$ by $\mathcal E$, and expand in the letters at site $m$.
Denote the vacuum coefficient by $D$ and the tangent coefficients by
$Y_a$, operators on the qubits of the first $m-1$ sites. On the direct sum these are
$D'=D\oplus\delta(\omega^{[m-1]})^{1/p}$ and $Y'_a=Y_a\oplus0$.
The outputs entering $W_{m-1}$ and $W_m$ are $D'+\sum_a\zeta_{m,a}Y'_a$ and
$D'\otimes\omega^{1/p}+\sum_aY'_a\otimes\mathcal E(\varepsilon_a)$,
up to the ordering of the blocks. Their base terms have equal
$p$-norm, at least $\delta$, and their linear terms vanish, in expectation
over $\zeta_m$ for the row and identically for the qubits. The local
comparison of the quadratic terms and Eq.~\eqref{eq:realtaylor} for
the remainders give
\begin{equation}
 W_m-W_{m-1}\le C_{p,\lambda,L,d_{\rm in}}\,
             \delta^{\,2-\varsigma}\sum_a\mathbb E\|Y_a\|_p^\varsigma .
 \label{eq:onestepreplacement}
\end{equation}

To sum this bound over $m$, we show that
$\mathbb E\|Y_a\|_p^\varsigma=O(N^{-\varsigma/2})$ uniformly in $m$.
Take an even integer $\bar p\ge p$, so that $\|\cdot\|_{\bar p}^{\bar p}$
is a trace polynomial.
For every density matrix $\rho_0$, Schatten H\"older gives
$\|\rho_0^{1/(2p)}A\rho_0^{1/(2p)}\|_p
 \le\|\rho_0^{1/(2\bar p)}A\rho_0^{1/(2\bar p)}\|_{\bar p}$.
Each $Y_a$ has this form with $\rho_0=\omega^{[m-1]}$, because every
image of $\mathcal E$ carries $\omega^{1/(2p)}$ on both sides, and it
contains a factor $N^{-1/2}$ from the tangent letter at site $m$.
Expand the $\bar p$th power of the right-hand norm as a trace and
average over $\zeta_{m+1},\ldots,\zeta_N$. Call a tangent letter
active if it sits at a site other than $m$. A term with $\ell$ active letters
carries the prefactor $N^{-\bar p/2-\ell/2}$. A site that appears only once gives zero,
since a single spin flip has zero trace against diagonal factors and
a row entry has zero mean. Hence at most $\lfloor \ell/2\rfloor$ site
indices vary freely, offsetting $N^{-\ell/2}$. At fixed $d_0$ the number of patterns and their local
matrix elements are bounded independently of $N$ and $m$.
The $\bar p$th moment is therefore $O(N^{-\bar p/2})$, and Jensen's
inequality gives $\mathbb E\|Y_a\|_p^\varsigma=O(N^{-\varsigma/2})$.
At $m=0$, $O_0$ is a scalar; independence of the rows and scalar
Parseval give $\mathbb E|O_0|^2=\|F_N\|^2$, so $W_0\le\|F_N\|^2+\delta^2$
because $p\ge2$.
At $m=N$, discarding the auxiliary block gives
$\|\mathcal E^{\otimes N}F_N\|_p^2\le W_N$.
Summing Eq.~\eqref{eq:onestepreplacement} over the $N$ steps adds at most
$C\delta^{2-\varsigma}N^{1-\varsigma/2}$, which vanishes as $N\to\infty$
because $\varsigma>2$. Taking $N\to\infty$ at fixed $\delta$ and then
letting $\delta\downarrow0$ gives
\begin{equation}
 \limsup_{N\to\infty}\|\mathcal E^{\otimes N}F_N\|_p\le\|f\| .
 \label{eq:spinbound}
\end{equation}

\emph{Identifying the oscillator limit.}
It remains to show that
\begin{equation}
 \frac{\|\mathcal Vf\|_p}{g_p}
 \le\liminf_{N\to\infty}\|\mathcal E^{\otimes N}F_N\|_p ,
 \label{eq:oscillatorlimit}
\end{equation}
which together with Eq.~\eqref{eq:spinbound} gives
$\|\mathcal Vf\|_p\le g_p\|f\|$. Only this lower bound is needed,
although the argument will give equality.
For the generating vector~\eqref{eq:fluctuationinput} with label $h$,
put $u_j=(Lh)_{-,j}/y_j$, $v_j=(Lh)_{+,j}/y_j$ and
$X_{s,j}=(u_j\sigma_{s,j}^\dagger+v_j\sigma_{s,j})/\sqrt{Nc_j}$, where $\sigma_{s,j}$
acts on the qubit of type $j$ (pair $j$) at site $s$. Then
$\mathcal E^{\otimes N}$ maps this vector to
$(\omega^{[N]})^{1/(2p)}A_N(\omega^{[N]})^{1/(2p)}$ with
$A_N=\prod_s(I+\sum_j X_{s,j})$.
Consider the simpler product $B_N=\prod_{s,j}(I+X_{s,j})$, which
factorizes over the types.
Every monomial in $B_N-A_N$ contains two distinct types at one site.
Let $\Delta_N$ be obtained from $B_N-A_N$ by applying
$\sum_\gamma f_\gamma\partial_h^\gamma/\sqrt{\gamma!}$ at $h=0$. Expand the
$\bar p$th moment
$\|(\omega^{[N]})^{1/(2\bar p)}\Delta_N(\omega^{[N]})^{1/(2\bar p)}\|_{\bar p}^{\bar p}$.
A term containing $\ell$ factors $X_{s,j}$ carries $N^{-\ell/2}$.
Every occupied site--type pair must occur at least twice
because a single flip has zero trace against diagonal factors.
There are at most $\lfloor \ell/2\rfloor$ distinct pairs, and two
of them share a site, leaving at most $\lfloor \ell/2\rfloor-1$
distinct site indices. The $\bar p$th moment is therefore $O(N^{-1})$.
Taking its $\bar p$th root and using the H\"older inequality above with
$\rho_0=\omega^{[N]}$ shows that the weighted $p$-norm
$\|(\omega^{[N]})^{1/(2p)}\Delta_N(\omega^{[N]})^{1/(2p)}\|_p$ tends to
zero, so replacing $A_N$ by $B_N$ leaves the limit unchanged.

For one type, drop the index $j$, write $\xi=\lambda^p$, so that
$c=(1-\xi)/(1+\xi)$, and put $\mu^2=uv/(Nc)$ and $\sigma_N=(Nc)^{-1/2}\sum_s \sigma_s$.
Since $\sigma^2=0$ and $\sigma\sigma^\dagger+\sigma^\dagger\sigma=I$, each $X_s$ squares to
$\mu^2I$, so $I+X_s=(1-\mu^2)^{1/2}\exp[(\atanh\mu/\mu)X_s]$. As different sites
commute, the formal logarithm of $\prod_s(I+X_s)$, a power series
in $u$ and $v$, is exactly
\[
 \tfrac N2\log(1-\mu^2)I+
       \frac{\atanh\mu}{\mu}(u \sigma_N^\dagger+v \sigma_N).
\]
For each fixed degree, the coefficients of the scalar term tend
to those of $-uv/(2c)$ and those of the factor $\atanh\mu/\mu$ tend to
those of $1$, with errors $O(N^{-1})$.

It remains to replace $\sigma_N$ by an oscillator annihilation operator $a$
in the weighted norm. The $N$ qubits of one type decompose into
sectors of total spin $J$, each the tensor product of a spin factor of
dimension $2J+1$, on which $\sum_s\sigma_s$ acts, and a multiplicity factor.
Let $n=0,\ldots,2J$ count the excitations $|1\rangle$ in the spin factor
above its least excited state. On the sector, $\omega^{\otimes N}$ is $w_J\rho_J$ tensored with the
normalized identity of the multiplicity factor, where $w_J$ is the
sector probability and
\[
 \rho_J=\frac{1-\xi}{1-\xi^{2J+1}}
           \sum_{n=0}^{2J}\xi^n|n\rangle\langle n|.
\]
Under $\omega^{\otimes N}$ the total number of excitations
$n_{\rm tot}=N/2-J+n$ is binomial with parameter
$\bar\xi=\xi/(1+\xi)$, while $\mathbb E[n\mid J]\le\xi/(1-\xi)$, the
mean of the untruncated geometric law. Since $c=1-2\bar\xi$ and
$2J/N-c=2(\bar\xi-n_{\rm tot}/N)+2n/N$,
\[
 \mathbb E|2J/N-c|
 \le 2\sqrt{\bar\xi(1-\bar\xi)/N}
       +\frac{2\xi}{N(1-\xi)} .
\]
As in the Holstein--Primakoff representation
\cite{HolsteinPrimakoff}, identify the spin basis $|n\rangle$ with the first
$2J+1$ number states.
The coefficient of $\sigma_N^\dagger$ from $|n\rangle$ to $|n+1\rangle$ is
$\sqrt{n+1}\sqrt{(2J-n)/(Nc)}$. For each fixed $n$ it converges in
probability under the sector law to $\sqrt{n+1}$, by the bound above,
and since $2J\le N$ it is
always bounded by $\sqrt{(n+1)/c}$.
The same holds for the adjoint. The operator selected by $f$ through
$\sum_\gamma f_\gamma\partial_h^\gamma/\sqrt{\gamma!}$ is a polynomial
of degree at most $d_0$ in $\sigma_N,\sigma_N^\dagger$, so it has bandwidth at
most $d_0$ in $n$. After weighting on
both sides by $\rho_J^{1/(2p)}$, its absolute entries are at most
$C(1+n+n')^{d_0}\xi^{(n+n')/(2p)}$, a summable bound independent
of $J,N$. The sum of the discarded absolute entries bounds the
trace norm of the tail. Taking a fixed cutoff first and then
removing it, we conclude that the weighted sector coefficient
converges in trace norm, in probability under the sector law, to
the weighted oscillator coefficient, with trace norm bounded
uniformly in $J$ and $N$. Since $\|\cdot\|_p\le\|\cdot\|_1$, the
$p$th power $\nu_{N,J}$ of its $p$-norm is uniformly bounded and
converges in probability to the oscillator value $\nu$.
The selected operator commutes with site permutations, so it is
block diagonal over the sectors and acts trivially on each
multiplicity factor. As the normalized identity has unit trace, the full
$p$th norm power is therefore $\sum_Jw_J\nu_{N,J}$.
Dominated convergence over the sector weights then gives
$|\sum_Jw_J\nu_{N,J}-\nu|\le\sum_Jw_J|\nu_{N,J}-\nu|\to0$, because
$|\nu_{N,J}-\nu|$ is uniformly bounded and tends to zero in
probability under the sector law. For
$k$ types the qubit weight state and the sector laws are products over $j$,
and the same summable bound applies in each occupation index.

Replacing $\sigma_N$ by $a$ in the logarithm above, the limiting one-type
expression is $e^{-uv/(2c)}e^{u a^\dagger+v a}$ coefficientwise.
Normal ordering by $e^{X+Y}=e^Xe^Ye^{-[X,Y]/2}$, with
$[u a^\dagger,v a]=-uv I$, gives
\begin{equation}
 e^{-uv/(2c)+uv/2}e^{u a^\dagger}e^{v a}
   =e^{-\xi uv/(1-\xi)}e^{u a^\dagger}e^{v a}.
 \label{eq:wicklimit}
\end{equation}
As $2J/N\to c$ in probability, $\rho_J$ tends to the thermal state
$\tau_\xi$, with $\tau_\theta$ as in Eq.~\eqref{eq:thermaltheta}. The
full oscillator weight state is therefore
$\tau_{\lambda^p}=\bigotimes_j\tau_{\lambda_j^p}$, and
$\tau_{\lambda^p}^{1/p}=T_{\Psi_\lambda}/g_p$, both being diagonal with
entries proportional to $\prod_j\lambda_j^{n_j}$.
The weights rescale creation operators,
$\tau_\xi^{1/(2p)}a^\dagger\tau_\xi^{-1/(2p)}=\xi^{1/(2p)}a^\dagger
=\sqrt\lambda\,a^\dagger$, so commuting the two weights $\tau_\xi^{1/(2p)}$
inward turns the weighted Eq.~\eqref{eq:wicklimit} into
$e^{-\xi uv/(1-\xi)}e^{\sqrt\lambda ua^\dagger}\tau_\xi^{1/p}e^{\sqrt\lambda va}$.
Because $T_{(X\otimes Y)\psi}=XT_\psi Y^{\top}$, the factor
$e^{\sqrt\lambda ua^\dagger}$ on the left acts on Alice's mode and
$e^{\sqrt\lambda va}$ on the right is $e^{\sqrt\lambda va^\dagger}$
on Bob's mode. For one squeezed pair,
\[
 S_\lambda e(z)=e^{-\lambda z_-z_+}
 \exp\!\left[\sqrt{1-\lambda^2}
       (z_- a_A^\dagger+z_+ a_B^\dagger)\right]|\Psi_\lambda\rangle .
\]
Matching the creation amplitudes $\sqrt\lambda\,u$ and $\sqrt\lambda\,v$
gives $z_-=yu$, $z_+=yv$, where $y^2=\lambda/(1-\lambda^2)$, that is,
$z=Lh$ by the definition of $u$ and $v$.
Besides $-\lambda z_-z_+$, the remaining scalar exponent is
$[\lambda^2/(1-\lambda^2)-\lambda^p/(1-\lambda^p)]uv=\varkappa z_-z_+$,
with $\varkappa$ as in Eq.~\eqref{eq:realQ}.
Multiplying over the pairs gives $e^{Q(Lh)}T_{S_\lambda e(Lh)}/g_p
=\mathcal Ve(h)/g_p$, by Eq.~\eqref{eq:realcoherentmap}, and the
selection by $f$ gives $\mathcal Vf/g_p$. The oscillator value is therefore
$\nu=(\|\mathcal Vf\|_p/g_p)^p$, and Eq.~\eqref{eq:oscillatorlimit}
holds with equality.

This proves $\|\mathcal Vf\|_p\le g_p\|f\|$ for every Fock polynomial $f$.
The map $v\mapsto e^{Q(Lv)}S_\lambda e(Lv)$ is entire with values in
the output Hilbert space, so the Taylor truncations of
Eq.~\eqref{eq:realcoherentmap}, the images of those of $e(v)$,
converge in $\mathcal S_2$, hence in $\mathcal S_p$.
Polynomial density gives the claimed bounded extension,
and the vacuum, mapped to $T_{\Psi_\lambda}$, gives norm exactly $g_p$.

Finally, replace zero ratios by positive ratios $\lambda_j'\to0$,
so that the corresponding matrices $\Xi'$ tend to $\Xi$, and set
$L'=L/\sqrt{1+\|\Xi'-\Xi\|}$.
Equation~\eqref{eq:coherentcontraction} survives the change, because
$\|L\|\le1$ bounds the extra term by $\|\Xi'-\Xi\|\|v\|^2$,
and the bounds pass to the limit on finite coherent
sums. An unpaired local mode is paired with an added
opposite-party vacuum mode, a pair with $\lambda_j=0$, and $L$
receives a zero row for the added mode;
this leaves the Schmidt probabilities and the original
rows of $L$ unchanged.
\end{proof}

\begin{lemma}[Stationary Gaussian input]
\label{lem:stationary}
Let $\mathcal G:\mathcal H_{\rm in}\to\mathcal H_A\otimes\mathcal H_B$
be bounded and Gaussian. For every real $p>2$, a stationary
point $f_0$ of $\|T_{\mathcal Gf}\|_p$ among normalized Gaussian
inputs, with $\mathcal Gf_0\ne0$, is a global maximizer among all
unit inputs.
\end{lemma}

\begin{proof}
Gaussian unitaries on the input and local ones on the output change
neither hypothesis nor conclusion. An input one takes $f_0$ to the
vacuum, and local output ones put
$\mathcal Gf_0$ in the Schmidt form
$c_0|\Psi_\lambda \rangle$, $c_0>0$, up to unpaired local
vacua \cite{BoteroReznik}. After these reductions,
the Gaussian kernel necessarily has the form
\begin{equation}
 \mathcal G e(v)=c_0\exp(b^{\top}v+\tfrac12v^{\top}\Sigma v)
                           S_\lambda e(Lv),\qquad \Sigma=\Sigma^{\top}.
 \label{eq:generalkernelnormalform}
\end{equation}
Indeed, $S_\lambda^\dagger\mathcal G$ is a Gaussian map sending
$|0\rangle$ to $c_0|0\rangle$, so its output-only terms $Z$ and $w$ vanish.
The matrix $L$ may be rectangular or singular.

We compute the first variation at $f_0=|0\rangle$. Put
$T=T_{\mathcal Gf_0}=c_0T_{\Psi_\lambda}$. For an output vector
$\chi$, the derivative of $\log\|T_{\mathcal Gf_0+\epsilon\chi}\|_p$
at $\epsilon=0$ is
\[
 \Lambda(\chi)=\frac{\Re\Tr\bigl[(TT^\dagger)^{p/2-1}T\,T_\chi^\dagger\bigr]}
                   {\|T\|_p^p} .
\]
It is real linear, with $\Lambda(c\,\mathcal Gf_0)=\Re c$ for complex
$c$, and $\Lambda(U\chi)=\Lambda(\chi)$ for every product of local
unitaries $U$ that fixes $|\Psi_\lambda\rangle$. Independent opposite phase
rotations of each pair and phase rotations of unpaired local vacua
are such unitaries. They commute with $S_\lambda$, and on
$S_\lambda$ applied to a creation monomial of degree one or two they
act by a nontrivial phase, except on the paired monomials
$a^\dagger_{A_j}a^\dagger_{B_j}|0\rangle$; averaging over the phases
shows that $\Lambda$ vanishes on all the other such vectors.
As $\partial_{r_j}S_\lambda=S_\lambda(a_{A_j}^\dagger a_{B_j}^\dagger
-a_{A_j}a_{B_j})$ and $a_{A_j}a_{B_j}|0\rangle=0$, the paired monomials give
$S_\lambda a^\dagger_{A_j}a^\dagger_{B_j}|0\rangle
=\partial_{r_j}|\Psi_\lambda\rangle$, whose coefficient operator is
real and diagonal in the number basis, so that
$\Lambda(i\,c_0\partial_{r_j}|\Psi_\lambda\rangle)=0$. With
$\lambda_j=\tanh r_j$ this gives, for every complex $c$,
\[
 \Lambda\bigl(c\,c_0\,\partial_{r_j}|\Psi_\lambda\rangle\bigr)
 =\Re(c)\,\partial_{r_j}\log g_p
 =-\Re(c)\,\frac{\lambda_j-\lambda_j^{p-1}}{1-\lambda_j^p}
 =-\Re(c)\,\varkappa_j .
\]

The input curves $e(\epsilon v)$ and
$\exp(\frac\epsilon2a^{\dagger\top}Ma^\dagger)|0\rangle$, with
$v\in\mathbb C^{d_{\rm in}}$ and $M$ complex symmetric, consist of
Gaussian vectors with norm $1+O(\epsilon^2)$, so normalizing them does
not change their first derivatives. Their tangents at
$\epsilon=0$ are $\partial_\epsilon e(\epsilon v)|_{\epsilon=0}$ and
$\frac12\sum_{il}M_{il}\partial_{v_i}\partial_{v_l}e(v)|_{v=0}$,
so stationarity requires that $\Lambda$ vanish on their images under
$\mathcal G$. Differentiating Eq.~\eqref{eq:generalkernelnormalform}
gives for the first image
$c_0[(b^{\top}v)|\Psi_\lambda\rangle
+S_\lambda(Lv)\cdot a^\dagger|0\rangle]$, so $\Re(b^{\top}v)=0$ for
all $v$, and $b=0$. With $b=0$ the second image is
\[
 c_0\Bigl[\tfrac12\Tr(M\Sigma)\,|\Psi_\lambda\rangle
   +\tfrac12S_\lambda\,a^{\dagger\top}(LML^{\top})a^\dagger|0\rangle\Bigr].
\]
The coefficient of $a^\dagger_{A_j}a^\dagger_{B_j}$ in
$\frac12a^{\dagger\top}(LML^{\top})a^\dagger$ is
$(LML^{\top})_{+j,-j}$, so by the definition of $\Xi$ the first variation
along $M$ is
\[
 \tfrac12\Re\Tr(M\Sigma)-\sum_j\varkappa_j\Re(LML^{\top})_{+j,-j}
 =\tfrac12\Re\Tr\bigl[M(\Sigma-L^{\top}\Xi L)\bigr].
\]
Both $\Sigma$ and $L^{\top}\Xi L$ are symmetric, so we may choose
$M=\overline{\Sigma-L^{\top}\Xi L}$, which turns the trace into a
squared Frobenius norm. Stationarity along all complex symmetric $M$
therefore gives
\begin{equation}
 \Sigma=L^{\top}\Xi L.
 \label{eq:quadraticstationarity}
\end{equation}

Boundedness gives
$c_0^2\exp[t^2(\|Lv\|^2+\Re v^{\top}\Sigma v)]=\|\mathcal Ge(tv)\|^2
\le\|\mathcal G\|^2e^{t^2\|v\|^2}$ for all $t>0$, and $t\to\infty$ yields
\[
 \|Lv\|^2+\Re(v^{\top}\Sigma v)\le\|v\|^2.
\]
By Eq.~\eqref{eq:quadraticstationarity} this
is Eq.~\eqref{eq:coherentcontraction}, and the same equation gives
$\tfrac12v^{\top}\Sigma v=Q(Lv)$, that is,
$T_{\mathcal Ge(v)}=c_0\mathcal Ve(v)$.
Lemma~\ref{lem:coherenttransfer} now proves
\begin{equation}
 \|T_{\mathcal Gf}\|_p\le c_0 g_p\|f\|
                  =\|T_{\mathcal Gf_0}\|_p\|f\|.
 \label{eq:stationaryglobal}
\end{equation}
The inequality holds for every input, because $T_{\mathcal G\,\cdot}$
and $c_0\mathcal V$ are bounded and agree on coherent vectors. Zero
ratios and unpaired modes are covered by
Lemma~\ref{lem:coherenttransfer}.
\end{proof}

\subsection{Existence by heat regularization}

For a nonzero map $\mathcal G$, damping high photon numbers prevents a
maximizing sequence of Gaussian inputs from escaping to infinite
squeezing or displacement. With
$\hat n=\sum_{j=1}^{d_{\rm in}}a_j^\dagger a_j$ the total input number
operator, set
\[
 \mathcal G_\epsilon=\mathcal G e^{-\epsilon\hat n},\qquad \epsilon>0.
\]
For a normalized Gaussian vector $|\psi_{\Gamma,\boldsymbol m}\rangle$
with covariance $\Gamma$ and mean $\boldsymbol m$,
\begin{equation}
 \norm{e^{-\epsilon\hat n}|\psi_{\Gamma,\boldsymbol m}\rangle}^2
 =c_\epsilon
 \frac{\exp[-\boldsymbol m^{\top}
       (\Gamma+\coth\epsilon\,\id)^{-1}\boldsymbol m]}
      {\sqrt{\det(\Gamma+\coth\epsilon\,\id)}},
 \label{eq:heatproper}
\end{equation}
where $c_\epsilon=[2/(1-e^{-2\epsilon})]^{d_{\rm in}}$.
This tends to zero when the squeezing or displacement
parameters escape compact sets, through the determinant or the exponent.
The continuous Gaussian objective
$\psi\mapsto\|T_{\mathcal G_\epsilon\psi}\|_p
\le\|T_{\mathcal G_\epsilon\psi}\|_2\le\|\mathcal G\|\,\|e^{-\epsilon\hat n}\psi\|$
therefore tends to zero at infinity, and it is not identically zero because
$\mathcal G_\epsilon e(v)=\mathcal G e(e^{-\epsilon}v)$ and coherent
vectors are total; hence it has a nonzero maximum, which is stationary.
Lemma~\ref{lem:stationary}, applied to $\mathcal G_\epsilon$, makes
this a maximum over all inputs.

Let $a_{\alpha}^{\rm G}(\mathcal G)$ denote the supremum over Gaussian inputs.
The heat filter maps Gaussian vectors to multiples of Gaussian vectors
and is a contraction, so, using the maximum just established,
\[
 a_{\alpha}(\mathcal G_\epsilon)=a_{\alpha}^{\rm G}(\mathcal G_\epsilon)\le a_{\alpha}^{\rm G}(\mathcal G)\le a_{\alpha}(\mathcal G).
\]
Strong convergence $\mathcal G_\epsilon f\to \mathcal Gf$, together with
$\|T_x-T_y\|_p\le\|x-y\|$, gives
$a_{\alpha}(\mathcal G)\le\liminf_{\epsilon\downarrow0}a_{\alpha}(\mathcal G_\epsilon)$.
Therefore
\begin{equation}
 a_{\alpha}(\mathcal G)=a_{\alpha}^{\rm G}(\mathcal G).
 \label{eq:exhaustion}
\end{equation}

\subsection{Tensor products}

Fix $\epsilon>0$ and write, until $\epsilon$ is restored below,
$\mathcal G_i$ for the regularized maps $\mathcal G_{i,\epsilon}$,
$i=1,2$. By the heat-regularization step they have normalized Gaussian
global maximizers $f_i$. Put
$v_i=\mathcal G_i f_i$, $T_i=T_{v_i}$, and
$F_i=\Tr(T_iT_i^\dagger)^\alpha$, the $2\alpha$-th power of the objective.
Stationarity on the whole unit sphere gives the Euler (Lagrange) equation
\begin{equation}
 \mathcal G_i^\dagger
 \bigl[(T_iT_i^\dagger)^{\alpha-1}\otimes\id\bigr]\mathcal G_i f_i
 =F_i f_i.
 \label{eq:euler}
\end{equation}
This equation factors under tensor products, because the coefficient
operator of $(\mathcal G_1\otimes\mathcal G_2)(f_1\otimes f_2)$ is
$T_1\otimes T_2$. Hence $f_1\otimes f_2$ is stationary for
$\mathcal G_1\otimes\mathcal G_2$ against all input variations,
including those coupling the two inputs. Being Gaussian, it is
globally optimal by Lemma~\ref{lem:stationary}; restoring $\epsilon$,
$a_\alpha(\mathcal G_{1,\epsilon}\otimes\mathcal G_{2,\epsilon})
=a_\alpha(\mathcal G_{1,\epsilon})a_\alpha(\mathcal G_{2,\epsilon})$.
As $(\mathcal G_1\otimes\mathcal G_2)_\epsilon=\mathcal G_{1,\epsilon}\otimes\mathcal G_{2,\epsilon}$, the two inequalities of the previous subsection give
$a_\alpha(\mathcal G_1\otimes\mathcal G_2)\le\liminf_{\epsilon\downarrow0}
a_\alpha(\mathcal G_{1,\epsilon})a_\alpha(\mathcal G_{2,\epsilon})
\le a_\alpha(\mathcal G_1)a_\alpha(\mathcal G_2)$, and product inputs
give the reverse inequality. This proves
\begin{equation}
 a_{\alpha}(\mathcal G_1\otimes\mathcal G_2)=a_{\alpha}(\mathcal G_1)a_{\alpha}(\mathcal G_2).
 \label{eq:generalmult}
\end{equation}

Finally, for every Hilbert--Schmidt operator,
\begin{equation}
 \norm T_\infty\le\norm T_{2\alpha}
 \le\norm T_\infty^{1-1/\alpha}\norm T_2^{1/\alpha}.
 \label{eq:uniformnorm}
\end{equation}
Since $\norm{T_{\mathcal Gf}}_2\le\norm{\mathcal G}$ on unit inputs,
$a_\infty(\mathcal G)\le a_\alpha(\mathcal G)\le a_\infty(\mathcal G)^{1-1/\alpha}\norm{\mathcal G}^{1/\alpha}$,
and likewise for $a_\alpha^{\rm G}$. Hence $a_\alpha\to a_\infty$ and
$a_\alpha^{\rm G}\to a_\infty^{\rm G}$ as $\alpha\to\infty$, and
Eqs.~\eqref{eq:exhaustion} and
\eqref{eq:generalmult} extend to $\alpha=\infty$. In summary:

\begin{theorem}[Gaussian optimality for Schatten norms]\label{thm:schatten}
For every bounded Gaussian map $\mathcal G$ and all $\alpha\in\cI$,
$a_\alpha(\mathcal G)=a_\alpha^{\rm G}(\mathcal G)$, and
$a_\alpha(\mathcal G_1\otimes\mathcal G_2)=a_\alpha(\mathcal G_1)a_\alpha(\mathcal G_2)$
for bounded Gaussian maps $\mathcal G_1,\mathcal G_2$.
\end{theorem}

\subsection{Gaussian channels and the von Neumann limit}
\label{sm:channels}

Norm multiplicativity fails for general quantum channels
\cite{HaydenWinter}. For Gaussian channels, the equal-exponent norm
formulas \cite{FrankLieb,HolevoNorm} and the sharp one-mode
attenuator and amplifier inequalities \cite{DPT} treat related
optimizations. A finite-mode Gaussian channel has a Gaussian Stinespring
isometry $J_{\mathcal N}$ with a pure Gaussian environment \cite{GaussianDilation}.
Convexity of the output Schatten norm allows restriction to
pure inputs, and
\[
 \norm{\mathcal N(\psi)}_{\alpha}
 =\norm{T_{J_{\mathcal N}|\psi\rangle}}_{2\alpha}^{\,2},
 \qquad \mathcal N(\rho)=\Tr_E(J_{\mathcal N}\rho J_{\mathcal N}^\dagger),
\]
with the coefficient operator taken across output versus
environment $E$. Since $\mathcal N(|x\rangle\langle y|)=T_{J_{\mathcal N}x}T_{J_{\mathcal N}y}^{\dagger}$,
Schatten H\"older with exponents $2\alpha,2\alpha$, applied to each term
of the singular-value decomposition of a trace-class input $X$, gives
$\|\mathcal N(X)\|_\alpha\le\|X\|_1\sup_\psi\|\mathcal N(\psi)\|_\alpha$.
Hence the supremum of $\|\mathcal N(\rho)\|_\alpha$ over states is the
induced $1\to\alpha$ norm, and Theorem~\ref{thm:schatten}, applied
to the Gaussian isometries $J_{\mathcal N}$ and
$J_{\mathcal N_1}\otimes J_{\mathcal N_2}$, gives
\begin{align}
 \|\mathcal N\|_{1\to\alpha}&:=\sup_\rho\|\mathcal N(\rho)\|_\alpha
   =\sup_{\psi\ \mathrm{pure\ Gaussian}}\|\mathcal N(\psi)\|_\alpha ,
 \label{eq:gaussianchannel}\\
 \|\mathcal N_1\otimes\mathcal N_2\|_{1\to\alpha}&=\|\mathcal N_1\|_{1\to\alpha}\|\mathcal N_2\|_{1\to\alpha}.
 \label{eq:channelproduct}
\end{align}

Since $S_\alpha(\omega)=-t_\alpha^{-1}\log\|\omega\|_\alpha$, it
follows that the minimum output R\'enyi entropy is additive and can be
attained or approached using pure Gaussian inputs, with unrestricted
input energy. This proves Corollary~\ref{cor:channels} for $\alpha\in\cI$.

\emph{The von Neumann limit.} Let $c_\alpha=1-\log\alpha/(\alpha-1)$,
which tends to zero as $\alpha\downarrow1$. A one-mode thermal state $\tau_\theta$ as in
Eq.~\eqref{eq:thermaltheta}, $0<\theta<1$, has
$S_\alpha=[\alpha\log(1-\theta)-\log(1-\theta^\alpha)]/(1-\alpha)$. In
terms of $\varphi(x)=-\log(1-\theta^x)$, $x>0$, this reads
$S_\alpha=[\alpha\varphi(1)-\varphi(\alpha)]/(\alpha-1)$, with
$S_1=\varphi(1)-\varphi'(1)$. Since
$\varphi''(x)=[\log\theta/(2\sinh(\tfrac x2\log\theta))]^2\le x^{-2}$,
because $|\sinh y|\ge|y|$, Taylor's formula with integral remainder gives
$S_1-S_\alpha=(\alpha-1)^{-1}\int_1^\alpha(\alpha-x)\varphi''(x)\,\dd x\le c_\alpha$.
The vacuum case is immediate. By Williamson's theorem an
$m$-mode Gaussian state is unitarily equivalent to a product of thermal
states, hence $S_1(\rho_{\rm G})\le S_\alpha(\rho_{\rm G})+mc_\alpha$ for every
$m$-mode Gaussian $\rho_{\rm G}$. For any input $\rho$ to an $m$-mode-output
Gaussian channel,
\[
 S_1[\mathcal N(\rho)]\ge S_\alpha[\mathcal N(\rho)]
 \ge\inf_{\psi\ \mathrm{pure\ Gaussian}}S_\alpha[\mathcal N(\psi)]
 \ge\inf_{\psi\ \mathrm{pure\ Gaussian}}S_1[\mathcal N(\psi)]-mc_\alpha ,
\]
where the last step uses that $\mathcal N(\psi)$ is Gaussian.
Letting $\alpha\downarrow1$ gives Eq.~\eqref{eq:minoutput} at $\alpha=1$.
Applying the same chain to $\mathcal N_1\otimes\mathcal N_2$ and using additivity
at order $\alpha$ gives
$S^{\min}_1(\mathcal N_1\otimes\mathcal N_2)\ge S^{\min}_1(\mathcal N_1)+S^{\min}_1(\mathcal N_2)-(m_1+m_2)c_\alpha$,
while product inputs give the reverse inequality. Hence the von Neumann
minimum output entropy is additive as well, completing the proof of
Corollary~\ref{cor:channels}.

\section{Covariance geometry}
\label{sm:geometry}

The cost of a seed $\psi_\Gamma$ in Eq.~\eqref{eq:Gaussianroof}, its
marginal entropy $S_\alpha(\psi_{\Gamma,A})$, is explicit.
Local Gaussian unitaries leave Schmidt spectra unchanged and bring any
seed to the normal form of pure Gaussian states recalled in the Letter
\cite{BoteroReznik}, a product of two-mode squeezed vacua with
squeezings $r_j\ge0$ and local vacua, so its cost is
\begin{equation}
 \begin{aligned}
 e_\alpha(\Gamma)&=S_\alpha(\psi_{\Gamma,A})=\sum_j f_\alpha(r_j),\\
 f_\alpha(r)&=\frac{\log(\cosh^{2\alpha}r-|\sinh r|^{2\alpha})}{\alpha-1},
 \quad f_\infty(r)=2\log\cosh r .
 \end{aligned}
 \label{eq:scalarcost}
\end{equation}
If one party holds a single mode, a $1{:}m$ cut with $m$ modes on the
other side, every pure seed has at most one entangled pair, and its
cost at order $\alpha$ is $f_\alpha$ of that pair's squeezing $r$.
The Schmidt probabilities $(1-\theta)\theta^k$ of the pair, with
$\theta=\tanh^2r$, have partial sums $1-\theta^n$ over the $n$ largest that decrease with
$r\ge0$. The Schmidt distribution for a larger $r$ is thus majorized
by that for any smaller $r$, and Schur concavity of R\'enyi entropies
makes every $f_\alpha$ increasing in $r$. A seed minimizing $r$
over the feasible set, the pure $\Gamma\le V$, is therefore optimal at every order at once.

We use the matrix geometric mean of Eq.~\eqref{eq:geometricmean}
together with its monotonicity in each argument, its homogeneity
$(aX)\#_u(bY)=a^{1-u}b^u\,X\#_uY$ for $a,b>0$, its congruence covariance
$C(X\#_uY)C^{\top}=(CXC^{\top})\#_u(CYC^{\top})$ for invertible $C$,
the inversion identity $(X\#_uY)^{-1}=X^{-1}\#_uY^{-1}$, and the
determinant identity $\det(X\#_uY)=(\det X)^{1-u}(\det Y)^u$
\cite{KuboAndo}.
The argument compares the squeezing spectra of two seeds with that of
a point on the geometric-mean path $u\mapsto\Gamma_0\#_u\Gamma_1$
joining them.

First, for positive $X,Y$ let $\ell(Y,X)$, the relative
eigenvalues of $Y$ with respect to $X$, list the eigenvalues of
$X^{-1/2}YX^{-1/2}$ in nonincreasing order.
For positive pairs $(X_j,Y_j)$, $j=0,1$, put
$X_u=X_0\#_u X_1$ and $Y_u=Y_0\#_u Y_1$. Then
\begin{equation}
 \log\ell(Y_u,X_u)
 \prec
 (1-u)\log\ell(Y_0,X_0)
   +u\log\ell(Y_1,X_1).
 \label{eq:relativemajorization}
\end{equation}
In the majorization order $\boldsymbol x\prec\boldsymbol y$, the
partial sums of the components of $\boldsymbol x$ in nonincreasing
order are bounded by those of $\boldsymbol y$, with equality for the
total sum. To prove Eq.~\eqref{eq:relativemajorization}, let
$a_j=\max\ell(Y_j,X_j)$ be the largest
relative eigenvalue. Then $Y_j\le a_jX_j$, and monotonicity and
homogeneity of the mean imply
$Y_u\le(a_0X_0)\#_u(a_1X_1)=a_0^{1-u}a_1^u X_u$, which is Eq.~\eqref{eq:relativemajorization} for the largest components.
Applying this to every exterior power $\bigwedge^k$, which commutes with the
geometric mean and whose largest relative eigenvalue is the product of
the $k$ largest ones, bounds each partial sum
\cite{AndoHiai,BhatiaMatrix}.
The determinant identity gives equality for the full product, proving
the log-majorization.

Second, geometric means preserve pure covariances.
Purity, $\Gamma\Omega\Gamma=\Omega$, is equivalent to
$\Gamma^{-1}=\Omega\Gamma\Omega^{\top}$. If $\Gamma_0$ and $\Gamma_1$
obey it, the inversion identity and congruence covariance give
$(\Gamma_0\#_u\Gamma_1)^{-1}
=(\Omega\Gamma_0\Omega^{\top})\#_u(\Omega\Gamma_1\Omega^{\top})
=\Omega(\Gamma_0\#_u\Gamma_1)\Omega^{\top}$.

Third, the pure-state cost $e_{\alpha}$ in
Eq.~\eqref{eq:scalarcost} is geodesically convex, that is, convex
along geometric-mean paths.
For real $\alpha>1$ and $r>0$, set $\theta=\tanh^2r$ and
$w=-\tfrac12\log\theta>0$.
Direct differentiation gives
\begin{equation}
 f_{\alpha}''(r)=\frac{4\alpha(1-\theta)\theta^{\alpha-1/2}}
                 {(\alpha-1)(1-\theta^{\alpha})^2}
 \left[\sinh((2\alpha-1)w)-(2\alpha-1)\sinh w\right]\ge0.
 \label{eq:realentropycurvature}
\end{equation}
The bracket is nonnegative because $\sinh$ is convex on the positive
axis with $\sinh0=0$, so $\sinh(cw)\ge c\sinh w$ for $c=2\alpha-1>1$.
Since $f_{\alpha}$ is even with $f_{\alpha}'(0)=0$,
it is convex on the whole real line, where
Eq.~\eqref{eq:relativecost} below evaluates it. The same holds for
$f_\infty(r)=2\log\cosh r$.

Let $d_A,d_B$ count the modes of $A,B$, and let $\Pi$, a local phase
rotation by $\pi$, change the sign of both quadratures of every $B$ mode.
In our position--momentum ordering,
$\Pi=P_B\oplus P_B$ with $P_B=\id_{d_A}\oplus(-\id_{d_B})$.
Local symplectic matrices commute with $\Pi$ and act on both arguments
of $\ell_k(\Pi\Gamma\Pi,\Gamma)$ by the same congruence, which
leaves relative eigenvalues unchanged, so for pure $\Gamma$ they can be
read off in the normal form. There $\Pi$ reverses the sign of the
cross-correlations of each entangled pair, and the pair with
squeezing $r_j$ contributes $e^{4r_j}$ and $e^{-4r_j}$, each twice;
all other relative eigenvalues are one.
Since $f_\alpha$ is even and $f_\alpha(0)=0$, it follows that
\begin{equation}
 e_{\alpha}(\Gamma)
 =\frac14\sum_k
 f_{\alpha}\!\left(\frac14
       \log\ell_k(\Pi\Gamma\Pi,\Gamma)\right).
 \label{eq:relativecost}
\end{equation}
Relative eigenvalues depend continuously on $\Gamma$, so
Eq.~\eqref{eq:relativecost} also shows that $e_\alpha$ is continuous
on pure covariances.
Now apply Eq.~\eqref{eq:relativemajorization} to the pairs
$(X_j,Y_j)=(\Gamma_j,\Pi\Gamma_j\Pi)$. Congruence covariance gives
$Y_u=\Pi\Gamma_u\Pi$ with $\Gamma_u=\Gamma_0\#_u\Gamma_1$, which is
pure by the second step. Because $f_\alpha$ is convex, the sum in
Eq.~\eqref{eq:relativecost} is a convex and Schur-convex function of
the vector of $\log\ell_k$, and therefore
\begin{equation}
 e_{\alpha}(\Gamma_0\#_u\Gamma_1)
 \le(1-u)e_{\alpha}(\Gamma_0)+u e_{\alpha}(\Gamma_1).
 \label{eq:geoconvex}
\end{equation}
These arguments extend the covariance-geometric methods of
Refs.~\cite{GA,Lami17,MishraWilde} to the present
costs.

\subsection{Minimizers, convexity, and strong superadditivity}

The feasible set in Eq.~\eqref{eq:Gaussianroof} is nonempty:
in Williamson's normal form
$V=S\operatorname{diag}(\boldsymbol\nu,\boldsymbol\nu)S^{\top}$ with
$S$ symplectic and all $\nu_j\ge1$ \cite{Williamson}, the pure
covariance $SS^{\top}$ lies below $V$.
The largest eigenvalue of a feasible $\Gamma$ is at most $\|V\|$,
and since purity gives $\|\Gamma^{-1}\|=\|\Gamma\|$, the smallest is
at least $\|V\|^{-1}$. The feasible set is closed and hence compact,
and $e_\alpha$ is continuous on it by Eq.~\eqref{eq:relativecost},
so the minimum exists.
No other pure covariance lies below a pure $\Gamma$: if $\Gamma'\le\Gamma$
are both pure, then $\Gamma^{-1}\le\Gamma'^{-1}$, that is,
$\Omega\Gamma\Omega^{\top}\le\Omega\Gamma'\Omega^{\top}$, so
$\Gamma=\Gamma'$. Hence $G_\alpha(\Gamma)=e_\alpha(\Gamma)$ for pure $\Gamma$.

For covariances $V_0,V_1$ and feasible pure $\Gamma_j\le V_j$,
monotonicity and the matrix arithmetic--geometric mean inequality give
\[
 \Gamma_0\#_u\Gamma_1
 \le V_0\#_uV_1
 \le(1-u)V_0+uV_1,
\]
so $\Gamma_0\#_u\Gamma_1$ is a pure seed feasible for
$(1-u)V_0+uV_1$.
Choosing $\Gamma_j$ optimal for $V_j$, Eq.~\eqref{eq:geoconvex}
proves convexity of $G_{\alpha}$ as a function of the covariance.
It is also nonincreasing in the Loewner order, since enlarging $V$
enlarges the feasible set. Hence $G_\alpha(V+\epsilon\id)\le G_\alpha(V)$.
For $0<\epsilon\le1$, minimizers $\Gamma_\epsilon$ for $V+\epsilon\id$ lie in the compact
feasible set of $V+\id$. Take $\epsilon_n\downarrow0$ with
$\Gamma_{\epsilon_n}$ converging to some $\Gamma_\star$, which is then
feasible for $V$. Continuity of $e_\alpha$ gives
$G_\alpha(V)\le e_\alpha(\Gamma_\star)=\lim_nG_\alpha(V+\epsilon_n\id)
\le G_\alpha(V)$, and since $G_\alpha(V+\epsilon\id)$ is monotone in
$\epsilon$, this proves the boundary continuity
\begin{equation}
 G_{\alpha}(V+\epsilon\id)\longrightarrow G_{\alpha}(V)
 \quad(\epsilon\downarrow0).
 \label{eq:boundarycontinuity}
\end{equation}

For strong superadditivity, group the modes as
$(A_1B_1):(A_2B_2)$ and let $P_2$ flip the signs of every
quadrature in the second group, a phase rotation by $\pi$ of each
mode in $A_2B_2$.
This is local for the entanglement cut
$A_1A_2:B_1B_2$, so $e_\alpha(P_2\Gamma P_2)=e_\alpha(\Gamma)$.
Given a feasible pure $\Gamma\le V$, form
\[
 \Gamma_{\rm bd}=\Gamma\#_{1/2}(P_2\Gamma P_2).
\]
It is pure, and since $X\#_{1/2}Y=Y\#_{1/2}X$ it is $P_2$-invariant,
hence block diagonal between
the two groups. After regrouping the coordinates so that each group's
positions and momenta are adjacent, monotonicity and the
arithmetic--geometric mean inequality give
\[
 \Gamma_{\rm bd}\le V\#_{1/2}(P_2VP_2)
 \le\tfrac12(V+P_2VP_2)
 =V_{A_1B_1}\oplus V_{A_2B_2}.
\]
The two blocks of $\Gamma_{\rm bd}$ are feasible pure covariances for the
inherited cuts $A_i:B_i$, and Eq.~\eqref{eq:geoconvex} at $u=1/2$
with the local invariance above gives
$e_{\alpha}(\Gamma_{\rm bd})\le e_{\alpha}(\Gamma)$.
The seed of $\Gamma_{\rm bd}$ is the product of the seeds of its
blocks, so additivity of pure-state entropy on product states and
minimization over $\Gamma$ prove
\begin{equation}
 G_{\alpha}(V_{A_1A_2:B_1B_2})
 \ge G_{\alpha}(V_{A_1:B_1})+G_{\alpha}(V_{A_2:B_2}).
 \label{eq:strongsuper}
\end{equation}
When $V=V_1\oplus V_2$, product seeds $\Gamma_1\oplus\Gamma_2$ give
the opposite inequality, so $G_\alpha$ is additive on direct sums.

Gaussian ensembles whose seed varies from component to component do
no better. If the components of an ensemble $\mu$ of $\rho_V$ are
Gaussian with covariances $\Gamma_\psi$
and means $\boldsymbol d_\psi$, then
$V=\int\Gamma_\psi\,\dd\mu+2\operatorname{Cov}_\mu(\boldsymbol d_\psi)$,
with $\operatorname{Cov}_\mu$ the covariance matrix of the means
under $\mu$. Since $e_\alpha(\Gamma_\psi)=G_\alpha(\Gamma_\psi)$,
convexity and monotonicity of $G_\alpha$ give
$\int e_{\alpha}(\Gamma_\psi)\,\dd\mu
\ge G_\alpha(\int\Gamma_\psi\,\dd\mu)\ge G_{\alpha}(V)$.

\subsection{An order-dependent optimal Gaussian seed}
\label{sm:orderseed}
A single source of position noise suffices to make the optimal
preparation depend on order.
Consider four oscillators ordered as
$(q_{A_1},q_{A_2},q_{B_1},q_{B_2},
p_{A_1},p_{A_2},p_{B_1},p_{B_2})$, with
$\Omega=\left(\begin{smallmatrix}0&\id_4\\-\id_4&0\end{smallmatrix}\right)$.
Set
\begin{equation}
 X_0=\begin{pmatrix}
 3/2&0&1/2&0\\0&21/2&0&19/2\\
 1/2&0&3/2&0\\0&19/2&0&21/2
 \end{pmatrix},\qquad
 v=\begin{pmatrix}1\\1\\-1\\-1\end{pmatrix},
 \label{sm:matrix}
\end{equation}
and $V=(X_0+vv^{\top})\oplus X_0^{-1}$.
The eigenvalues of $X_0$ are $1,1,2,20$. The pure covariance
$X_0\oplus X_0^{-1}$ describes the ground state of
$H=(\boldsymbol p^{\top}\boldsymbol p+
\boldsymbol q^{\top}X_0^{-2}\boldsymbol q)/2$, where
$\boldsymbol q,\boldsymbol p$ are the four position and
momentum vectors. A shared displacement
$\boldsymbol q\mapsto\boldsymbol q+\xi v$, with $\xi$ a
centered normal variable of variance $1/2$, adds $vv^{\top}$ to the
position block and prepares $V$.

Global time reversal
$P_{\rm T}=\id_4\oplus(-\id_4)$ is antisymplectic, yet it preserves
$V$ and purity. Since it commutes with $\Pi$, conjugating both
arguments in Eq.~\eqref{eq:relativecost} by $P_{\rm T}$ gives
$e_\alpha(P_{\rm T}\Gamma P_{\rm T})=e_\alpha(\Gamma)$.
For every feasible pure $\Gamma$, the
geometric midpoint $\Gamma\#_{1/2}(P_{\rm T}\Gamma P_{\rm T})$ is then
pure, feasible because it lies below $V\#_{1/2}V=V$, and by
Eq.~\eqref{eq:geoconvex} no more costly at any order. The midpoint is invariant
under $P_{\rm T}$, hence block diagonal between positions and
momenta, and purity forces the form $X\oplus X^{-1}$.
Feasibility, $X\le X_0+vv^{\top}$ and $X^{-1}\le X_0^{-1}$, means
$X_0\le X\le X_0+vv^{\top}$. As $0\le X-X_0\le vv^{\top}$, the range
of $X-X_0$ lies in the span of $v$, so $X=X_0+s vv^{\top}$ with
$0\le s\le1$, proving the global reduction
\begin{equation}
 \Gamma_s=X_s\oplus X_s^{-1},\quad
 X_s=X_0+s vv^{\top},\quad0\le s\le1.
 \label{eq:seedpath}
\end{equation}
Because $v^{\top}X_0^{-1}v=4$, the matrix $X_0^{-1/2}X_1X_0^{-1/2}$ has
the single nontrivial eigenvalue $5$, so $X_s=X_0\#_uX_1$ with
$u=\log(1+4s)/\log5$, and, as the mean acts blockwise on direct sums, the inversion
identity gives
$\Gamma_s=\Gamma_0\#_u\Gamma_1$.
By Eq.~\eqref{eq:geoconvex}, at each order the cost
$e_\alpha(s)=e_\alpha(\Gamma_s)$ along this path is convex in $u$, which is itself an
increasing function of $s$.

Let $\nu_+\ge\nu_-\ge1$ be the symplectic eigenvalues
of Alice's reduced covariance $(\Gamma_s)_A$, namely the
positive moduli of the eigenvalues of
$i\Omega_A(\Gamma_s)_A$, with $\Omega_A$ the symplectic form of $A$. Put
$\varpi(s)=\nu_+\nu_-$ and $\upsilon(s)=\nu_+^2+\nu_-^2$.
Direct evaluation gives
\begin{equation}
 \varpi^2=\frac{9(16s+21)^2}{640(4s+1)},\qquad
 \upsilon=\frac{88s^2+1266s+531}{80(4s+1)}.
 \label{sm:invariants}
\end{equation}
For arbitrary real $\alpha>1$, evaluate the cost from
$\nu_\pm=\sqrt{(\upsilon\pm\sqrt{\upsilon^2-4\varpi^2})/2}$ and
Eq.~\eqref{eq:scalarcost}, with $\nu_\pm=\cosh(2r_\pm)$.
The minimizing seed coordinate $s_{\alpha}$ generates
Fig.~\ref{fig:optimizer}, with $s_1\simeq0.60$ in the von Neumann limit.
At orders two and infinity the costs take the forms
\begin{equation}
 e_2(s)=\log\varpi,\qquad
 e_\infty(s)=\log\frac{1+\varpi+\sqrt{\upsilon+2\varpi}}4.
 \label{sm:costs}
\end{equation}
Differentiation with respect to $s$ gives
\begin{equation}
 e_2'(s)=\frac{2(16s-13)}{(16s+21)(4s+1)}.
 \label{sm:derivative}
\end{equation}
Thus $e_2$ has its unique minimum on $[0,1]$ at $s_2=13/16$, with
$G_2(V)=\log(3\sqrt{170}/20)$. At $s_2$, $\varpi'=0$ and
Eq.~\eqref{sm:invariants} gives $\upsilon'<0$, so $e_\infty'(s_2)<0$,
while $e_\infty'(1)>0$. Convexity in $u$
and analyticity yield a unique minimum $s_\infty\simeq0.89$.
Here $e^{-e_\infty(s_\infty)}\simeq0.68$, in agreement with the
iteration of Sec.~\ref{sm:fidelityprogram} applied to
Eq.~\eqref{eq:fidelityprogram}.
Suppose a Gaussian seed $\Gamma$ were optimal at both orders $2$ and
$\infty$. Its midpoint with its time reversal would then also be
optimal at both orders and would equal $\Gamma_s$ for some $s$ on the
path of Eq.~\eqref{eq:seedpath}; uniqueness of the minimizers on the
path would force $s_2=s=s_\infty$, a contradiction.
The separation persists for $V+\epsilon\id$ with sufficiently small
$\epsilon>0$. Otherwise, common minimizers for a sequence
$\epsilon_n\downarrow0$ would have, by compactness, a limit point
feasible for $V$, and continuity of $e_\alpha$ together with
Eq.~\eqref{eq:boundarycontinuity} would make it a common minimizer
for $V$.

The rate $\mathcal R$ at which a given order $\alpha$ attains the
supremum in Eq.~\eqref{eq:generalexponent} is read off from the
optimal seed's Schmidt
distribution. A pair with symplectic eigenvalue $\nu$ has Schmidt
probabilities $(1-\theta)\theta^k$ with
$\theta=\tanh^2r=(\nu-1)/(\nu+1)$; write $\theta_\pm$ for the two
pairs of $\Gamma_s$. The Schmidt distribution tilted to order $\alpha$
has probabilities proportional to the $\alpha$th powers of the
original ones. For each pair it is again geometric, with ratio
$\theta^\alpha$, that is, thermal with $\theta^\alpha/(1-\theta^\alpha)$
mean photons. With $g(x)=(1+x)\log(1+x)-x\log x$, $0\log0=0$, the
entropy of a thermal mode with $x$ mean photons, the tilted
distribution of the seed therefore has entropy
$H_\alpha=\sum_{\pm}g(\theta_\pm^\alpha/(1-\theta_\pm^\alpha))$.
At a fixed seed $\Gamma_s$,
$\partial_\alpha\{t_\alpha[e_\alpha(\Gamma_s)-\mathcal R]\}
=(H_\alpha-\mathcal R)/\alpha^2$, and the variation of $s_\alpha$ does
not contribute at the minimum, so stationarity in $\alpha$ selects
$\mathcal R=H_\alpha$.
For $\alpha=2$ at $s_2$ this gives $\mathcal R=0.36$ nats ($0.53$
ebits) per copy.

\subsection{Computing the closest separable state}
\label{sm:fidelityprogram}

For any physical covariance $X$, define its pure part by
\begin{equation}
 P(X)=X\#_{1/2}(\Omega X^{-1}\Omega^{\top}),
 \label{eq:purepart}
\end{equation}
with $\Omega$ the symplectic form on the modes of $X$.
In Williamson coordinates
$X=S\operatorname{diag}(\boldsymbol\nu,\boldsymbol\nu)S^{\top}$,
symplecticity of $S$ makes the
two arguments of the geometric mean
$S\operatorname{diag}(\boldsymbol\nu,\boldsymbol\nu)^{\pm1}S^{\top}$.
Thus $P(X)=SS^{\top}\le X$. The pure Gaussian state with covariance
$P(X)$ is a leading eigenvector of the Gaussian state $\rho_X$:
the Gaussian unitary implementing $S^{-1}$ turns $\rho_X$ into a product of thermal states,
each with the vacuum as its leading eigenvector.

For $X$ in the set $\mathcal C(V)$ of Eq.~\eqref{eq:fidelityprogram},
write $h(X_A)=\sum_j\log[(1+\nu_j(X_A))/2]$ for the objective there,
so that $e^{-h(X_A)}$ is the vacuum
probability, and hence the largest eigenvalue, of the Gaussian
marginal $\rho_{X_A}$ in Williamson coordinates.
The pure--mixed overlap formula
$\Tr(\rho_{X_A}\psi_Y)=\det[(X_A+Y)/2]^{-1/2}$ for centered Gaussian
states \cite{Banchi,SerafiniBook} gives the variational form
\begin{equation}
 h(X_A)=\min_{Y\in\mathcal C(V_A)}
 \frac12\log\det\!\left(\frac{X_A+Y}{2}\right).
 \label{eq:auxiliaryfidelity}
\end{equation}
For pure $Y$ the objective is $-\log\Tr(\rho_{X_A}\psi_Y)\ge h(X_A)$,
since the overlap is at most the largest eigenvalue of $\rho_{X_A}$, and the
leading eigenvector attains the minimum at
$Y=P(X_A)\le X_A\le V_A$. Allowing mixed $Y$ gives the same answer,
since replacing $Y$ by $P(Y)\le Y$, which stays in $\mathcal C(V_A)$,
does not increase the determinant.
The same monotonicity shows that $h$ does not increase when $X$ is
replaced by $P(X)\le X$, which stays in $\mathcal C(V)$.
Consequently a minimizing $X_*$ in Eq.~\eqref{eq:fidelityprogram}
gives a pure minimizer $\Gamma=P(X_*)$. For pure $\Gamma$,
$h(\Gamma_A)=S_\infty(\psi_{\Gamma,A})=e_\infty(\Gamma)$, so the
minimum in Eq.~\eqref{eq:fidelityprogram} equals $G_\infty(V)$ and
$\Gamma$ is an optimal seed at $\alpha=\infty$.
In Eq.~\eqref{eq:separablecovariance} the leading Schmidt product
$\phi$ of this seed
has covariance $\Gamma_\phi=P(\Gamma_A)\oplus P(\Gamma_B)$, with the modes
regrouped by party. Since the argument applies equally to $B$,
either party can be used for the minimization.

The feasible set $\mathcal C(V)$ is compact and closed under
geometric means: for positive $X$, the uncertainty relation
$X+i\Omega\ge0$ is equivalent to $X\ge\Omega X^{-1}\Omega^{\top}$, and
inversion, congruence, and monotonicity of $\#_u$ preserve this
inequality and the upper bound $X\le V$. The determinant objective of
Eq.~\eqref{eq:auxiliaryfidelity} is jointly geodesically convex in
$(X,Y)$, that is, convex along pairs of geometric-mean paths
$(X\#_uX',Y\#_uY')$.
For $X,X'\in\mathcal C(V)$ and $Y,Y'\in\mathcal C(V_A)$, the
transformer inequality $(X\#_uX')_A\le X_A\#_uX'_A$ for the compression
to the $A$ diagonal block and the joint
concavity of the geometric mean \cite{KuboAndo}, in its superadditive
form, give
\[
 (X\#_u X')_A+Y\#_u Y'
 \le (X_A+Y)\#_u(X'_A+Y').
\]
Taking log determinants with the determinant identity proves the
joint convexity. Since $Y\#_uY'\in\mathcal C(V_A)$, choosing the minimizing $Y$ and $Y'$ at the
two endpoints proves geodesic convexity of $h(X_A)$.

The minimum can be found by a sequence of ordinary semidefinite
programs. Start at $X=V$, put $D_X=\tfrac12[X_A+P(X_A)]^{-1}$,
the gradient of $Z_A\mapsto\frac12\log\det[(Z_A+Y)/2]$ at $Z_A=X_A$
for $Y=P(X_A)$,
and replace $X$ by any solution $Z$ of
\begin{equation}
 \min_{Z\in\mathcal C(V)}\Tr(D_X Z_A).
 \label{eq:fidelitylinearstep}
\end{equation}
A real SDP solver can impose $Z+i\Omega\ge0$ as
$\left(\begin{smallmatrix}Z&-\Omega\\ \Omega&Z\end{smallmatrix}\right)\ge0$
using the same real decision matrix $Z$.
To see convergence, keep $Y=P(X_A)$ in Eq.~\eqref{eq:auxiliaryfidelity}
and use the ordinary concavity of $\log\det$ to obtain
\[
 h(Z_A)\le h(X_A)+\Tr[D_X(Z_A-X_A)].
\]
Interchanging $X$ and $Z$ bounds $h(Z_A)-h(X_A)$ below by
$\Tr[D_Z(Z_A-X_A)]$, and continuity of $D_X$ then shows that
the first variation along a feasible path is $\Tr(D_X\dot X_A)$.
Each step lowers the objective by at least the linear optimization
gap $\delta=\Tr(D_X X_A)-\min_{Z\in\mathcal C(V)}\Tr(D_X Z_A)$.
Since $h\ge0$, these nonnegative gaps tend to zero. By compactness and
continuity each accumulation point $X$ has zero gap. Every point
$X\#_uZ$ of a geometric-mean path toward a feasible $Z$ is feasible,
so zero gap gives $\Tr[D_X(X\#_uZ-X)_A]\ge0$, and every such path
has nonnegative initial derivative. Geodesic convexity then gives
$h(Z_A)\ge h(X_A)$, which proves global optimality.
This argument also applies on the physical boundary.
For general finite-dimensional targets, alternating optimizations
can approximate geometric entanglement \cite{StreltsovAlgorithm}.
For Gaussian targets the error can be bounded from covariances alone.
With $r_V=2\log\|V\|$ and $\delta$ the gap at the current $X$, every
$0<u\le1$ gives the stopping bound
\begin{equation}
 0\le h(X_A)-G_\infty(V)
 \le \frac{\delta+d_A(e^{ur_V}-1-ur_V)}{u}.
 \label{eq:fidelityiterationerror}
\end{equation}
Uncertainty, $X\ge\Omega X^{-1}\Omega^{\top}\ge\Omega V^{-1}\Omega^{\top}$,
bounds every feasible matrix between
$\|V\|^{-1}\id$ and $\|V\|\id$. For a global minimizer $Z_*$ of $h$, the matrix
$\Delta_*=\log(X^{-1/2}Z_* X^{-1/2})$ therefore obeys $\|\Delta_*\|\le r_V$.
Since $0\le e^x-1-x\le e^{ur_V}-1-ur_V$ for $|x|\le ur_V$,
the feasible path $\gamma_u=X^{1/2}e^{u\Delta_*}X^{1/2}=X\#_uZ_*$
satisfies $0\le\gamma_u-X-u\dot\gamma_0\le(e^{ur_V}-1-ur_V)X$.
Geodesic convexity bounds $h(X_A)-h(Z_{*,A})$ by
$-\Tr[D_X(\dot\gamma_0)_A]$. Applying the gap inequality
$\Tr(D_X\gamma_{u,A})\ge\Tr(D_XX_A)-\delta$ to $\gamma_u$
and using $\Tr(D_X X_A)\le\frac12\Tr\id_{2d_A}=d_A$, which follows
from $X_A\le X_A+P(X_A)$, gives Eq.~\eqref{eq:fidelityiterationerror}.
A dual-feasible lower bound in Eq.~\eqref{eq:fidelitylinearstep}
can replace the linear minimum when computing an upper bound on
$\delta$. Exponentiating the resulting interval for $G_\infty$ gives
both bounds on the fidelity.

\section{A positive Gaussian operator}
\label{sm:operator}

The covariance minimum supplies a supporting matrix $W$ for the
convex function $G_\alpha$ (Sec.~\ref{sm:weighted}). To compare
non-Gaussian preparations we need an operator $K$ such that, for
Gaussian vectors $\psi$ of covariance $\Gamma$,
$\log\langle\psi|K|\psi\rangle$ has the same first variation in
$\Gamma$ at the optimal seed as $t_\alpha\Tr(W\Gamma)$. By
Eq.~\eqref{eq:slack} below, $W$ vanishes on the directions in which
the seed is displaced to prepare $\rho_V$, so it may be singular, and
we allow this.

Let $\boldsymbol r$ be a canonical quadrature vector obeying
the convention of the Letter, $[r_j,r_k]=i\Omega_{jk}$, with vacuum
covariance $\id$, and let $C=C^{\top}\ge0$ be real, possibly singular.
Define the positive quadratic form
\begin{equation}
 Q_C(v)
 =\mathbb E_{\ell\sim N(0,C)}
 e^{-|\ell|^2/4}\,
 \norm{e^{\ell^{\top}\boldsymbol r/2}v}^{\,2},
 \label{eq:positiveform}
\end{equation}
where the norm is $+\infty$ if $v$ is not in the domain of the
unbounded positive operator $e^{\ell^{\top}\boldsymbol r/2}$.
The real Gaussian law $N(0,C)$ may be singular, supported on
$\operatorname{range}C$. The form domain, where $Q_C(v)<\infty$,
contains the coherent vectors: by
Eq.~\eqref{eq:GaussianQ} below with $\tilde\Gamma=\id$, a normalized
coherent vector with mean $\boldsymbol m$ has
$Q_C=e^{\boldsymbol m^{\top}C\boldsymbol m/2}$.
Each integrand is lower semicontinuous in $v$, so Fatou's lemma makes
$Q_C$ lower semicontinuous, hence a closed form \cite{Kato}; its domain
is dense, since finite combinations of coherent vectors are dense.
Kato's representation theorem then gives a unique positive
self-adjoint operator $K_C$ with
$Q_C(v)=\|K_C^{1/2}v\|^2$ on the form domain, which is the domain of
$K_C^{1/2}$. For pure-state
projectors we abbreviate $Q_C(\psi)=Q_C(|\psi\rangle)$.

\begin{lemma}[Positive Gaussian inverse]
\label{lem:positiveinverse}
Set
\begin{align}
 D&=C(2\id+C)^{-1},&
 Y&=iD^{1/2}\Omega D^{1/2},\nonumber\\
 \Theta&=2D^{1/2}\chi(Y)D^{1/2},&
 a_0&=\det(\id+C/2)^{-1/2},\nonumber\\
 a_D&=\det(\id-Y^2)^{1/4}.&&
 \label{eq:operatorparameters}
\end{align}
Here $Y$ is Hermitian, and $\chi(Y)$ is defined by spectral
calculus from $\chi(x)=\atanh(x)/x$ with $\chi(0)=1$.
Then
\begin{equation}
 K_C=\frac{a_0}{a_D}e^{\boldsymbol r^{\top}\Theta\boldsymbol r/2},
 \qquad
 K_C^{-1/2}
 =\sqrt{\frac{a_D}{a_0}}e^{-\boldsymbol r^{\top}\Theta\boldsymbol r/4}.
 \label{eq:positivegaussian}
\end{equation}
In particular, $K_C\ge a_0\id$, so $K_C^{-1/2}$ is a bounded
Gaussian operator, and $K_C^{\pm1/2}$ map Gaussian vectors to
Gaussian vectors whenever the image has finite nonzero norm.
\end{lemma}

\begin{proof}
Since $\norm{e^{\ell^{\top}\boldsymbol r/2}v}^2
=\langle v|e^{\ell^{\top}\boldsymbol r}|v\rangle$ and the weight
$e^{-|\ell|^2/4}$ turns the law $N(0,C)$ into $a_0$ times the law
$N(0,2D)$, Eq.~\eqref{eq:positiveform} becomes the exact
positive-form identity
\begin{equation}
 K_C=a_0\,
 \mathbb E_{\ell\sim N(0,2D)}e^{\ell^{\top}\boldsymbol r}.
 \label{eq:completedmeasure}
\end{equation}
Averaging $\ell$ and $-\ell$ replaces $e^{\ell^{\top}\boldsymbol r}$
by $\cosh(\ell^{\top}\boldsymbol r)\ge\id$, which gives the lower bound.

To evaluate Eq.~\eqref{eq:completedmeasure}, write $\ell=\sqrt2D^{1/2}g$ with $g$
standard normal, so that
$\ell^{\top}\boldsymbol r=\sqrt2g^{\top}D^{1/2}\boldsymbol r$, where the
observables $D^{1/2}\boldsymbol r$ have
commutator matrix $iD^{1/2}\Omega D^{1/2}$. An orthogonal change of
$g$, which leaves its law invariant, reduces the real
skew matrix $D^{1/2}\Omega D^{1/2}$ into nonzero
$d_j\Omega_1$ blocks and zero blocks, where
$\Omega_1=\left(\begin{smallmatrix}0&1\\-1&0\end{smallmatrix}\right)$,
choosing the basis of the zero blocks adapted to $\ker D$.
The corresponding linear observables from the nonzero blocks,
divided by $\sqrt{d_j}$, form canonical pairs.
The remaining independent nonzero observables, the free quadratures,
commute with these pairs and with one another. Symplectic Gram--Schmidt completes
them to canonical coordinates, in which their conjugates do not appear.
The form thus splits into independent oscillator,
free-quadrature, and zero blocks, and every oscillator parameter satisfies $0<d_j<1$ because
$d_j\le\norm{D^{1/2}\Omega D^{1/2}}\le\norm D<1$.

For an oscillator block, let $\hat n$ be its number operator.
Rotational invariance of the law of $\ell$, matched by the phase
rotations $e^{i\varphi\hat n}$, makes the integral diagonal in the
number basis.
The elementary Laguerre integral gives
\[
 \mathbb E_{\ell\sim N(0,2d\id_2)}
 e^{\ell_q q+\ell_p p}
 =(1-d^2)^{-1/2}
 e^{2\atanh(d)(\hat n+1/2)}.
\]
For a commuting free quadrature the spectral theorem gives
$\mathbb E e^{\ell q}=e^{bq^2}$ for $b\ge0$ and
$\ell\sim N(0,2b)$. A zero block gives $\id$.
Expanding in the number basis, or in the spectral representation of
$q$, Tonelli's theorem identifies these
integrals with the closed positive forms of the operators on the right.
Tensoring the blocks and transforming back proves
Eq.~\eqref{eq:positivegaussian}, since $\chi(Y)$ acts as
$\atanh(d_j)/d_j$ on each oscillator block and as $1$ on the zero
blocks, and $a_D^{-1}=\prod_j(1-d_j^2)^{-1/2}$.
The matrix in the exponent can also be identified from the
convergent, symplectically covariant series
\[
 \Theta=2\sum_{k\ge0}
       \frac{D(i\Omega D)^{2k}}{2k+1}.
\]
In the canonical coordinates of the block reduction, use Bargmann
coordinates for oscillator blocks and
position coordinates for free quadratures. There $e^{c\hat n}$ acts
by the Bargmann scaling $f(z)\mapsto f(e^cz)$ and $e^{bq^2}$ by
Gaussian multiplication, which proves the
last statement.
\end{proof}

For a normalized Gaussian vector $|\psi_{\tilde\Gamma,\boldsymbol m}\rangle$
with covariance $\tilde\Gamma$ and mean $\boldsymbol m$ in the
coordinates $\boldsymbol r$,
$\langle\psi|e^{\ell^{\top}\boldsymbol r}|\psi\rangle
=\exp(\ell^{\top}\boldsymbol m+\frac14\ell^{\top}\tilde\Gamma\ell)$, so
Eq.~\eqref{eq:positiveform} gives
\begin{equation}
 Q_C(\psi_{\tilde\Gamma,\boldsymbol m})
 =\mathbb E_{\ell\sim N(0,C)}
 \exp\!\left[
   \ell^{\top}\boldsymbol m+
   \frac14\ell^{\top}(\tilde\Gamma-\id)\ell
 \right].
 \label{eq:GaussianQ}
\end{equation}
At the vacuum, $Q_C(\psi_{\id,0})=1$.

\section{The weighted norm bound and entanglement of formation}
\label{sm:weighted}

Fix $\alpha\in\cI$, and first suppose
$V+i\Omega>0$, so that $V$ is an interior point of the convex set of
physical covariances, on which $G_\alpha$ is convex
(Sec.~\ref{sm:geometry}).
It therefore has a subgradient at $V$, which we write as $-W$ with $W$ real symmetric:
$G_\alpha(V')\ge G_\alpha(V)-\Tr[W(V'-V)]$ for every physical $V'$.
Since $G_\alpha$ is
nonincreasing in the covariance order, $\Tr[W(V'-V)]\ge0$ for every
$V'\ge V$, so $W\ge0$.
If $\Gamma_*$ is a minimizer in Eq.~\eqref{eq:Gaussianroof} and $e_*=G_{\alpha}(V)$, then
\begin{align}
 e_{\alpha}(\Gamma)+\Tr(W\Gamma)
 &\ge e_*+\Tr(W\Gamma_*)
 \quad\hbox{for every pure }\Gamma,
 \label{eq:support}\\
 \Tr[W(V-\Gamma_*)]&=0.
 \label{eq:slack}
\end{align}
Indeed, at a pure $V'=\Gamma$, where $G_{\alpha}(\Gamma)=e_{\alpha}(\Gamma)$
(Sec.~\ref{sm:geometry}), the subgradient inequality reads
$e_\alpha(\Gamma)+\Tr(W\Gamma)\ge e_*+\Tr(WV)$. At $\Gamma=\Gamma_*$
it gives $\Tr[W(V-\Gamma_*)]\le0$, while $W\ge0$ and
$V-\Gamma_*\ge0$ give the opposite sign. This proves
Eq.~\eqref{eq:slack}, and inserting $\Tr(WV)=\Tr(W\Gamma_*)$
proves Eq.~\eqref{eq:support}.

Write $\Gamma_*=SS^{\top}$ with $S$ symplectic. In the
canonical coordinates $\boldsymbol r=S^{-1}\boldsymbol R$, the seed has
covariance $S^{-1}\Gamma_*S^{-\top}=\id$, so it is the
vacuum. Set $W_0=S^{\top}WS$, $C=4t_\alpha W_0$, and $K=K_C$.
For a normalized Gaussian vector $\psi$ of covariance $\Gamma$, its
covariance in these coordinates is $\tilde\Gamma=S^{-1}\Gamma S^{-\top}$.
Jensen's inequality in
Eq.~\eqref{eq:GaussianQ}, with $\mathbb E\ell=0$ and
$\mathbb E\ell\ell^{\top}=C$, gives
\[
 \log Q_C(\psi)\ge\tfrac14\Tr[C(\tilde\Gamma-\id)]
             =t_\alpha\Tr[W(\Gamma-\Gamma_*)].
\]
By Eq.~\eqref{eq:support}, a normalized Gaussian $\psi$ therefore
obeys $\|\psi_A\|_\alpha=e^{-t_\alpha e_\alpha(\Gamma)}
\le e^{-t_\alpha e_*}Q_C(\psi)$, and since both sides scale as
$\|\psi\|^2$, so does every Gaussian
vector. To extend this inequality, let $\mathcal G=K^{-1/2}$.
This bounded Gaussian map (Lemma~\ref{lem:positiveinverse}) satisfies
$Q_C(\mathcal Gf)=\|f\|^2$ and maps Gaussian $f$
to Gaussian vectors. With $T_v$ the coefficient operator of
Sec.~\ref{sm:norm}, so that $\|T_v\|_{2\alpha}^2=\|v_A\|_\alpha$, the bound
gives $\|T_{\mathcal Gf}\|_{2\alpha}^2\le e^{-t_\alpha e_*}\|f\|^2$
for Gaussian $f$. Equation~\eqref{eq:exhaustion}, which equates the
supremum over all inputs with the supremum over Gaussian inputs, and
its infinite-order limit extend this to every $f$.
Taking $f=K^{1/2}|\psi\rangle$, so that $\mathcal Gf=|\psi\rangle$
and $\|f\|^2=Q_C(\psi)$, proves the bound on
the form domain,
\begin{equation}
 \norm{\psi_A}_\alpha=\norm{T_{|\psi\rangle}}_{2\alpha}^2
                 \le e^{-t_\alpha e_*}Q_C(\psi).
 \label{eq:fullweightedbound}
\end{equation}
This is Eq.~\eqref{eq:weightedbound} of the End Matter, with
$K=K_C$. The Schmidt norms throughout refer to the original $A:B$ cut.

To compute the expectation of $K$ in the target state, note that in
$\boldsymbol r$ coordinates the random displacement that turns the
seed into $\rho_V$ has covariance
\[
 \frac12\Delta_0,\qquad
 \Delta_0=S^{-1}(V-\Gamma_*)S^{-\top}.
\]
Equation~\eqref{eq:slack} gives $\Tr(W_0\Delta_0)=0$, and positivity
of $W_0$ and $\Delta_0$ then implies
$W_0\Delta_0=0$.
Thus the displacement $\boldsymbol\xi$ lies in
$\operatorname{range}\Delta_0\subseteq\ker C$, while
$\ell\sim N(0,C)$ lies in $\operatorname{range}C$. Displacing $v$ by
$\boldsymbol\xi$ multiplies
$\norm{e^{\ell^{\top}\boldsymbol r/2}v}^2$ by $e^{\ell^{\top}\boldsymbol\xi}=1$,
so the form in Eq.~\eqref{eq:positiveform} is invariant under
these displacements. Every displaced seed therefore has the value
$1$ of the vacuum, and
\begin{equation}
 \Tr(\rho_VK)=1.
 \label{eq:meanweight}
\end{equation}

Let $\mu$ be any pure-state ensemble of $\rho_V$, as in
Eq.~\eqref{eq:arithroof}, and let
$K\wedge c\,\id=\min\{K,c\,\id\}$, defined by spectral calculus, be the
bounded truncation of $K$. Linearity gives
$\int\langle\psi|K\wedge c\,\id|\psi\rangle\,\dd\mu
=\Tr[\rho_V(K\wedge c\,\id)]$ for every $c$, independently of $\mu$.
As $c\uparrow\infty$, monotone convergence of
form expectations takes the left side to $\int Q_C(\psi)\,\dd\mu$
and, evaluated with the Gaussian displacement ensemble, the right side
to $1$ by Eq.~\eqref{eq:meanweight}. Hence
\[
 \int Q_C(\psi)\,\dd\mu(\psi)=1.
\]
Almost every component is therefore in the form domain.
Averaging Eq.~\eqref{eq:fullweightedbound} and taking the supremum
over $\mu$ proves
\[
 \ensnorm_{\alpha}(\rho_V)\le e^{-t_{\alpha}G_{\alpha}(V)}.
\]
For every ensemble Jensen's inequality for the concave logarithm also gives
$\int S_\alpha(\psi_A)\,\dd\mu\ge
-t_\alpha^{-1}\log\int\|\psi_A\|_\alpha\,\dd\mu\ge G_\alpha(V)$.

The Gaussian displacement ensemble, whose components all share the
Schmidt spectrum of the seed $\psi_{\Gamma_*}$, attains both bounds,
proving Eq.~\eqref{eq:mainformation} for strictly physical
covariances.

For a boundary covariance, apply independent local Gaussian
displacements of covariance $\epsilon\id/2$ to the components of any
ensemble of $\rho_V$ to prepare
$\rho_{V+\epsilon\id}$.
They preserve every component's Schmidt spectrum.
Hence
\[
 E^{\rm av}_{F,\alpha}(\rho_V)
 \ge E^{\rm av}_{F,\alpha}(\rho_{V+\epsilon\id}),\qquad
 \ensnorm_{\alpha}(\rho_V)\le\ensnorm_{\alpha}(\rho_{V+\epsilon\id}).
\]
Since $V+\epsilon\id$ is strictly physical, the right sides are
$G_\alpha(V+\epsilon\id)$ and $e^{-t_\alpha G_\alpha(V+\epsilon\id)}$,
so $E^{\rm av}_{F,\alpha}(\rho_V)$ and $E^{\rm c}_{F,\alpha}(\rho_V)$
are at least $G_\alpha(V+\epsilon\id)$, which tends to $G_\alpha(V)$
by Eq.~\eqref{eq:boundarycontinuity}. Together with the Gaussian upper
bound $E^{\rm c}_{F,\alpha}\le E^{\rm av}_{F,\alpha}\le G_\alpha(V)$
of the Letter, this finishes the proof.

For any second
covariance $V'$, the equality just proved applies to the Gaussian state
$\rho_V\otimes\rho_{V'}$ with covariance $V\oplus V'$, and hence to
all its ensembles, including those that correlate the copies.
Since $G_\alpha(V\oplus V')=G_\alpha(V)+G_\alpha(V')$
(Sec.~\ref{sm:geometry}), additivity follows, and
Eq.~\eqref{eq:strongsuper} gives strong superadditivity.
Theorem~\ref{thm:formation} is thus proved at every stated order.
We record for later use the elementary uniform bound
\begin{equation}
 t_\alpha S_\alpha(\rho)\le S_\infty(\rho)\le S_\alpha(\rho),
 \label{eq:entropysqueeze}
\end{equation}
which follows from
$\lambda_{\max}^\alpha\le\Tr\rho^\alpha\le\lambda_{\max}^{\alpha-1}\Tr\rho=\lambda_{\max}^{\alpha-1}$,
with $\lambda_{\max}$ the largest eigenvalue of $\rho$.
\section{Dilution fidelity and the entanglement cost}
\label{sm:dilution}

A purification argument determines the optimal dilution fidelity
directly on bosonic Hilbert spaces, extending the finite-dimensional
relations of Refs.~\cite{Streltsov,BowenDatta,BuscemiDatta,TheurerFangGour}.

\subsection{Schmidt-rank fidelity bound}

For a pure state, the Ky Fan norm $\|\psi_A\|_{(M)}$ of the Letter is
the sum of its $M$ largest Schmidt probabilities.
For every state $\rho$ and every $\sigma\in\SN_M$,
\begin{equation}
 \Fid(\rho,\sigma)
 \le
 \sup_{\rho=\int\psi\,\dd\mu}
       \int \|\psi_A\|_{(M)}\,\dd\mu(\psi).
 \label{eq:uhlmannbound}
\end{equation}
To prove it, first take
$\sigma=\sum_iw_i\phi_i$ with finitely many unit vectors $\phi_i$ of
Schmidt rank at most $M$.
Fix the purification
$\sum_i\sqrt{w_i}\ket{\phi_i}\ket i$ in a countable ancilla,
padding by zero weights.
Any purification $\sum_i\sqrt{p_i}\ket{\psi_i}\ket i$ of $\rho$ on this
ancilla gives an exact countable ensemble $\{p_i,\psi_i\}$.
Using the maximal rank-$M$ overlap
$|\langle\psi_i|\phi_i\rangle|^2\le\|\psi_{i,A}\|_{(M)}$ and
Cauchy--Schwarz,
\begin{align}
 \left|\sum_i\sqrt{p_iw_i}
                  \langle\psi_i|\phi_i\rangle\right|^2
 &\le
 \left(\sum_i\sqrt{p_iw_i\,\|\psi_{i,A}\|_{(M)}}\right)^2
 \nonumber\\
 &\le\sum_i p_i \|\psi_{i,A}\|_{(M)}.
\end{align}
By Uhlmann's theorem on separable Hilbert spaces
\cite{HouQi}, $\Fid(\rho,\sigma)$ is the supremum of the left side over
these purifications, which proves Eq.~\eqref{eq:uhlmannbound} for
finite mixtures.
Fidelity continuity extends it to $\SN_M$.

Conversely, LOCC preparations from $\Phi_M$ attain the supremum. Let $\{p_i,\psi_i\}$ be a
countable ensemble of $\rho$, write $c_i=\|\psi_{i,A}\|_{(M)}$, and let
$\phi_i$ be the normalized truncation of $\psi_i$ to its $M$ leading
Schmidt terms, so that $\langle\psi_i|\phi_i\rangle=\sqrt{c_i}$.
Alice samples $i$ with probability $w_i=p_ic_i/Z$, where
$Z=\sum_ip_ic_i$, and tells Bob. They then convert $\Phi_M$ into
$\phi_i$ by LOCC, since the uniform Schmidt vector of $\Phi_M$ is
majorized by every Schmidt vector of rank at most $M$
\cite{Nielsen,VidalJonathanNielsen}.
Uhlmann's theorem with the purifications
$\sum_i\sqrt{p_i}\ket{\psi_i}\ket i$ of $\rho$ and
$\sum_i\sqrt{w_i}\ket{\phi_i}\ket i$ of the output
$\sigma=\sum_iw_i\phi_i$ gives
\[
 \Fid(\rho,\sigma)\ge\Bigl(\sum_i\sqrt{p_iw_ic_i}\Bigr)^2
 =\frac1Z\Bigl(\sum_ip_ic_i\Bigr)^2
 =\sum_ip_i\|\psi_{i,A}\|_{(M)} ,
\]
and finitely many outcomes approach this value by fidelity continuity.
For an arbitrary ensemble $\mu$, choose a countable dense list of unit
vectors of Schmidt rank at most $M$. For each $\psi$, take the first
vector $\phi(\psi)$ on the list whose squared overlap $b(\psi)$ with
$\psi$ exceeds $(1-\delta)\|\psi_A\|_{(M)}$, with $0<\delta<1$. This
choice is measurable since $\psi\mapsto\|\psi_A\|_{(M)}$ is continuous.
Preparing $\phi(\psi)$ with weight $b(\psi)\,\dd\mu(\psi)/Z$, where now
$Z=\int b\,\dd\mu$, is a countable mixture of the same kind, as $\phi$
takes countably many values. With
$L^2(\mu)$ as the purifying system, $\psi\mapsto\ket\psi$ purifies
$\rho$ and $\psi\mapsto\sqrt{b(\psi)/Z}\,\ket{\phi(\psi)}$ purifies the
output; once the phases are aligned, their overlap is
$\int\sqrt{b/Z}\sqrt b\,\dd\mu=\sqrt Z$, and
$Z\ge(1-\delta)\int\|\psi_A\|_{(M)}\dd\mu$ with $\delta$ arbitrarily
small. Finally, LOCC cannot increase the Schmidt number, so every LOCC
output from a Schmidt-rank-$k$ resource lies in the closed set $\SN_k$,
limits of protocols included. Applied to $\rho^{\otimes n}$
with $M=k$, Eq.~\eqref{eq:uhlmannbound}
and this construction therefore give the optimal fidelity
$F^*_\rho(n,k)$ for preparing $\rho^{\otimes n}$ from $\Phi_k$ by LOCC,
\begin{equation}
 F^*_\rho(n,k)=\sup_{\int\psi\,\dd\mu=\rho^{\otimes n}}
                 \int \|\psi_A\|_{(k)}\,\dd\mu(\psi).
 \label{eq:generalrankfidelity}
\end{equation}
For a Gaussian target, $F^*_{\rho_V}(n,M)$ is the dilution fidelity
$F^*_V(n,M)$ of the Letter, and $F^*_\rho(1,1)=F_{\Sep}(\rho)$ for every
state, since $\SN_1=\Sep$.

For $\alpha\in\cI$, H\"older's inequality
$\|\psi_A\|_{(M)}\le M^{t_{\alpha}}\norm{\psi_A}_{\alpha}$ and
Theorem~\ref{thm:formation}, which evaluates the averaged norm
$\ensnorm_\alpha$ of Eq.~\eqref{eq:conditionalroof}, turn
Eq.~\eqref{eq:uhlmannbound} into
\begin{equation}
 \Fid(\rho_V,\sigma)
 \le M^{t_{\alpha}}\ensnorm_{\alpha}(\rho_V)
 =M^{t_{\alpha}}e^{-t_{\alpha}G_{\alpha}(V)},
 \qquad \sigma\in\SN_M.
 \label{eq:schmidtfidelity}
\end{equation}
Applying it to $\rho_V^{\otimes n}$, with $G_\alpha$ additive, and
optimizing over $\alpha\in\cI$ gives
Eq.~\eqref{eq:dilution}.

\subsection{The closest separable state}

At $M=1$, $\SN_1=\Sep$ and $\|\psi_A\|_{(1)}=\|\psi_A\|_\infty$, so
Eq.~\eqref{eq:generalrankfidelity} with $n=k=1$ gives
$F_{\Sep}(\rho)=\ensnorm_\infty(\rho)$ for every state, and
Theorem~\ref{thm:formation} evaluates it as $e^{-G_\infty(V)}$ for
$\rho_V$. The Gaussian separable state $\sigma_*$ of
Eq.~\eqref{eq:separablecovariance} attains this value by the
joint-concavity argument of the Letter, which proves
Corollary~\ref{cor:closest}. At $M=1$ the bound \eqref{eq:dilution}
reads $e^{-nG_\infty(V)}$, because $t_\alpha G_\alpha\le G_\infty$
[Eq.~\eqref{eq:entropysqueeze}] puts the infimum at $\alpha=\infty$.
Since $\Phi_1$ is a product state, LOCC prepares
$\sigma_*^{\otimes n}$ from it, and by multiplicativity its fidelity is
exactly this value, the zero-resource dilution fidelity.
The geometric entanglement $1-F_\Sep$ and the Bures distance quoted
in the Letter are functions of $F_\Sep$, as is the separable minimum
of the order-$1/2$ sandwiched R\'enyi relative entropy
\begin{equation}
 \widetilde D_{1/2}(\rho\|\sigma)=-\log\Fid(\rho,\sigma),
 \label{eq:sandwichedhalf}
\end{equation}
namely $-\log F_\Sep(\rho_V)$, which equals $G_\infty(V)$.
Equation~\eqref{eq:fidproduct} follows from Corollary~\ref{cor:closest}
applied to $\rho_V\otimes\rho_{V'}=\rho_{V\oplus V'}$, together with
additivity of $G_\infty$ on $V\oplus V'$.

\subsection{The order-one limit}

For fixed finite $V$, its feasible pure covariances (pure $\Gamma\le V$)
form a compact set. Their ordinary eigenvalues are uniformly
bounded above and away from zero, so the symplectic eigenvalues of
their $A$ blocks lie in a compact interval $[1,c_V]$.
The R\'enyi entropy of a single-mode Gaussian state with symplectic
eigenvalue $\nu$,
\[
 S_{\alpha}(\nu)=\frac1{\alpha-1}\log\left[
       \left(\frac{\nu+1}{2}\right)^\alpha-
       \left(\frac{\nu-1}{2}\right)^\alpha\right]
\]
extends jointly continuously to $\alpha=1$, $\nu\ge1$, with
value $g((\nu-1)/2)$, with $g$ the thermal-mode entropy of Sec.~\ref{sm:orderseed}.
The convergence is uniform on that
compact interval, and hence on the feasible set for the seed costs,
which are sums of such terms. Their minima
therefore obey
\begin{equation}
 \lim_{\alpha\to1^+}G_{\alpha}(V)=G_1(V)=E_C(\rho_V),
 \label{eq:orderonelimit}
\end{equation}
where the entanglement cost $E_C$ is the infimum of the rates of
shared entanglement at
which the target can be prepared by LOCC with fidelity tending to
one \cite{GA,Yamasaki}. The argument
also covers covariances on the boundary of the uncertainty relation
$V+i\Omega\ge0$.

\subsection{The strong converse exponent of dilution for an arbitrary bipartite target}

This subsection proves the exponent formula for every bipartite target:
Eq.~\eqref{eq:dyadicpreparation} sandwiches the averaged norms
$\ensnorm_\alpha$ between dilution fidelities, with dyadic ranks $2^j$
on the upper side, and Legendre inversion expresses the exponent through
their regularized rates, which Theorem~\ref{thm:formation} makes
single-letter for Gaussian targets.

Let $\rho$ be any bipartite state, with optimal fidelities
$F^*_\rho(n,k)$ given by Eq.~\eqref{eq:generalrankfidelity}.
Measure rates in nats and set $M_n=\lfloor e^{n\mathcal R}\rfloor$, $\mathcal R\ge0$.
Product protocols give
$F^*_\rho(n+m,k\ell)\ge F^*_\rho(n,k)F^*_\rho(m,\ell)$, and
$M_nM_m\le M_{n+m}$. Since $F^*_\rho$
is nondecreasing in $k$, the sequence $-\log F^*_\rho(n,M_n)$
is subadditive in $n$, and Fekete's lemma \cite{Fekete} gives the limit
\[
 \mathscr E(\mathcal R)=\lim_{n\to\infty}-\frac1n\log F^*_\rho(n,M_n)
     =\inf_n-\frac1n\log F^*_\rho(n,M_n).
\]
It is finite since $\mathscr E\le-\log F^*_\rho(1,1)$ and
$F^*_\rho(1,1)>0$. For $\ell\ge k$ and every state $\omega$,
$\|\omega\|_{(k)}\le\|\omega\|_{(\ell)}\le(\ell/k)\|\omega\|_{(k)}$.
By Eq.~\eqref{eq:generalrankfidelity} the same inequalities hold for
$F^*_\rho(n,\cdot)$, so
$\mathscr E$ is nonincreasing and
$1$-Lipschitz, including at $\mathcal R=0$. Time sharing between two
rates gives convexity first
for rational proportions and then, by continuity, for all proportions.
Also $\mathscr E(\log k)\le-\log F^*_\rho(1,k)\to0$ as $k\to\infty$, by
the infimum formula at $n=1$.

For $0<t\le1$, put $\alpha=1/(1-t)$, so that $t=t_\alpha$, with
$\alpha=\infty$ at $t=1$.
H\"older's inequality gives $k^{-t}\|\psi_A\|_{(k)}\le\|\psi_A\|_\alpha$.
Let $p_1\ge p_2\ge\dots$ be the Schmidt probabilities of $\psi$.
In the block $2^j\le i<2^{j+1}$ each $p_i$ is at most
$2^{-j}\|\psi_A\|_{(2^j)}$, so that block has $\ell_\alpha$ norm at most
$2^{-jt}\|\psi_A\|_{(2^j)}$.
Summing the block norms and optimizing over
ensembles gives
\begin{equation}
 \sup_{k\ge1} k^{-t}F^*_\rho(n,k)\ \le\ \ensnorm_\alpha(\rho^{\otimes n})
 \ \le\ \sum_{j\ge0}2^{-jt}F^*_\rho(n,2^j).
 \label{eq:dyadicpreparation}
\end{equation}
Both sides have the same exponent $a(t)=\inf_{\mathcal R\ge0}[\mathscr E(\mathcal R)+t\mathcal R]$.
The left side allows $k=M_n$ for every $\mathcal R$. On the right,
each term is at most $e^{-na(t)}$ by the infimum formula for $\mathscr E$ at
$\mathcal R=j\log2/n$, where $M_n=2^j$. For $C$ with
$Ct\log2>\mathscr E(0)$, the first $\lfloor Cn\rfloor+1$ terms
contribute only a linear prefactor, and the remaining geometric tail,
at most $2^{-Cnt}/(1-2^{-t})$, decays faster than
$e^{-n\mathscr E(0)}\le e^{-na(t)}$.
Consequently $a(t)=\lim_n-n^{-1}\log\ensnorm_\alpha(\rho^{\otimes n})$
is $t$ times the regularized conditional R\'enyi entanglement of
formation of Eq.~\eqref{eq:conditionalroof}.

The endpoints are $a(1)=\mathscr E(0)$ and $a(0)=\lim_{t\downarrow0}a(t)=0$,
since $\mathscr E$ has slopes in $[-1,0]$ and tends to zero at high rates.
Being convex with these slopes, $\mathscr E$ is the upper
envelope of its supporting lines of slope $-t$, $0\le t\le1$, whose
intercepts are $a(t)$; hence
\begin{equation}
 \mathscr E(\mathcal R)=\sup_{0\le t\le1}\{a(t)-t\mathcal R\}.
 \label{eq:generalpreparationduality}
\end{equation}
For a Gaussian target Theorem~\ref{thm:formation} and additivity give
$a(t)=tG_{1/(1-t)}(V)$ for $t>0$, with $F^*_{\rho_V}=F_V^*$ and
$\mathscr E=\mathscr E_V$. This proves Eq.~\eqref{eq:generalexponent};
the supremum over $\alpha\in\cI$ there equals the one over
$0\le t\le1$ here, since the omitted endpoint $t=0$ contributes $a(0)=0$ and $a$ is continuous there.
Since $G_\alpha\le G_1$ and $G_\alpha\to G_1$ as $\alpha\downarrow1$
[Eq.~\eqref{eq:orderonelimit}], every term
$t_\alpha[G_\alpha(V)-\mathcal R]$ is nonpositive for
$\mathcal R\ge G_1(V)$, while terms with $\alpha$ near one are
positive for $\mathcal R<G_1(V)$. Thus the exponent vanishes exactly at and above
the cost, and $a(1)=\mathscr E(0)$ gives $\mathscr E_V(0)=G_\infty(V)$.
The one-copy covariance optimization therefore determines the full exponent. Rates and exponents in bits
are obtained by dividing both axes by $\log2$.

\subsection{One Gaussian seed attains the strong converse exponent}

An explicit protocol attains the exponent of
Eq.~\eqref{eq:generalexponent} with one Gaussian seed, truncated to
its most likely Schmidt strings; together with the bound
\eqref{eq:dilution} it gives a constructive proof of that equation.
Fix a rate $\mathcal R\ge0$ and $M_n=\lfloor e^{n\mathcal R}\rfloor$.
We first choose a seed adapted to this rate and then truncate
the Schmidt expansion of its $n$th tensor power.

Let $\mathcal K_V$ be the compact set of feasible pure covariances
$\Gamma\le V$,
and let $p_\Gamma$ be the Schmidt probability distribution of
$\psi_\Gamma$. For $0\le t\le1$ write
\[
 \phi_\Gamma(t)=-\log\|p_\Gamma\|_{1/(1-t)},\qquad
 h_V(t)=\min_{\Gamma\in\mathcal K_V}\phi_\Gamma(t),
\]
so that $\phi_\Gamma(t)=t\,e_\alpha(\Gamma)$ for $t>0$, with
$\alpha=1/(1-t)$ and $e_\alpha$ the seed cost of
Eq.~\eqref{eq:scalarcost}.
Here $\phi_\Gamma(0)=0$ and $\phi_\Gamma(1)=-\log\max_i p_{\Gamma,i}$;
hence $h_V(0)=0$ and $h_V(1)=G_\infty(V)$, while
$h_V(t)=tG_{1/(1-t)}(V)$ for $0<t<1$.
Interpolation of sequence norms ($\log\|p\|_{1/s}$ is convex in $s$)
makes $\phi_\Gamma$ concave in $t$, and
nondecreasing since $\|p\|_\alpha$ decreases with $\alpha$.
It is jointly continuous in $(\Gamma,t)$, including at the endpoints,
where Eq.~\eqref{eq:entropysqueeze} bounds $\phi_\Gamma(t)$ and
$\phi_\Gamma(1)-\phi_\Gamma(t)$ by $t$ and $1-t$ times
$S_1(\psi_{\Gamma,A})$, which is bounded on $\mathcal K_V$.

Consider the average of $\phi_\Gamma(t)-t\mathcal R$ under a probability
measure $\mu$ on $\mathcal K_V$. It is affine and weakly continuous in
$\mu$ and concave in $t$. Sion's minimax theorem \cite{Sion} therefore
interchanges the minimum over $\mu$ and the maximum over $t$; at fixed
$t$, point masses give the minimum $h_V(t)-t\mathcal R$.
A single seed also achieves the minimum over $\mu$ of the maximum over
$t$. Indeed, for a finite mixture with weights $w_i$,
combine its seeds $\Gamma_i$ successively by geometric means,
weighting the $i$th by $u_i=w_i/\sum_{j\le i}w_j$. The result $\Gamma$
is feasible, as geometric means preserve purity and the bound $\le V$,
and Eq.~\eqref{eq:geoconvex} gives
$\phi_\Gamma(t)\le\int\phi_{\Gamma'}(t)\,\dd\mu(\Gamma')$ for every $t$.
For an arbitrary $\mu$, approximate it weakly by finite mixtures and
take a convergent subsequence of their seeds; joint continuity passes
the inequality to the limit. Thus
\[
 \min_{\Gamma\in\mathcal K_V}\max_{0\le t\le1}
    [\phi_\Gamma(t)-t\mathcal R]
 =\max_{0\le t\le1}[h_V(t)-t\mathcal R].
\]

Let $t_*$ maximize the right-hand side and let $\Gamma_*$ attain the
left-hand side. Since $h_V\le\phi_{\Gamma_*}$, the chain
$h_V(t_*)-t_*\mathcal R\le\phi_{\Gamma_*}(t_*)-t_*\mathcal R
\le\max_t[\phi_{\Gamma_*}(t)-t\mathcal R]$ begins and ends at the
common value of the two sides above. Hence $t_*$ also
maximizes $\phi_{\Gamma_*}(t)-t\mathcal R$, and
$\phi_{\Gamma_*}(t_*)=h_V(t_*)$.
If $0<t_*<1$, $\Gamma_*$ thus attains $G_\alpha(V)$ at
$\alpha=1/(1-t_*)$, and the single-seed function
$\phi_{\Gamma_*}(t)-t\mathcal R$, differentiable for $t<1$, is
stationary at $t_*$, so that $\phi'_{\Gamma_*}(t_*)=\mathcal R$ whether
or not $h_V$ is differentiable there. The derivative is the entropy of the
tilted distribution, as shown below, and it tends to zero as $t\uparrow1$
because the Schmidt distribution of a Gaussian seed has a unique largest
term. By concavity, maximality at $t=1$ requires
$\mathcal R\le\phi'_{\Gamma_*}(t)$ for all $t<1$, hence $\mathcal R\le0$,
so the endpoint
$t_*=1$ occurs only at $\mathcal R=0$. There $\Gamma_*$ is an optimal
seed of order $\infty$ and the exponent is $h_V(1)=G_\infty(V)$. At
$t_*=0$, which forces $\mathcal R\ge G_1(V)$, the exponent is zero and
one can choose any seed minimizing the
von Neumann entropy.

Fix a seed chosen in this way for the rate $\mathcal R$, with Schmidt
distribution $p$.
The normalized truncation of its $n$th tensor power to the $M_n$
largest Schmidt terms has
squared overlap $\|p^{\otimes n}\|_{(M_n)}$ with the untruncated
$n$-copy seed.
Alice and Bob obtain this truncation from $\Phi_{M_n}$ by LOCC; for example,
Alice prepares it locally and teleports its $B$ part, which is
supported on the span of $M_n$ Schmidt vectors.
Applying the same shared random
displacements that prepare $\rho_V^{\otimes n}$ from the $n$-copy seed
cannot decrease the fidelity, by joint concavity of the root fidelity.
Hence
$F_V^*(n,M_n)\ge \|p^{\otimes n}\|_{(M_n)}$.
Finite shared-randomness mixtures approximate the displaced truncation
in trace norm, giving the same supremum.

The decay of $\|p^{\otimes n}\|_{(M_n)}$ is the strong converse
exponent of fixed-length source coding of the i.i.d.\ source $p$,
derived for pure-state quantum sources in Ref.~\cite{HayashiExponents};
we include the short argument for the present spectra, in which $p$ is
a finite product of geometric distributions.
If it is not a point mass, its distribution tilted to order $\alpha>1$,
$p_i^{(\alpha)}=p_i^\alpha/\sum_j p_j^\alpha$, has Shannon entropy
$H_\alpha=H(p^{(\alpha)})$ decreasing continuously from $H(p)$
to zero as $\alpha$ runs from $1$ to $\infty$. At $t=t_\alpha$,
differentiation gives
$\phi'_\Gamma(t)=H_\alpha$ and
$D(p^{(\alpha)}\|p)=\phi_\Gamma(t)-tH_\alpha$, where $D$ is classical relative
entropy. When $H_\alpha<\mathcal R$, fix $0<\delta<\mathcal R-H_\alpha$
and let the typical set $T_n$ contain the strings whose information
$-n^{-1}\log p^{(\alpha)\otimes n}_i$ and log-likelihood ratio
$n^{-1}\log(p^{(\alpha)\otimes n}_i/p^{\otimes n}_i)$ lie within
$\delta$ of $H_\alpha$ and $D(p^{(\alpha)}\|p)$. It has at most
$e^{n(H_\alpha+\delta)}\le M_n$ elements for large $n$, and
$p^{\otimes n}(T_n)\ge e^{-n[D(p^{(\alpha)}\|p)+\delta]}
p^{(\alpha)\otimes n}(T_n)$, where $p^{(\alpha)\otimes n}(T_n)\to1$ by
the law of large numbers, since both quantities have finite variance
for a geometric spectrum. Thus
$\|p^{\otimes n}\|_{(M_n)}\ge e^{-n[D(p^{(\alpha)}\|p)+o(1)]}$.
For $0<\mathcal R\le H(p)$, let $H_\alpha$ approach $\mathcal R$ from
below. Then $D(p^{(\alpha)}\|p)$ tends to the value of the concave
function $\phi_\Gamma(t)-t\mathcal R$ at its stationary point, so the
decay exponent is at most
$\max_t[\phi_\Gamma(t)-t\mathcal R]$, which for the chosen seed equals
$\max_t[h_V(t)-t\mathcal R]$. Above $H(p)$ the ordinary typical
set gives exponent zero. At $\mathcal R=0$ the largest probability is
$(\max_i p_i)^n$. If $p$ is a point mass, the seed is a product and
prepares the separable Gaussian target exactly at every rate. Hence a single
displaced Gaussian seed, truncated to its $M_n$ most likely Schmidt
strings, attains the exponent~\eqref{eq:generalexponent} for every
finite-mode Gaussian target.

\section{Channel simulation}
\label{sm:simulation}

We prove Theorem~\ref{thm:simulation} against adaptive tests, evaluate
the optimum in closed form, and derive the finite-brightness benchmark
of the End Matter.

\subsection{Adaptive tests for optical simulation}

A test $\mathcal T$ of $n$ channel calls, a quantum comb
\cite{ChiribellaCombs}, prepares an input and a reference that it
keeps, turns each output and its memory into the next input by an
arbitrary quantum operation, and retains its final quantum memory,
last output included. For an $n$-round simulator $\mathcal S$, an LOCC
protocol consuming the maximally entangled state $\Phi_M$ of Schmidt
rank $M$, write $\rho_{\mathcal T}^{\mathcal S}$ for this final state,
and $\rho_{\mathcal T}^{\mathcal N}$ for the one with $n$ ideal calls to
$\mathcal N$. The simulation fidelity is then
\begin{equation}
 F^{\rm sim}_{n,M}(\mathcal N)=
 \sup_{\mathcal S\in\mathrm{LOCC}(\Phi_M)}
 \inf_{\mathcal T}
 \Fid(\rho_{\mathcal T}^{\mathcal N},\rho_{\mathcal T}^{\mathcal S}).
\end{equation}
Parallel reference-assisted inputs are included as tests,
and the infimum has no input-energy restriction.

For the converse, a test may place a pure Gaussian probe and its
reference on the sender's side before the protocol.
Across the cut between sender and receiver, the initial Schmidt rank is
that of the resource, at most $M$, and LOCC cannot increase it.
After discarding all sender systems other than the
reference, the reference--output state therefore remains in $\SN_M$.
Equation~\eqref{eq:uhlmannbound} bounds its fidelity
with the ideal reference--output state for each finite-energy Gaussian
probe. For a measure-and-prepare simulator the output is separable, and
the bound applies with $M=1$.

\subsection{Optimality of the truncated optical resource}
Let
$|\Psi_\theta\rangle=\sqrt{1-\theta}
\sum_{k\ge0}\theta^{k/2}|k,k\rangle$ be the two-mode squeezed vacuum
whose arms are in the state $\tau_\theta$ of
Eq.~\eqref{eq:thermaltheta}; it is the vector $\Psi_\lambda$ of
Sec.~\ref{sm:norm} with $\lambda=\sqrt\theta$.
Braunstein--Kimble teleportation with amplitude gain
$\sqrt\eta$ and resource $\Psi_\eta$ implements
$\mathcal L_\eta$ exactly. Gain $\sqrt\kappa$ and
$\Psi_{1/\kappa}$ implement $\mathcal A_\kappa$
\cite{BraunsteinKimble,LiuzzoScorpoTeleportation,LaurenzaFinite,WildeCost}.
The same teleportation
circuit can be supplied with this exact resource or a truncated one.
Truncating $\Psi_\theta^{\otimes n}$ to its $M$ largest
Schmidt probabilities gives a normalized vector of rank at
most $M$, preparable from a maximally entangled rank-$M$
state by LOCC \cite{Nielsen}. Its squared overlap with
the ideal resource is $\|\tau_\theta^{\otimes n}\|_{(M)}$.
For a fixed adaptive test, the teleportation rounds and the test form
one channel, mapping the ideal resource to
$\rho_{\mathcal T}^{\mathcal N}$ and the truncated one to
$\rho_{\mathcal T}^{\mathcal S}$, so fidelity monotonicity proves
achievability uniformly for all tests.

For the converse, let $\rho_{\eta,q}$ be the reference--output state
obtained by sending one arm of $\Psi_q$, with mean photon number $N$
per arm and $q=N/(N+1)$, through $\mathcal L_\eta$. With
$\theta_N=\eta q$ and $\tanh^2r_N=\theta_N$, it has the exact
representation
\begin{equation}
 \rho_{\eta,q}=U_{r_N}
 [\omega_{(1-\eta)N}\otimes|0\rangle\langle0|]U_{r_N}^\dagger,
 \label{sm:support}
\end{equation}
where $U_r=\exp[r(a_R^\dagger a_B^\dagger-a_Ra_B)]$ is the two-mode
squeezing unitary on the reference mode $R$ and the output mode $B$, and
$\omega_x$, on the reference, is thermal
with mean occupation $x$. Almost every vector in a pure-state ensemble of
$\rho_{\eta,q}^{\otimes n}$ lies in its support and thus, as $\omega_x$
has full support,
has the form $U_{r_N}^{\otimes n}(|f\rangle\otimes|0\rangle^{\otimes n})$
with $f$ a vector of the $n$ reference modes.
Its reference marginal is the output for input $f$ of the
quantum-limited amplifier of gain $\cosh^2r_N$. The multimode
output-majorization theorem \cite{GHM}, by which the vacuum output
$\tau_{\theta_N}^{\otimes n}$ majorizes all others, therefore bounds
the sum of its $M$ largest
Schmidt probabilities by $\|\tau_{\theta_N}^{\otimes n}\|_{(M)}$.
This bound is attained in $\SN_M$ at finite $N$. Since $\omega_x$ is
a mixture of coherent states and $U_{r_N}$ turns a displacement of the
reference into local displacements, randomly displacing
$U_{r_N}|0,0\rangle$ prepares $\rho_{\eta,q}$. Apply the
same displacements to the normalized rank-$M$ truncation of the
$n$-copy seed. By joint concavity, its fidelity with
$\rho_{\eta,q}^{\otimes n}$ is at least
$\|\tau_{\theta_N}^{\otimes n}\|_{(M)}$, and the bound just proved
gives equality. Since the reference--output state of every LOCC
simulator supplied with $\Phi_M$ lies in $\SN_M$,
Eq.~\eqref{eq:uhlmannbound} bounds its fidelity by this value. Letting
$N\to\infty$, so that $\theta_N\to\eta$, proves the loss
converse.

For the amplifier the output on reference $R$ and system
$B$ is $U_{r_N}[|0\rangle\langle0|_R\otimes
\omega_{(\kappa-1)(N+1),B}]U_{r_N}^\dagger$, now with
$\tanh^2r_N=N/[\kappa(N+1)]$. For each vector in this support the
$B$ marginal is the output of the quantum-limited amplifier of gain
$\cosh^2r_N$. The same majorization bound holds with the parties
interchanged and $\theta_N=\tanh^2r_N\to1/\kappa$.
Product probes are admissible adaptive tests, so this completes the
proof of Theorem~\ref{thm:simulation} against adaptive tests.

\subsection{Finite sums and asymptotic rates}
The optimal fidelity in Theorem~\ref{thm:simulation} is the total
probability of the $M$ most likely photon-number strings of $n$ ideal
resources, which we now evaluate exactly. For either optical channel, a
string $(k_1,\ldots,k_n)$ with $j=\sum_i k_i$ photons has probability $(1-\theta)^n\theta^j$, and there
are $d_j=\binom{n+j-1}{j}$ such strings. The cumulative count
$D_k=\sum_{j\le k}d_j=\binom{n+k}{k}$ of strings with at most $k$
photons, with $D_{-1}=0$, first reaches $M$ at $k=k_*$.
The optimal simulator keeps all strings
with fewer than $k_*$ photons and fills the remaining rank with strings
of exactly $k_*$ photons, so
\begin{equation}
 \|\tau_\theta^{\otimes n}\|_{(M)}=(1-\theta)^n
 \left[\sum_{j=0}^{k_*-1}d_j\theta^j+(M-D_{k_*-1})\theta^{k_*}\right].
 \label{sm:finiterank}
\end{equation}
This finite sum gives every point in Fig.~\ref{fig:simulation}.
For $M\simeq e^{n\mathcal R}$ below the entropy rate
$h_\theta=S_1(\tau_\theta)=g(\theta/(1-\theta))$, with $g$ the thermal-mode
entropy of Sec.~\ref{sm:orderseed},
Stirling's formula gives the exact exponent
\begin{equation}
 \mathscr E_\theta(\mathcal R)=
 -\log(1-\theta)-\mathcal R-x_{\mathcal R}\log\theta,
 \quad g(x_{\mathcal R})=\mathcal R,
 \label{sm:closedexponent}
\end{equation}
equivalent to Eq.~\eqref{eq:exactexponent}. Here $x_{\mathcal R}$
is the mean occupation of the thermal mode with entropy $\mathcal R$;
the dominant retained strings carry about $nx_{\mathcal R}$ photons.
If $M_\epsilon(n)$ is the smallest rank giving fidelity
at least $1-\epsilon$, $0<\epsilon<1$, the corresponding
source-coding normal approximation \cite{HayashiSecondOrder} is
\begin{equation}
 \log M_\epsilon(n)=nh_\theta+
 \sqrt{2nv_\theta}\,\operatorname{erfc}^{-1}(2\epsilon)
 +O(\log n),
 \label{sm:secondorder}
\end{equation}
where $v_\theta=\theta(\log\theta)^2/(1-\theta)^2$ is the variance of
$-\log[(1-\theta)\theta^k]$ under $\tau_\theta$, and
$\operatorname{erfc}^{-1}$ is the inverse complementary error function.

For the geometric distribution Eq.~\eqref{sm:secondorder} can be
derived directly. The total photon
number $K_n=k_1+\dots+k_n$ is a sum of $n$ independent geometric
variables with mean $\bar x_\theta=\theta/(1-\theta)$, variance
$\theta/(1-\theta)^2$ and finite third moment. Since the retained strings
are those with fewest photons, $\|\tau_\theta^{\otimes n}\|_{(M)}$ lies
between $\Pr[K_n<k_*]$ and $\Pr[K_n\le k_*]$. Stirling's formula gives
$\log D_k=ng(k/n)+O(\log n)$ for $k/n$ in compact subsets of
$(0,\infty)$, and $\log M=\log D_{k_*}+O(1)$, since
$D_{k_*-1}<M\le D_{k_*}$ and $D_k/D_{k-1}=1+n/k=O(1)$. With
$g(\bar x_\theta)=h_\theta$
and $g'(\bar x_\theta)=-\log\theta$, a first-order expansion of $g$
shows that the choice
$\log M=nh_\theta+z\sqrt{nv_\theta}$ gives
$k_*=n\bar x_\theta+z\sqrt{n\theta}/(1-\theta)+O(\log n)$, so $k_*$
lies $z+O(n^{-1/2}\log n)$ standard deviations above the mean of $K_n$,
and the
Berry--Esseen theorem \cite{BerryEsseen,Esseen} yields
\begin{equation}
 \|\tau_\theta^{\otimes n}\|_{(M)}=\tfrac12\operatorname{erfc}(-z/\sqrt2)+O(n^{-1/2}\log n),
 \label{sm:berryesseen}
\end{equation}
where the leading term is the standard normal distribution function.
Setting it equal to $1-\epsilon$ recovers Eq.~\eqref{sm:secondorder}. At the entropy
rate ($z=0$) the fidelity tends to $1/2$, and the transition in the rate
$n^{-1}\log M$ has width of order $n^{-1/2}$. The convergence is slow:
for $\theta=1/2$, where $h_\theta=2\log2$ (two ebits) and $M=4^n$,
Eq.~\eqref{sm:finiterank} gives $0.786$, $0.641$ and $0.560$ at
$n=10$, $100$ and $1000$.

\subsection{A finite-energy measure-and-prepare benchmark}

As before, $\Psi_q$ is the two-mode squeezed vacuum with mean photon
number $N$ per arm, $q=N/(N+1)$, and
$\rho_{\eta,q}=({\rm id}_R\otimes\mathcal L_\eta)(\Psi_q)$.
Average fidelities over coherent-state inputs give optical benchmarks
for classical strategies, including lossy links
\cite{HammererBenchmark,NamikiKoashiImoto}, and for devices of bounded
Schmidt number \cite{NamikiBenchmark}. A squeezed probe instead
retains a reference and tests the correlations of that reference with
the output.

We prove Eq.~\eqref{eq:finiteenergyEB}, including optimality
over non-Gaussian channels and collective processing.
Every output $\sigma_{RB}=({\rm id}\otimes\mathcal M)(\Psi_q)$
of a measure-and-prepare channel is separable and preserves
the reference marginal $\tau_q$, given by Eq.~\eqref{eq:thermaltheta}
with $\theta=q$.

\emph{Opposite local filtering.} Let $\rho,\sigma$ be states on $RB$
with $\sigma_R=\rho_R$, and let $\Pi_R$ be a bounded
positive injective operator on $R$, with
\[
 Z_\Pi=\Tr(\rho_R \Pi_R^{-1})<\infty,\qquad
 W_\Pi=\Tr(\rho_R \Pi_R)<\infty.
\]
Filtering $\rho$ by $\Pi_R^{-1/2}$ and $\sigma$ by the opposite filter
$\Pi_R^{1/2}$ gives the states
\[
 \rho_\Pi=\frac{\Pi_R^{-1/2}\rho \Pi_R^{-1/2}}{Z_\Pi},
 \qquad
 \sigma_\Pi=\frac{\Pi_R^{1/2}\sigma \Pi_R^{1/2}}{W_\Pi},
\]
where the operators on $R$ are tensored with the identity on $B$, and
$\sigma_\Pi$ has unit trace because $\sigma_R=\rho_R$.
The inverse in the first expression is understood through
its positive trace-class form. The operators
$X=\Pi_R^{-1/2}\sqrt\rho$ and $Y=\Pi_R^{1/2}\sqrt\sigma$
are Hilbert--Schmidt amplitudes, $XX^\dagger=Z_\Pi\rho_\Pi$ and
$YY^\dagger=W_\Pi\sigma_\Pi$, with $X^\dagger Y=\sqrt\rho\sqrt\sigma$.
The amplitude formula
$\sqrt{\Fid(A,B)}=\sup_{XX^\dagger=A,\,YY^\dagger=B}\|X^\dagger Y\|_1$,
applied with these filters and with the reversed ones, therefore gives
\begin{equation}
 \Fid(\rho,\sigma)=Z_\Pi W_\Pi\Fid(\rho_\Pi,\sigma_\Pi)
 \le Z_\Pi W_\Pi F_{\Sep}(\rho_\Pi).
 \label{eq:oppositefilter}
\end{equation}
The inequality uses separability of $\sigma_\Pi$, preserved
by the bounded local filter.

Take $N>0$, let $\hat n_R$ be the reference photon-number
operator, and set $\Pi_R=(q/\xi)^{\hat n_R}$ with $q\le\xi<1$.
For $\rho=\rho_{\eta,q}$, $\Pi_R^{-1/2}$ maps $\Psi_q$ to a multiple of
$\Psi_\xi$ and commutes with $\mathcal L_\eta$, so
$\rho_\Pi=\rho_{\eta,\xi}$, and geometric sums over $\tau_q$ give
\begin{equation}
 Z_\Pi=\frac{1-q}{1-\xi},\qquad
 W_\Pi=\frac{1-q}{1-q^2/\xi}.
\end{equation}
By Eq.~\eqref{sm:support} with $q\to\xi$, the pure-loss seed of
$\rho_{\eta,\xi}$ has $\tanh^2r_*=\eta\xi$, and the argument following
that equation, at $n=M=1$, gives
$F_{\Sep}(\rho_{\eta,\xi})=1-\eta\xi$.
For every measure-and-prepare output $\sigma$ and $\xi\in[q,1)$,
Eq.~\eqref{eq:oppositefilter} yields
\begin{equation}
 \Fid(\rho_{\eta,q},\sigma)
 \le \overline F(\xi):=
 \frac{(1-q)^2\xi(1-\eta\xi)}{(1-\xi)(\xi-q^2)}.
 \label{eq:filteredEBbound}
\end{equation}
For $0<\eta<1$, this bound has a unique minimum on $[q,1)$, at
\begin{equation}
 \xi_*=
 \frac{q}{\eta q+\sqrt{(1-\eta)(1-\eta q^2)}}.
\end{equation}
Indeed, the sign of $\overline F'(\xi)$ is that of the quadratic
$[1-\eta(1+q^2)]\xi^2+2\eta q^2\xi-q^2$, which is negative at $\xi=q$,
positive at $\xi=1$, and vanishes at $\xi_*$.
Substituting $\xi_*$ and rationalizing gives the
right side of Eq.~\eqref{eq:finiteenergyEB}. At $\eta=1$,
the limit $\xi\uparrow1$ gives $(1-q)/(1+q)=1/(2N+1)$, the ideal-link
value of the End Matter, while at $\eta=0$ the
choice $\xi=q$ gives one, as does the limit $N\to0$.

\emph{Gaussian attainment.} Let $\mathcal M_\gamma$ heterodyne
the input and prepare $|\gamma\beta\rangle$ for its outcome
$\beta$, where $\gamma\ge0$. Its reference--output state is
the coherent-product mixture
\begin{equation}
 \sigma_\gamma=\int_{\mathbb C}\frac{\dd^2z}{\pi N}
 e^{-|z|^2/N}
 |z\rangle\langle z|_R\otimes
 |\gamma z^*/\sqrt q\rangle\langle \gamma z^*/\sqrt q|_B,
 \label{eq:heterodyneEBstate}
\end{equation}
where $z=\sqrt q\,\beta^*$ is the coherent amplitude to which the
outcome $\beta$ projects the reference.
The reference--output covariance of $\sigma_\gamma$ has local
blocks $(2N+1)\id_2$ and $[1+2\gamma^2(N+1)]\id_2$ and
correlation block $2\gamma\sqrt{N(N+1)}\operatorname{diag}(1,-1)$.
The Gaussian fidelity formula \cite{Banchi} gives
\begin{equation}
 \Fid(\rho_{\eta,q},\sigma_\gamma)
 =\frac{(1-q)^2}
 {[\sqrt{1+\gamma^2-2\gamma\sqrt\eta q}-q\sqrt{1-\eta}]^2}.
 \label{eq:gainfidelity}
\end{equation}
The bracket in the denominator is smallest at $\gamma=\sqrt\eta q$,
where it equals $\sqrt{1-\eta q^2}-q\sqrt{1-\eta}$, positive because
$1-\eta q^2-q^2(1-\eta)=1-q^2$. This gain attains
Eq.~\eqref{eq:finiteenergyEB} with Bob's coherent
amplitude $\sqrt\eta\,q\beta$, as in the End Matter.

\emph{Collective channels.} On $n$ product probes, every
collective measure-and-prepare output is separable across
$R^n:B^n$ with reference marginal $\tau_q^{\otimes n}$.
In Eq.~\eqref{eq:oppositefilter}, the filter $\Pi_R^{\otimes n}$ raises
$Z_\Pi$ and $W_\Pi$ to the $n$th power, and multiplicativity of
$F_{\Sep}$, Eq.~\eqref{eq:fidproduct}, gives the upper bound
$\overline F(\xi)^n$, optimized at $\xi_*$. The product of the
attaining Gaussian channels has fidelity $F^{\rm EB}_\eta(N)^n$,
proving equality.

\section{Equilibrium oscillators and Gaussian field modes}
\label{sm:physics}

\subsection{The thermal covariance and the even--odd cut}
\label{sm:chain}

Here we derive the thermal-ring results stated in the End
Matter, Eqs.~\eqref{em:ringz} and~\eqref{em:ringfamily}, together
with the threshold asymptotics.

For unit-mass oscillators with
$H=(\boldsymbol p^{\top}\boldsymbol p+
\boldsymbol q^{\top}\Lambda\boldsymbol q)/2$, the eigenvalues of the positive
matrix $\Lambda$ are the squared normal frequencies. At temperature $T$, a normal
oscillator of frequency $\omega$ has covariance entries (doubled position and momentum
variances) $\coth(\omega/2T)/\omega$ and $\omega\coth(\omega/2T)$,
in units $k_B=1$. Transforming back
to the site coordinates gives
\begin{equation}
 V_T=\bigl[\Lambda^{-1/2}\coth(\Lambda^{1/2}/2T)\bigr]\oplus
         \bigl[\Lambda^{1/2}\coth(\Lambda^{1/2}/2T)\bigr].
 \label{eq:thermalcovariance}
\end{equation}
For two disjoint sets of sites $A$ and $B$, the reduced state is Gaussian, with
covariance $V_{AB}$ the principal submatrix of $V_T$ on their positions and momenta \cite{AudenaertHarmonic,
AndersWinter,PeschelEisler}. Since $\coth(\omega/2T)$ increases with $T$, so does $V_{AB}$
in matrix order, and a larger covariance admits more pure $\Gamma\le V_{AB}$ in Eq.~\eqref{eq:Gaussianroof}. Every $G_\alpha(V_{AB})$ therefore decreases or stays
constant as $T$ grows, and $F_{\Sep}=e^{-G_\infty}$ increases or stays constant. All $G_\alpha$
vanish at the same separability boundary: $G_\alpha(V_{AB})=0$
requires a feasible product seed, that is, a pure product covariance below $V_{AB}$, and the existence of one is equivalent
to Gaussian separability \cite{WernerWolfBound}. Hence every
$G_\alpha$ is strictly positive on entangled Gaussian states, including
those with positive partial transpose. This is consistent with $F_{\Sep}<1$ for every entangled state.

For a ring of $2L$ oscillators, impose the periodic boundary condition $q_{2L}=q_0$ and take
\begin{equation}
 H=\frac12\sum_{i=0}^{2L-1}
 \left[p_i^2+\omega_0^2 q_i^2+\gamma(q_{i+1}-q_i)^2\right],
 \qquad \omega_0,\gamma>0 .
 \label{eq:ringham}
\end{equation}
As in the End Matter, $A$ holds the even sites and $B$ the odd ones.
Group the sites into $L$ cells $(2l,2l+1)$, $l=0,\ldots,L-1$, and list
the even sites before the odd ones. The diagonal blocks of $\Lambda$
are then $(\omega_0^2+2\gamma)\id_L$ and the off-diagonal block is $-\gamma(\id_L+C)$,
where $C$ is the cyclic shift by one cell, $(C\boldsymbol y)_l=y_{l-1\bmod L}$. Since $C$ has eigenvalues $e^{2\pi ij/L}$, the singular values of
$\id_L+C$ are $2|\cos(\pi j/L)|$, $j=0,\ldots,L-1$. Its real singular
value decomposition $\id_L+C=O_e\Delta O_o^{\top}$ defines orthogonal transformations $O_e$ and $O_o$ of
the two sublattices. Applied to positions and momenta alike, they
form a local change of canonical coordinates after which $\Lambda$ couples only the $j$th even and the $j$th odd coordinate, so the Gibbs
state factors into $L$ independent pairs with squared normal
frequencies
\begin{equation}
 \omega_{j,\pm}^2=\omega_0^2+2\gamma\ \pm\ 2\gamma|\cos(\pi j/L)|.
 \label{eq:ringfrequencies}
\end{equation}
The low- and high-frequency coordinates in each pair are the symmetric
and antisymmetric combinations of its even and odd coordinates.

Fix one pair, suppress the label $j$, and write
$X_\pm=\coth(\omega_\pm/2T)/\omega_\pm$ and
$P_\pm=\omega_\pm\coth(\omega_\pm/2T)$ for the position and momentum covariance entries of its two normal modes. Then $X_+\le X_-$ and
$P_-\le P_+$, because $\coth x/x$ decreases and $x\coth x$ increases
for $x>0$. Partial transposition flips the odd-site momentum, which in normal
coordinates interchanges $P_-$ and $P_+$. The partially transposed symplectic eigenvalues are thus $\sqrt{X_-P_+}$ and the smaller one,
$z=\sqrt{X_+P_-}$. Since the smallest symplectic eigenvalue of a positive matrix $D$
is at least $\nu$ precisely when $D+i\nu\Omega\ge0$, it increases in matrix order;
and partial transposition acts on covariances by a congruence, so it preserves $\Gamma\le V$. A pure seed with pair squeezing $r$ has smallest partially transposed symplectic eigenvalue $e^{-2r}$, and a feasible seed must therefore
satisfy $e^{-2r}\le z$.

This bound can be attained. If $z<1$, choose the high-frequency
pure mode with covariance entries $(X_+,1/X_+)$ and the low-frequency pure mode
with covariance entries $(1/P_-,P_-)$. Each fits below the thermal entries $(X_\pm,P_\pm)$ of the same normal mode because
$X_\pm P_\pm\ge1$. Undoing the symmetric and antisymmetric change of
coordinates gives a pure pair whose single-site covariance has symplectic eigenvalue
$\tfrac12[(X_++1/P_-)(P_-+1/X_+)]^{1/2}=(z+z^{-1})/2=\cosh(2r)$, so $r=-\tfrac12\log z$. If $z\ge1$,
choose the same pure-mode covariance entries $(x,1/x)$ in both normal coordinates,
with $1/P_-\le x\le X_+$. This covariance is invariant under the rotation back to even and odd sites, so the seed is a product across
them. In terms of the original pair frequencies, the
optimal squeezing is $r_j=\max\{0,-\frac12\log z_j\}$ with $z_j$ given
by Eq.~\eqref{em:ringz}, which is $z^2=X_+P_-$ restored to pair $j$.
Each pair is a $1{:}1$ cut, and every $f_\alpha$ in Eq.~\eqref{eq:scalarcost} increases with $r$, as shown after that equation, so this seed is optimal at every order. Additivity, from
Theorem~\ref{thm:formation} for $\alpha\in\cI$ and from
Refs.~\cite{AdessoEoF,GA} at order one, then
proves the first equality in Eq.~\eqref{em:ringfamily} against arbitrary decompositions. Equations~\eqref{eq:separablefidelity} and~\eqref{eq:fidproduct}, with $e^{-f_\infty(r)}=\operatorname{sech}^2r$, prove the second against arbitrary
separable competitors.

\subsection{Approach to the separability temperature}

The index $j$ labels the cell quasi-momentum $j/L$ in units
where the Brillouin zone has length one. Let $x\in[-1/2,1/2]$ be its representative modulo one, which leaves $|\cos(\pi j/L)|$ unchanged, and treat $x$ as continuous when $L\to\infty$.
Write $z(x,T)$ for
Eq.~\eqref{em:ringz} with $j/L$ replaced by $x$. It is smallest at $x=0$,
where the two frequencies are furthest apart. Indeed, $z^2=X_+P_-$, where
$P_-$ decreases as $\omega_-$ decreases and $X_+$ decreases
as $\omega_+$ increases. Both factors also increase strictly with temperature, as $\coth(\omega/2T)$ does,
and since $z(0,0)=(\omega_-/\omega_+)^{1/2}<1$ while $z(0,T)\to\infty$ at high temperature,
the condition $z(0,T_*)=1$ fixes a unique positive $T_*$. This is the
even--odd partial-transpose threshold studied in
Ref.~\cite{CavalcantiThermal}. Here the explicit product seed constructed above for $z\ge1$ also
gives a separable preparation at and above $T_*$, where $z(x,T)\ge1$ for all $x$.

Let $r(x,T)=\max\{0,-\log z(x,T)\}/2$. In the thermodynamic limit, the Riemann sum $L^{-1}\sum_jf_\alpha(r_j)$ from Eq.~\eqref{em:ringfamily} gives
the entanglement per cell (a pair of even and odd sites) at order $\alpha\in[1,\infty]$,
$\bar G_\alpha(T)=\lim_{L\to\infty}G_\alpha(V_T)/L
=\int_{-1/2}^{1/2}f_\alpha(r(x,T))\,\dd x$.
By Refs.~\cite{AdessoEoF,GA}, $\bar G_1$ is the entanglement cost per
cell. At $\alpha=\infty$ it is the negative log separable fidelity
per cell,
\begin{equation}
 \bar G_\infty(T)=-\lim_{L\to\infty}L^{-1}\log F_{\Sep}(\rho_{V_T})
       =\int_{-1/2}^{1/2}2\log\cosh r(x,T)\,\dd x .
 \label{eq:chaindensity}
\end{equation}
At any fixed $T<T_*$, $z(x,T)<1$ near $x=0$, so $\bar G_\infty(T)>0$ and the best global
separable fidelity decays exponentially in the number of cells.

To approach the threshold from below, set $\epsilon=1-T/T_*$.
The cosine in Eq.~\eqref{eq:ringfrequencies} has a quadratic maximum
at zero, and $-\log z(x,T)$ is smooth in $x^2$ and $T$, with nonzero partial
derivatives at $(0,T_*)$. Indeed, $\dd\cos(\pi x)/\dd(x^2)=-\pi^2/2$ at
$x=0$, $\omega_\pm^2$ is affine in the cosine, and the $\omega$-derivatives
of $\coth(\omega/2T)/\omega$ and $\omega\coth(\omega/2T)$ and the
$T$-derivative of $\coth(\omega/2T)$ do not vanish; by the monotonicity
described above, the two factors of $z^2=X_+P_-$ change in the same
direction. A first-order expansion
therefore gives, for some $a,b>0$,
\[
 -\log z(x,T)=a\epsilon-bx^2+
                  O(\epsilon^2+\epsilon x^2+x^4).
\]
Since $r(x,T)=0$ where this is negative, only $|x|=O(\sqrt\epsilon)$ contributes. Substituting
$x=\sqrt{a\epsilon/b}\,w$ in Eq.~\eqref{eq:chaindensity} and using
$2\log\cosh r=r^2+O(r^4)$ gives
\begin{equation}
 \bar G_\infty(T)\sim \frac{4a^{5/2}}{15\sqrt b}\,\epsilon^{5/2},
 \label{eq:chaincritical}
\end{equation}
since $\int_{-1}^1(1-w^2)^2\,\dd w=16/15$.
For fixed $1<\alpha<\infty$, Eq.~\eqref{eq:scalarcost} gives
$f_\alpha(r)=\alpha r^2/(\alpha-1)+o(r^2)$, since $\sinh^{2\alpha}r=o(r^2)$, proving
$\bar G_\alpha/\bar G_\infty\to\alpha/(\alpha-1)$. At order one,
$f_1(r)=g(\sinh^2r)=r^2[1-\log r^2]+O(r^4|\log r|)$, with $g$ the thermal-mode entropy of Sec.~\ref{sm:orderseed}.
For each fixed $|w|<1$, $2r=a\epsilon(1-w^2)+O(\epsilon^2)$, so $\log r^2=2\log(a\epsilon/2)+2\log(1-w^2)+o(1)$.
The integrand is supported on $|w|\le1+O(\epsilon)$, and dominated convergence, with the
integrable majorant $C(1-w^2)^2\bigl(1+\bigl|\log|1-w^2|\bigr|\bigr)$,
handles the edges. Using
$\int_{-1}^1(1-w^2)^2\log(1-w^2)\,\dd w=\frac{32}{15}\log2-\frac{376}{225}$,
we obtain
\begin{equation}
 \frac{\bar G_1(T)}{\bar G_\infty(T)}
 =2|\log\epsilon|+\frac{62}{15}-2\log(2a)+o(1).
 \label{eq:chainlog}
\end{equation}
For $\omega_0=1/2$ and $\gamma=1$, $a\simeq0.685$ and the additive
constant is about $3.50$. Numerically, the ratio is $12.7$ at
$\epsilon=10^{-2}$ and $17.3$ at $\epsilon=10^{-3}$, against
$2|\log\epsilon|=9.2$ and $13.8$. Taking the order-one limit before
$\epsilon\to0$ therefore produces an extra factor that grows as
$2|\log\epsilon|$.

\subsection{Opposite momenta in a decohering field}
\label{sm:primordial}

The two-mode results also give the exact separable fidelity of a band
of opposite-momentum pairs in a homogeneous Gaussian field, and show
how weak per-pair entanglement accumulates across the band.
Consider a homogeneous, parity-symmetric Gaussian field in a finite volume, so that momenta are discrete. In the radiation-era
canonical modes used by Campo and Parentani \cite{CampoParentaniEntropy08},
let $a_k$ annihilate a quantum of momentum $k$. The band $\mathcal B$ is a
finite set of nonzero momenta containing one representative $k$ of each
pair $(k,-k)$ it involves, and the bipartition separates the modes $k\in\mathcal B$ from their partners $-k$. By homogeneity, $\langle a_ka_{k'}\rangle$ vanishes unless $k'=-k$ and $\langle a_k^\dagger a_{k'}\rangle$ unless $k'=k$, so distinct
pairs are uncorrelated and the Gaussian band state is
$\rho_{\mathcal B}=\bigotimes_{k\in\mathcal B}\rho_{k,-k}$. Define the mode
occupation $n_k=\langle a_k^\dagger a_k\rangle=n_{-k}$ and, for $n_k>0$, the decoherence
parameter $\delta_k$ by
$|\langle a_k a_{-k}\rangle|^2=n_k(n_k+1-\delta_k)$, so that $\delta_k=0$ for a two-mode squeezed vacuum. Vacuum pairs, with $n_k=0$ and $z_k=1$ in Eq.~\eqref{eq:fieldq} below, are
included by continuity.
Put $c_k=2|\langle a_k a_{-k}\rangle|$. After local phase
rotations making $\langle a_ka_{-k}\rangle$ positive, the pair has no position--momentum correlations, its position block is
$\left(\begin{smallmatrix}2n_k+1&c_k\\c_k&2n_k+1\end{smallmatrix}\right)$,
and its momentum block is
$\left(\begin{smallmatrix}2n_k+1&-c_k\\-c_k&2n_k+1\end{smallmatrix}\right)$.
In the symmetric and antisymmetric coordinates the smaller position and momentum covariance
entries, the analogs of $X_+$ and $P_-$, are both $2n_k+1-c_k$, and the two-mode argument of Sec.~\ref{sm:chain}
applies with
\begin{equation}
 z_k=2n_k+1-2\sqrt{n_k(n_k+1-\delta_k)},\qquad
 r_k=\max\{0,-\tfrac12\log z_k\}.
 \label{eq:fieldq}
\end{equation}
For a single symmetric pair, the Bures optimum over Gaussian separable states is known
\cite{MarianBures}. Equations~\eqref{eq:separablefidelity}
and~\eqref{eq:fidproduct} give the optimum over all separable
states of the band, including those correlated between different momenta,
\begin{equation}
 F_{\Sep}(\rho_{\mathcal B})=\prod_{k\in\mathcal B}\operatorname{sech}^2r_k,
 \qquad
 \operatorname{sech}^2r_k=
 \begin{cases}4z_k/(1+z_k)^2,&z_k<1,\\1,&z_k\ge1.\end{cases}
 \label{eq:fieldband}
\end{equation}
For $n_k>0$, $z_k<1$ exactly when $\delta_k<1$, so the separability threshold is $\delta_k=1$.
At large occupation and fixed $0<\delta_k<1$ the normalized
pair correlation $|\langle a_k a_{-k}\rangle|^2/n_k^2
=1+(1-\delta_k)/n_k$ approaches one, its value at the threshold.
Yet $z_k\to\delta_k$, so the single-pair fidelity tends to
$4\delta_k/(1+\delta_k)^2<1$. Near the threshold, a band of identical
pairs with common value $z_k=z$ slightly below one has
$-|\mathcal B|^{-1}\log F_{\Sep}(\rho_{\mathcal B})=(1-z)^2/4+O((1-z)^3)$,
so weak per-pair entanglement can accumulate across a large band.
The mode choice and bipartition enter this statement explicitly
\cite{CampoParentaniEntropy08,MartinVenninFields21,BelfiglioCosmology25}.

More generally, let a quadratic Hamiltonian generate a symplectic evolution $S(t)$, with $S(t)\Omega S(t)^{\top}=\Omega$, acting on an initial
covariance $V_0$ as $V_0\mapsto S(t)V_0S(t)^{\top}$.
If the rectangular matrix $P_{AB}$ selects the quadratures of the canonical modes assigned to $A$ and $B$, and a Gaussian detector acts on their covariance as
$D\mapsto XDX^{\top}+Y$, the
covariance entering Theorem~\ref{thm:formation} is
\begin{equation}
 V_{AB}(t)=XP_{AB}S(t)V_0S(t)^{\top}P_{AB}^{\top}
                   X^{\top}+Y.
 \label{eq:fieldcovariance}
\end{equation}
Here $P_{AB}\Omega P_{AB}^{\top}=\Omega_{AB}$, the symplectic form of the selected modes. This formulation
also covers pairs of smooth canonical wave packets of finite
covariance in free bosonic fields \cite{SteaneIshizaka2026} and analog
Hawking pairs \cite{AgulloHawking22}. Sharp momentum modes of a
continuum field have infinite covariance and are excluded.

\section{Non-Gaussian states and second moments}
\label{sm:nongaussian}

At order one the covariance-matched Gaussian value lower-bounds the
entanglement of formation and the entanglement cost of an arbitrary
state \cite{AdessoEoF,GA}. At every order above one this lower bound fails, as we show first.
The rest of the section turns to what can still be certified: entanglement from second moments near a pure Gaussian comparison state,
the sharp bound for an imperfect source with a measured fidelity to a Gaussian target, and a fidelity bound for every covariance incompatible with separability.

\subsection{Two states with the same covariance on opposite sides of the Gaussian value}

Let $U_r=\exp[r(a_A^\dagger a_B^\dagger-a_Aa_B)]$ be the two-mode
squeezing unitary of Sec.~\ref{sm:simulation}, now acting on the modes
of $A$ and $B$. Fix $r>0$, and consider
\[
 |\psi_\epsilon\rangle
 =U_r\frac{|0,0\rangle+\epsilon|2,2\rangle}{\sqrt{1+\epsilon^2}},
 \qquad \epsilon\in\mathbb R .
\]
Because $U_r$ maps quadratures linearly to quadratures, it suffices to compute the moments of the unsqueezed superposition. A linear or quadratic quadrature observable changes the total photon number by at most two, while $|0,0\rangle$ and $|2,2\rangle$ differ by four, so the cross terms vanish. The superposition thus has zero means and covariance $v\id$, where
$v=1+4\epsilon^2/(1+\epsilon^2)$ reflects the mean occupation $2\epsilon^2/(1+\epsilon^2)$ of each mode, and both signs of
$\epsilon$ give zero means and the same covariance $vV_r$, where
$V_r$ is the covariance of $U_r|0,0\rangle$. For small $|\epsilon|$, the
least squeezing of a feasible Gaussian seed is $r'=r-\frac12\log v$.
Indeed, the smallest symplectic eigenvalue after partial
transposition increases in covariance order (Sec.~\ref{sm:chain}), so every feasible seed obeys
$e^{-2r_{\rm seed}}\le ve^{-2r}$, and $V_{r'}\le vV_r$ attains
equality. Consequently $G_\alpha(vV_r)=f_\alpha(r')=
f_\alpha(r)+O(\epsilon^2)$.

Since $U_r$ preserves the
photon-number difference, $\psi_\epsilon=\sum_nw_n|n,n\rangle$ with real amplitudes $w_n(\epsilon)$, whose squares are the Schmidt probabilities. Fix a finite $\alpha>1$ and write
$Z_\alpha(\epsilon)=\Tr\psi_{\epsilon,A}^{\alpha}=\sum_nw_n^{2\alpha}$. As the normalization has zero derivative at $\epsilon=0$,
$Z_\alpha'(0)=2\alpha\sum_nw_n(0)^{2\alpha-1}\langle n,n|U_r|2,2\rangle$. At $\epsilon=0$ the amplitudes are $w_n(0)=\lambda^n/\cosh r$, with $\lambda=\tanh r$, so their
$(2\alpha-1)$th powers are proportional, with a positive factor, to the
amplitudes of $U_{r_\alpha}|0,0\rangle$, where $\tanh r_\alpha=\lambda^{2\alpha-1}$ and hence $r_\alpha<r$. Because $U_{r_\alpha}^\dagger U_r=U_{r-r_\alpha}$, $Z_\alpha'(0)$ has the sign of
\[
 \langle 0,0|U_{r-r_\alpha}|2,2\rangle
 =\frac{\tanh^2(r-r_\alpha)}{\cosh(r-r_\alpha)}>0 .
\]
The entropy $(1-\alpha)^{-1}\log Z_\alpha$ therefore decreases at
first order in $\epsilon$. At $\alpha=\infty$, the gap $w_0(0)>w_1(0)$ keeps the
vacuum pair the largest Schmidt component for small $|\epsilon|$, and
\[
 F_{\Sep}(\psi_\epsilon)=w_0^2
 =\frac{(1+\epsilon\lambda^2)^2}
        {(1+\epsilon^2)\cosh^2r}.
\]
Its negative logarithm, $S_\infty(\psi_{\epsilon,A})$, again decreases at first order. As $G_\alpha(vV_r)$ differs from $f_\alpha(r)=S_\alpha(\psi_{0,A})$ only at second order, for each fixed
$\alpha\in\cI$ and sufficiently small $\epsilon>0$ we have proved
\begin{equation}
 S_\alpha(\psi_{\epsilon,A})<G_\alpha(vV_r)
                         <S_\alpha(\psi_{-\epsilon,A}).
 \label{eq:ngstraddle}
\end{equation}
At order one the Gaussian value is a lower bound at fixed covariance
\cite{GA}. The general Gaussian extremality theorem
\cite{WolfExtremality}, which would give such a bound, assumes strong superadditivity on arbitrary
states, whereas Theorem~\ref{thm:formation} establishes it only on Gaussian states.

\subsection{A photon-number witness near a pure comparison state}
\label{sm:ngwitness}

Let $\omega$ be any state with covariance $V_\omega$ on $d$ modes.
Choose a pure Gaussian comparison state $\psi_\Gamma$ with covariance $\Gamma$
and the same first moments as $\omega$. A displacement and a symplectic $S$ with $S\Gamma S^{\top}=\id$ map $\psi_\Gamma$ to the
vacuum. In these coordinates the total photon-number operator $\hat n$ obeys
$\hat n\ge\id-|0\rangle\langle0|$, and its expectation in $\omega$ is
$\Tr(\Gamma^{-1}V_\omega)/4-d/2$, since $\Gamma^{-1}=S^{\top}S$. The photonic fidelity witness
\cite{AolitaWitness}, $\Fid(\omega,\psi_\Gamma)=\langle0|\omega|0\rangle\ge1-\Tr(\omega\hat n)$, therefore gives
\begin{equation}
 \Fid(\omega,\psi_\Gamma)\ge f_\Gamma(V_\omega)
 :=\left[1+\frac d2-\frac14\Tr(\Gamma^{-1}V_\omega)\right]_+ ,
 \label{eq:ngmomentwitness}
\end{equation}
where $[x]_+=\max(0,x)$. Write $c_\Gamma=\|\psi_{\Gamma,A}\|_\infty=F_{\Sep}(\psi_\Gamma)$
for the leading Schmidt probability of $\psi_\Gamma$. Let
$\vartheta(\rho,\sigma)=\arccos\sqrt{\Fid(\rho,\sigma)}$ be the Bures angle, a metric on states,
and $\vartheta_{\Sep}(\omega)=\inf_{\sigma\in\Sep}\vartheta(\omega,\sigma)$ the separable Bures angle. For every separable $\sigma$, the triangle inequality, also used for fidelity estimates in
Ref.~\cite{Kondra}, gives $\arccos\sqrt{c_\Gamma}\le\vartheta(\psi_\Gamma,\sigma)\le\arccos\sqrt{f_\Gamma(V_\omega)}+\vartheta(\omega,\sigma)$. Taking the infimum over $\sigma$ and optimizing the comparison state,
\[
 \vartheta_{\Sep}(\omega)\ge
 \sup_{\Gamma\ {\rm pure}}\!
 \left[\arccos\sqrt{c_\Gamma}
           -\arccos\sqrt{f_\Gamma(V_\omega)}\right]_+ .
\]
For the two signs in Eq.~\eqref{eq:ngstraddle}, choose $\psi_\Gamma=
U_r|0,0\rangle$, so that $\Gamma=V_r$. Then $f_\Gamma=2-v=1-4\epsilon^2/(1+\epsilon^2)$,
whereas $c_\Gamma=1/\cosh^2r<1$. For small $|\epsilon|$, $f_\Gamma>c_\Gamma$ and the bound is positive, so the same second moments certify entanglement in both states.

\subsection{A sharp bound for an imperfect source}

The same triangle inequality, now with a Gaussian target $\rho_V$ in place of $\psi_\Gamma$ and a measured fidelity in place of $f_\Gamma$, proves Eq.~\eqref{eq:imperfectfidelity} of the Letter. It also gives the two-qubit
fidelity-neighborhood formula of Ref.~\cite{Kondra}. Let $c=F_{\Sep}(\rho_V)$, the analog of $c_\Gamma$, and suppose that a source state $\omega$ has
$\Fid(\omega,\rho_V)\ge f$. Then
\[
 \vartheta_{\Sep}(\omega)\ge
 \max\{0,\arccos\sqrt c-\arccos\sqrt f\},
\]
since $\vartheta_{\Sep}(\rho_V)=\arccos\sqrt c$ and, for every separable $\sigma$, $\arccos\sqrt c\le\vartheta(\rho_V,\sigma)\le\arccos\sqrt f+\vartheta(\omega,\sigma)$.

The bound is attained for all $c$ and $f$. Let $\sigma_*$ be the
closest separable state of Corollary~\ref{cor:closest}. If $f\le c$, $\sigma_*$ is itself an admissible output.
For $c<f<1$, choose purifications of $\rho_V$ and $\sigma_*$ with
overlap $\sqrt c$ by Uhlmann's theorem \cite{HouQi}. On the shorter
great-circle arc between them take the unit vector at angle
$\arccos\sqrt f$ from the purification of $\rho_V$, and let $\omega$ be its reduced state.
Tracing out the purifying system can only shorten each of the two
angles, since the fidelity of reduced states is at least the squared overlap of the vectors. Their sum was already $\vartheta(\rho_V,\sigma_*)$, so the triangle
inequality forces equality on both segments. This $\omega$ has
fidelity $f$ with the target and $\vartheta(\omega,\sigma_*)=\arccos\sqrt c-\arccos\sqrt f$, so it attains the bound. At $f=1$ the
output is the target itself. Since $1-F_{\Sep}=\sin^2\vartheta_{\Sep}$ and, for $f>c$,
$\sin(\arccos\sqrt c-\arccos\sqrt f)=\sqrt{f(1-c)}-\sqrt{c(1-f)}$, squaring the sine of the angle bound gives
Eq.~\eqref{eq:imperfectfidelity}. For example, a two-mode squeezed
vacuum with one photon per arm on average has photon-number
probabilities $2^{-(k+1)}$, so $c=1/2$; a source reaching fidelity
$f=0.9$ with it has geometric entanglement at least $1/5$, whatever
its noise.

\subsection{A fidelity bound for every covariance incompatible with separability}
\label{sm:momentfilter}
Let $\rho$ be any bipartite state with measured covariance $V$ on
$d=d_A+d_B$ modes. Local displacements set its first moments to zero.
Choose a real matrix $Z=Z^{\top}>0$ and $\tau>0$. The quadratic
Hamiltonian $H=\boldsymbol R^{\top}Z\boldsymbol R/2$ has mean
$\Tr(\rho H)=\Tr(ZV)/4$, and since $H>0$ its unnormalized Gibbs operator, the filter $\Pi_\tau=e^{-\tau H}$, is a
measurement effect, $0<\Pi_\tau\le\id$. Define
\begin{equation}
 p=e^{-\tau\Tr(ZV)/4},\qquad
 b=\sup_{\|\phi_A\|=\|\phi_B\|=1}\langle \phi_A,\phi_B|\Pi_\tau|\phi_A,\phi_B\rangle ,
 \label{eq:momentfilterpb}
\end{equation}
where $|\phi_A,\phi_B\rangle=|\phi_A\rangle_A|\phi_B\rangle_B$.
Jensen's inequality for the convex function $e^{-\tau x}$ gives $\Tr(\rho\Pi_\tau)\ge p$, while $b$ is the largest
acceptance probability of the filter for a separable state, as separable states are mixtures of product vectors. Quantitative entanglement estimates from witness data are well
established \cite{GuhneWitness}, and Gaussian operator witnesses have been studied for specific mode
configurations \cite{ChenWitness}. Here both the separable threshold
and a bound on dilution fidelity follow from the Gaussian norm
theorem, Theorem~\ref{thm:schatten}.

For $\bar b\in[b,1]$ with $p>\bar b$, put
\begin{equation}
 \zeta=\left(\sqrt{p\bar b}+\sqrt{(1-p)(1-\bar b)}\right)^2<1.
 \label{eq:momentfilterc}
\end{equation}
For every $n,M\ge1$ and every state $\sigma\in\SN_M$ across
$A^n:B^n$,
\begin{equation}
 \Fid(\rho^{\otimes n},\sigma)\le\min\{1,M \zeta^n\}.
 \label{eq:momentfiltercopies}
\end{equation}
Measured second moments alone thus bound the fidelity of even a
single non-Gaussian source with all separable states, and for independent
copies they give an exponential preparation bound at every rate below
$-\log\zeta$. They also bound the R\'enyi entanglements of formation themselves.
For each pure state $\psi$, $\Tr\psi_A^\alpha\le\|\psi_A\|_\infty^{\alpha-1}$ gives
$\|\psi_A\|_\alpha\le\|\psi_A\|_\infty^{t_\alpha}$. Jensen's inequality for the concave function
$x^{t_\alpha}$, together with Eq.~\eqref{eq:generalrankfidelity} at $k=1$, $F_{\Sep}(\omega)=\sup_\mu\int\|\psi_A\|_\infty\,\dd\mu$, then yields
$\ensnorm_\alpha(\omega)\le F_{\Sep}(\omega)^{t_\alpha}$ for any state $\omega$.
Consequently, using Eq.~\eqref{eq:momentfiltercopies} at $M=1$ in the last step,
\[
 E^{\rm av}_{F,\alpha}(\rho^{\otimes n})
 \ge E^{\rm c}_{F,\alpha}(\rho^{\otimes n})
 \ge-\log F_{\Sep}(\rho^{\otimes n})\ge -n\log\zeta .
\]

\emph{Evaluating the threshold.} Maximize
$|\langle \phi_A,\phi_B|\Pi_\tau^{1/2}|\psi\rangle|^2$ over unit vectors $\phi_A,\phi_B,\psi$ in either order. Maximizing first over $\psi$ gives $b$; maximizing first over $\phi_A,\phi_B$ gives the largest squared Schmidt coefficient of $\Pi_\tau^{1/2}\psi$, whose supremum over $\psi$ is
$a_\infty(\Pi_\tau^{1/2})^2$ in the notation of Sec.~\ref{sm:norm}. Hence $b=a_\infty(\Pi_\tau^{1/2})^2$, and Eq.~\eqref{eq:exhaustion}, valid at $\alpha=\infty$ by Theorem~\ref{thm:schatten}, allows
Gaussian $\psi$. The leading Schmidt vectors of the Gaussian vector
$\Pi_\tau^{1/2}\psi$ are Gaussian, so Gaussian products suffice in
Eq.~\eqref{eq:momentfilterpb}. Write the Williamson decomposition as
$Z=S^{-\top}D S^{-1}$, where $S$ is symplectic and
$D=\operatorname{diag}(\nu_1,\ldots,\nu_d,\nu_1,\ldots,\nu_d)$, with
$\nu_j=\nu_j(Z)$ the symplectic eigenvalues of $Z$. In the
normal modes with quadratures $S^{-1}\boldsymbol R$, $H$ is a sum of oscillators of frequencies $\nu_j$ and
ground energy $\sum_j \nu_j/2$, and $\Tr\Pi_\tau=\prod_j[2\sinh(\tau \nu_j/2)]^{-1}$. The normalized Gibbs state has covariance $S\coth(\tau D/2)S^{\top}$, with inverse
$\Theta_\tau=S^{-\top}\tanh(\tau D/2)S^{-1}$, and its overlap with a
pure Gaussian product is largest at zero displacement. The pure--mixed
overlap formula of Sec.~\ref{sm:fidelityprogram}, applied to this
normalized Gibbs state, gives
$\Tr(\Pi_\tau\psi_\Gamma)=\Tr\Pi_\tau\,\det[(\Theta_\tau^{-1}+\Gamma)/2]^{-1/2}$
for centered $\psi_\Gamma$, and hence
\begin{align}
 b&=\frac{h}{\displaystyle\min_{\Gamma\ {\rm pure\ product}}
       \sqrt{\det(\id+\Theta_\tau^{1/2}\Gamma\Theta_\tau^{1/2})}},\nonumber\\
 b&\le \bar b_\tau:=\frac{h}{\sqrt{1+m(\Theta_\tau)}},\qquad
 h=\prod_{j=1}^d\operatorname{sech}(\tau \nu_j/2).
 \label{eq:momentfilterexplicit}
\end{align}
Here $\Gamma=\Gamma_A\oplus\Gamma_B$ and
$m(\Theta_\tau)=\min_{\Gamma\ {\rm pure\ product}}\Tr(\Theta_\tau\Gamma)$.
Only the principal blocks $\Theta_{\tau,A}$ and $\Theta_{\tau,B}$ of
$\Theta_\tau$ on $A$ and $B$ enter $\Tr(\Theta_\tau\Gamma)$, and Williamson's theorem
\cite{Williamson,SerafiniBook} gives
$m(\Theta_\tau)=2\sum_{j=1}^{d_A}\nu_j(\Theta_{\tau,A})
+2\sum_{j=1}^{d_B}\nu_j(\Theta_{\tau,B})$; this is four times the sum of
the two local quadratic ground energies. The bound $b\le\bar b_\tau$ uses
$\det(\id+X)\ge1+\Tr X$ for $X=\Theta_\tau^{1/2}\Gamma\Theta_\tau^{1/2}\ge0$. One may thus choose
$\bar b=\bar b_\tau$, which requires only symplectic eigenvalues, or $\bar b=b$
for a tighter bound.

\emph{Validity against collective competitors.}
Equation~\eqref{eq:generalmult} at $\alpha=\infty$ gives $a_\infty((\Pi_\tau^{1/2})^{\otimes k})^2=b^k$, so by the argument above $b^k$ is the largest expectation of $\Pi_\tau^{\otimes k}$ in a product vector across $A^k{:}B^k$, even one with arbitrary correlations among the local
copies. A unit vector of Schmidt rank at most $M$ is a sum of at most $M$ product
vectors whose coefficients, by Cauchy--Schwarz, have sum at most
$\sqrt M$. Applying $(\Pi_\tau^{\otimes k})^{1/2}$ and then the triangle inequality
bounds its expectation by $M b^k$.
Convexity and trace-norm continuity extend the bound to $\SN_M$,
a class that tracing out whole copies also preserves.

For $s\ge0$ set $Y_s=(\id+s\Pi_\tau)^{\otimes n}$. Expanding the tensor
product gives terms $s^{|J|}\Pi_\tau^{\otimes J}\otimes\id$ for subsets $J$ of the copies, each with expectation at most $M(sb)^{|J|}$ in $\sigma$, so $\Tr(\sigma Y_s)\le M(1+s\bar b)^n$. On the other side, convexity gives
$(1+sx)^{-1}\le1-sx/(1+s)$ for $0\le x\le1$. Applied to $\Pi_\tau$ on each copy, with $\Tr(\rho\Pi_\tau)\ge p$, this yields
$\Tr(\rho^{\otimes n}Y_s^{-1})\le[1-sp/(1+s)]^n$.
The Cauchy--Schwarz inequality for the trace norm, applied to
$(\sqrt{\rho^{\otimes n}}Y_s^{-1/2})(Y_s^{1/2}\sqrt\sigma)$, gives
$\Fid(\rho^{\otimes n},\sigma)\le
\Tr(\rho^{\otimes n}Y_s^{-1})\Tr(\sigma Y_s)$, and
minimizing $(1+s\bar b)[1-sp/(1+s)]$ over $s\ge0$ yields
Eq.~\eqref{eq:momentfilterc}.

\emph{Covariances that give a strict bound.}
Suppose $V$ is incompatible with every separable state, that is, no pure product covariance lies below $V$. Separation from the
closed convex set of separable covariances gives a quadratic witness
$Z\ge0$ with $\Tr(ZV)<\inf_{\Gamma\ {\rm pure\ product}}\Tr(Z\Gamma)$
\cite{WernerWolfBound,HyllusEisert}, and adding a sufficiently small
multiple of $\id$ makes $Z$ positive definite while preserving the
strict inequality. For this $Z$, the Williamson formula for $\Theta_\tau$ and $x-x^3/3\le\tanh x\le x$ for $x\ge0$ bound $\Theta_\tau$
between $(1-O(\tau^2))\tau Z/2$ and $\tau Z/2$ in matrix order.
As $m$ is monotone and homogeneous, $m(\Theta_\tau)=\tau m(Z)/2+O(\tau^3)$, where $m(Z)$ is the infimum in the witness inequality; with $h=1-O(\tau^2)$, this gives
$\bar b_\tau=1-\tau m(Z)/4+O(\tau^2)$. Since
$p=1-\tau\Tr(ZV)/4+O(\tau^2)$, every sufficiently small $\tau>0$ gives
$p>\bar b_\tau$. Conversely, every covariance above some pure product
covariance has a separable Gaussian realization, obtained by random
local displacements of that product.

\end{document}